\documentclass[referee]{aa}

\usepackage{times}
\usepackage{graphicx}
\usepackage{xspace}
\usepackage{epsfig}
\usepackage{natbib}
\usepackage{rotating}
\usepackage{dcolumn}

\begin{document}

\title{A Chandra and Spitzer census of the young star cluster \\ in the reflection nebula NGC\,7129}


\author{B. Stelzer \inst{1} \and A. Scholz \inst{2}}

\offprints{B. Stelzer}

\institute{INAF - Osservatorio Astronomico di Palermo,
  Piazza del Parlamento 1,
  I-90134 Palermo, Italy \\ \email{B. Stelzer, stelzer@astropa.unipa.it} \and
  SUPA, School of Physics \& Astronomy, University of St. Andrews, North Haugh, St.Andrews KY\,16\,9SS, United Kingdom} 

\titlerunning{Paper Test}

\date{Received $<$date$>$ / Accepted $<$date$>$}

\abstract
{The reflection nebula NGC\,7129 has long been known to be a site of recent star formation
as evidenced, e.g., by the presence of deeply embedded protostars and HH objects. However, studies of the
stellar population produced in the star formation process have remained rudimentary. A major step forward
was made with recent Spitzer imaging of the region. 
}
{This study represents the next step towards a systematic assessment of the pre-main sequence population
in NGC\,7129. Completeness of the pre-main sequence sample is necessary for studying key features that 
allow to understand the star forming process, such as disk evolution, dynamical evolution and mass function. 
At a presumed age of $\sim 3$\,Myr, NGC\,7129 is in the critical range where disks around young stars disappear.  
}
{We make use of X-ray and IR imaging observations to identify the pre-main sequence stars in NGC\,7129. 
We define a sample of Young Stellar Objects based on color-color diagrams composed from IR photometry between
$1.6$ and $8\,\mu$m, from 2\,MASS and Spitzer, and based on X-ray detected sources from a Chandra observation. 
}
{This sample is composed of $26$ Class\,II and $25$\,Class\,III candidates.  
It has been selected from infrared sources in the Chandra field 
($287$ objects with photometry in all four Spitzer/IRAC bands, $811$ objects with near-IR photometry) 
and the $59$ X-ray sources detected with Chandra. The sample is estimated to be complete down to $\sim 0.5\,{\rm M_\odot}$. 
The most restricted and least biased sub-sample of pre-main sequence stars is composed of lightly absorbed
($A_{\rm V} < 5$\,mag) stars in the cluster core. This sample comprises $7$ Class\,II and $14$ Class\,III sources,
it has a disk fraction of $33^{+24}_{-19}$\,\%,
and a median X-ray luminosity of $\log{L_{\rm x}}\,{\rm [erg/s]} = 30.3$.
}
{
Despite the various uncertainties related to the sample selection, absorption, mass distribution, distance and,
consequently, the computation of disk fraction and X-ray luminosities, 
the data yield consistent results. In particular, we confirm the age of $\sim 3$\,Myr for the NGC\,7129 cluster. The 
derived disk fraction is similar to that of $\sigma$\,Orionis, smaller than that of Cha\,I ($\sim 2$\,Myr), 
and larger than that of Upper Sco ($5$\,Myr). The X-ray luminosity function is similar to that of NGC\,2264
($2$\,Myr) but fainter than that of the Orion Nebula Cluster ($1$\,Myr). 
The pre-main sequence census should be further refined and extended with optical photometric and spectroscopic
searches for cluster members.
}

\keywords{X-rays: stars -- Infrared: stars -- stars: pre-main sequence, formation}

\maketitle

\section{Introduction}\label{sect:intro}

Young stars undergo various phases on their way to the main-sequence. 
Their different evolutionary stages are recognized by specific observational features, such as
excess emission over the expectation for a stellar photosphere that arises from circumstellar matter. 
Initially, Young Stellar Objects (YSOs) are embedded in envelopes of cold and dusty material
 that emit prominently at infrared (IR) and sub-millimeter wavelengths. The youngest, protostellar
phases are termed Class\,0 and Class\,I in a classification scheme based on the spectral energy
distribution (SED) \citep{Lada87.1}. 
As their evolution proceeds, disk accretion and reprocessing of stellar light in the disk 
become the identifying feature and the YSO is of Class\,II. 
Finally, when all circumstellar matter has dispersed or accreted onto the star a Class\,III object 
is left that is unconspicuous at IR bands as its SED is purely photospheric. 

Low sensitivity of mid-IR detectors have long hampered systematic searches for the signatures of 
circumstellar matter revealing young stars in their accretion and/or embedded phases. The launch of 
{\em Spitzer} has dramatically increased the census of YSOs in various star forming regions
\citep[see e.g.][and references therein]{Evans09.1}.
However, as suggested above, IR observations have no potential for discriminating young stars
after the disk has dispersed from evolved stars. Therefore, a YSO census based on IR data alone
is highly biased and other techniques are needed to reveal the Class\,III population. Diskless
stars can, indeed, be identified on basis of their strong X-ray emission, a prevalent characteristic 
for the whole pre-main sequence (pre-MS) evolution. X-ray and IR observations provide thus complementary 
tools to identify the complete pre-MS population. Knowing all members
of a given star forming region is essential for understanding fundamental problems such as 
disk evolution, dynamical evolution and the initial mass function. 

We present here a {\em Chandra} X-ray and {\em Spitzer} IR census of the star cluster in 
NGC\,7129. According to \cite{Racine68.1}, 
the NGC\,7129 reflection nebula in Cepheus is located at a distance of $\sim 1$\,kpc. 
\cite{Shevchenko89.1} published a value of $1260 \pm 50$\,pc. Both distance estimates are
based on optical photometry of few stars and ought to be considered quite uncertain.  
NGC\,7129 is dominated by the young B-type stars BD\,+65\,1637 (SVS\,7) and BD\,+65\,1638 (SVS\,8)
in a cavity surrounded 
by dense molecular ridges that form the interface with the molecular cloud \citep{Miskolczi01.1}. 
A third Herbig Be star, LkH$\alpha$\,234 (SVS\,12), is located on the eastern end of the ridge. 
SVS\,12 was presumably formed in the ridge following
compression of material \citep{Bechis78.1}. 
Further signs of very recent star formation in NGC\,7129 are 
deeply embedded protostars identified in IR images \citep{Weintraub94.1, Cabrit97.1}, 
and a large outflow initially thought to originate from SVS\,12, 
but later assigned to an embedded pre-MS object IRS\,6 located $3^{\prime\prime}$ of the
Herbig star \citep{Cabrit97.1, Fuente01.1}.  
The region also contains the probably youngest intermediate-mass object known at present,
the Class\,0 star FIRS-2 \citep{Eiroa98.1, Fuente05.1}. 
Several Herbig-Haro outflows \citep{Hartigan85.1, Strom86.1, Miranda93.1} and a 
cluster of low-mass pre-MS stars are associated with NGC\,7129. 
The most recent census of optical emission line stars counts $22$ objects \citep{Magakian04.1}. 

{\em Spitzer} imaging observations for NGC\,7129 have been reported by \cite{Gutermuth04.1}
and \cite{Muzerolle04.2}. 
With a combination of IR photometry from ground-based observations, 2\,MASS and IRAC 
a total of $84$ objects with circumstellar disks
could be identified, half of them in the $0.5$\,pc wide cluster core. The longer wavelength images 
obtained with MIPS are dominated in the central region of NGC\,7129 by nebular emission. 
Therefore, MIPS could detect YSOs only in the outer parts of the reflection
nebula. The discovery of over $10$ protostars in the outskirts of the cluster led \cite{Muzerolle04.2}
to suggest that the area of active star formation in NGC\,7129 extends over a large area of $\sim 3$\,pc.

NGC\,7129 has now been studied at virtually all wavelengths from the sub-millimeter \citep{Font01.1}
to optical bands but a systematic investigation of X-ray emission from the pre-MS population
is absent from the literature. We close this important gap with a {\em Chandra} observation
of NGC\,7129 that we combine with {\em Spitzer} photometry and data at other wavelengths from 
the literature for the few previously known cluster members. 
This new assessment of the cluster population is a solid basis for future optical photometric and spectroscopic
observations with the aim to confirm the new members. 
Sect.~\ref{sect:chandra} describes the X-ray data analysis.   
We give a brief description of the {\em Spitzer} photometry provided to us by R.Gutermuth
(Sect.~\ref{sect:spitzer}) and of the optical catalog of NGC\,7129 (Sect.~\ref{sect:optical}). 
We present our selection
criteria for YSOs and the resulting census of NGC\,7129 members in Sect.~\ref{sect:census}. 
In Sect.~\ref{sect:results} the results are presented. We discuss selected SEDs, derive the disk
fraction and the X-ray luminosity function for the pre-MS cluster, and add a note on the X-ray
emission from protostars. 
A summary and conclusions are given in Sect.~\ref{sect:conclusions}.

\section{{\em Chandra} data analysis}\label{sect:chandra}

A $22$\,ksec long {\em Chandra} observation targeting the Herbig star SVS\,12 
was carried out on Mar 11, 2006 (start of observation UT 14h29m18s). The observation was
obtained within a survey for X-ray emission from Herbig stars, and the data for the three 
B-type stars in the field-of-view were presented by \cite{Stelzer09.1}. 
The prime instrument for this observation was ACIS-S3 because of the higher sensitivity
at low energies of the ACIS spectroscopic array with respect to the imaging array. 
ACIS-S2 and all four ACIS-I chips were turned on as well. 
 
\begin{figure*}
\begin{center}
\includegraphics[width=18cm]{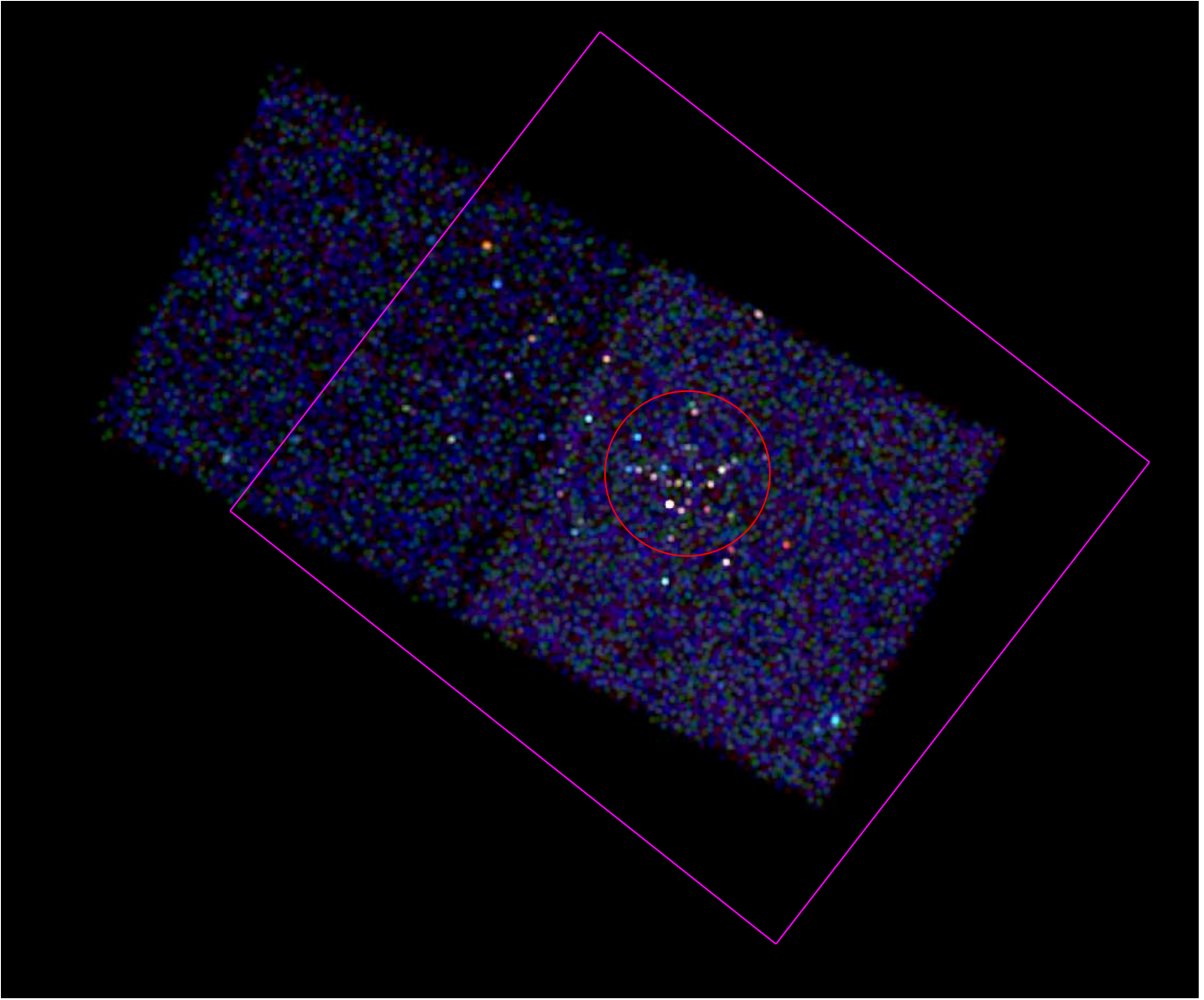}
\caption{{\em Chandra} false-color image of NGC\,7129 (data shown is from ACIS-S2 and ACIS-S3 chips).  
The large rectangle is the IRAC field-of-view (circa $12.6^\prime \times 14.5^\prime$), and the circle shows the
$\sim 1.7^{\prime}$ cluster core as defined by \protect\cite{Gutermuth04.1}. Evidently, the
bulk of X-ray sources is concentrated in the area of the cluster core.}
\label{fig:acis_fov}
\end{center}
\end{figure*}

SVS\,12 
lies at the heart of the NGC\,7129 reflection nebula giving serendipitously 
access to study for the first time the X-ray emission from the associated young star cluster. 
The cluster core was estimated by \cite{Gutermuth04.1} from $K_{\rm s}$ band star counts 
to comprise a $0.5$\,pc radius,
corresponding to less than $2^{\prime}$ at the distance of $1$\,kpc. Therefore,
most X-ray sources are expected to be located on a single ACIS chip. 
We analysed only data from ACIS-S3 and ACIS-S2 that overlap (partially)
with the {\em Spitzer} survey. 
Fig.~\ref{fig:acis_fov} shows the 
Chandra ACIS-S2 and ACIS-S3 image with the Spitzer/IRAC field overlaid. 
The cluster core of $\sim 1.7^{\prime}$ radius \citep{Gutermuth04.1} is marked as red circle.

The data analysis was performed with the CIAO software 
package\footnote{CIAO is made available by the CXC and can be downloaded 
from \\ http://cxc.harvard.edu/ciao/download/} version 4.0. 
We started our analysis with the level\,1 events file provided by the
{\em Chandra} X-ray Center (CXC). 
In the process of converting the level\,1 events file to a level\,2 events file
for each of the observations we performed the following steps: 
We removed the pixel randomization which is automatically applied by the CXC pipeline
in order to optimize the spatial resolution. 
We filtered the events file for event grades
(retaining the standard grades $0$, $2$, $3$, $4$, and $6$), 
and applied the standard good time interval file. 
Events flagged as cosmic rays were not removed in our analysis. In principle, such events 
can lead to the detection of spurious sources. However, if identified on the position of a
bright X-ray source, the flag is often erroneous (as a result of the event pattern used for 
the identification of cosmic rays). 

Source detection was carried out in the $0.3-10$\,keV band for the ACIS-S3 and ACIS-S2 chips 
with the {\sc wavdetect} algorithm \citep{Freeman02.1} using an image with spatial resolution 
of $0.5^{\prime\prime}$/pixel and a congruent, monochromatic exposure map for $1.5$\,keV.
The {\sc wavdetect} algorithm correlates the data with a mexican hat function
to search for deviations from the background. This method 
is well suited for separating closely spaced point sources.  
We used wavelet scales between $1$ and $8$ in steps of $\sqrt{2}$. 
The detection significance was set to $10^{-6}$ to avoid spurious detections.

We cleaned the resulting source list on basis of the individual signal-to-noise (S/N) 
of each detection. First, the point-spread-function (PSF) was computed for each X-ray 
position. A circular source photon extraction region was defined as the area that contains 
$90$\,\% of the PSF. 
The background was
extracted individually from a squared region centered on the source extraction area and 
several times larger than the latter one.
If detected X-ray sources are in the selected background area a circular area around them 
was eliminated from the background region. The S/N was computed from the counts summed in
the source and background areas, respectively, after applying the appropriate 
area scaling factor to the background counts. In general, the background is negligibly low.  
We removed all objects with S/N $< 3$ as likely spurious from the {\sc wavdetect} X-ray source list. 
In Table~\ref{tab:xraytab} we provide the final list of $59$ X-ray sources. 

\begin{figure*}
\begin{center}
\parbox{18cm}{
\parbox{6cm}{
\includegraphics[width=6cm]{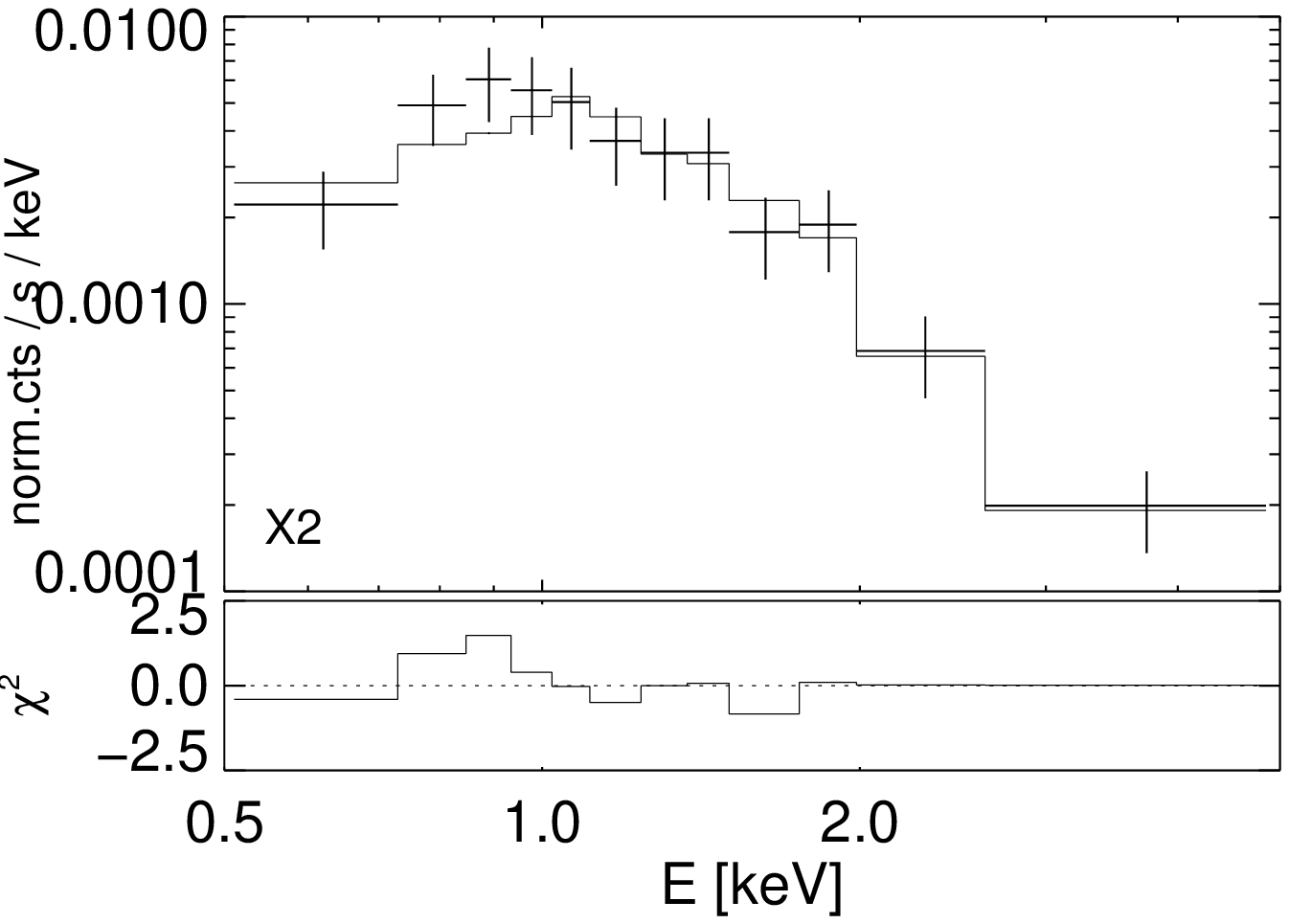}
}
\parbox{6cm}{
\includegraphics[width=6cm]{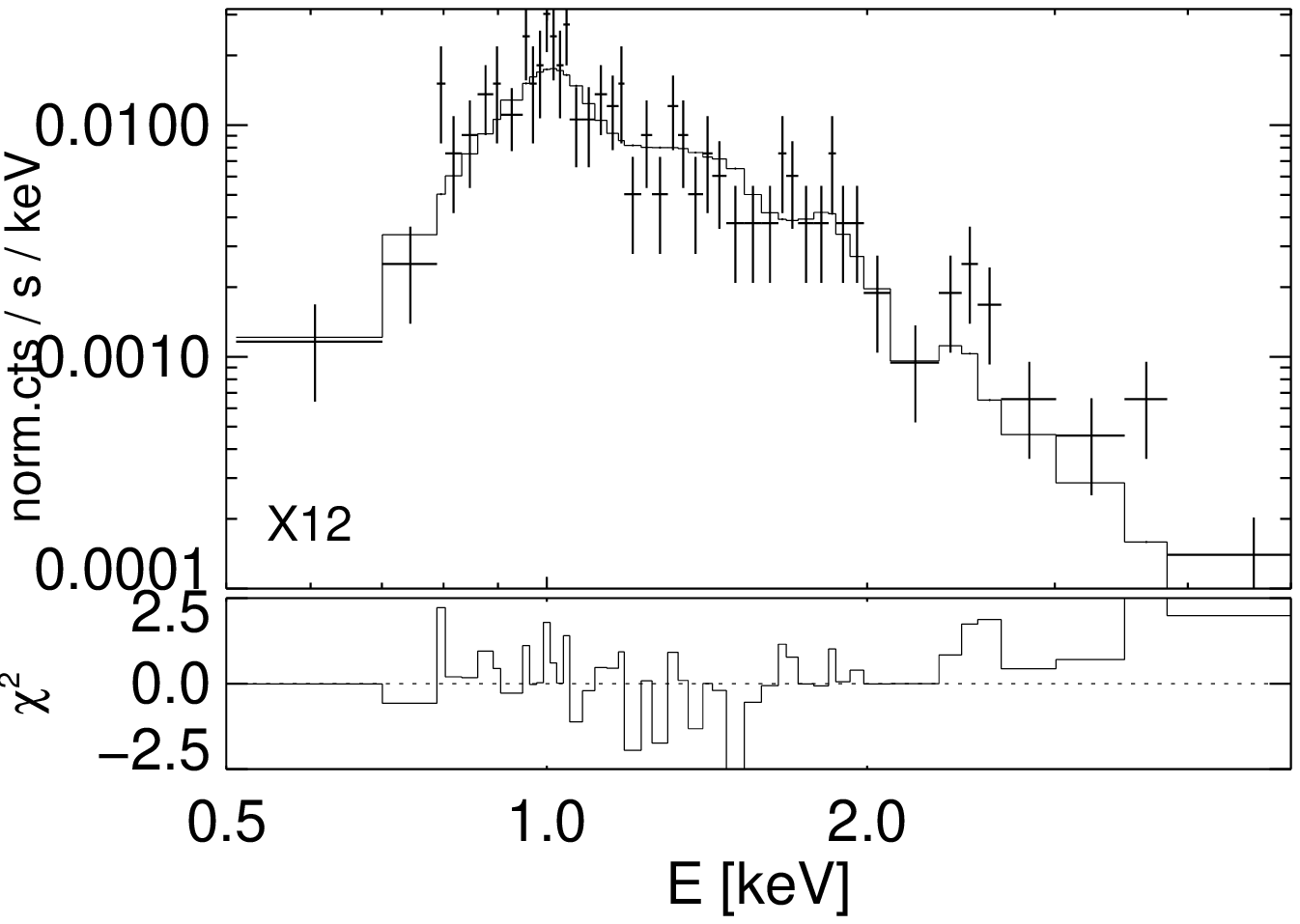}
}
\parbox{6cm}{
\includegraphics[width=6cm]{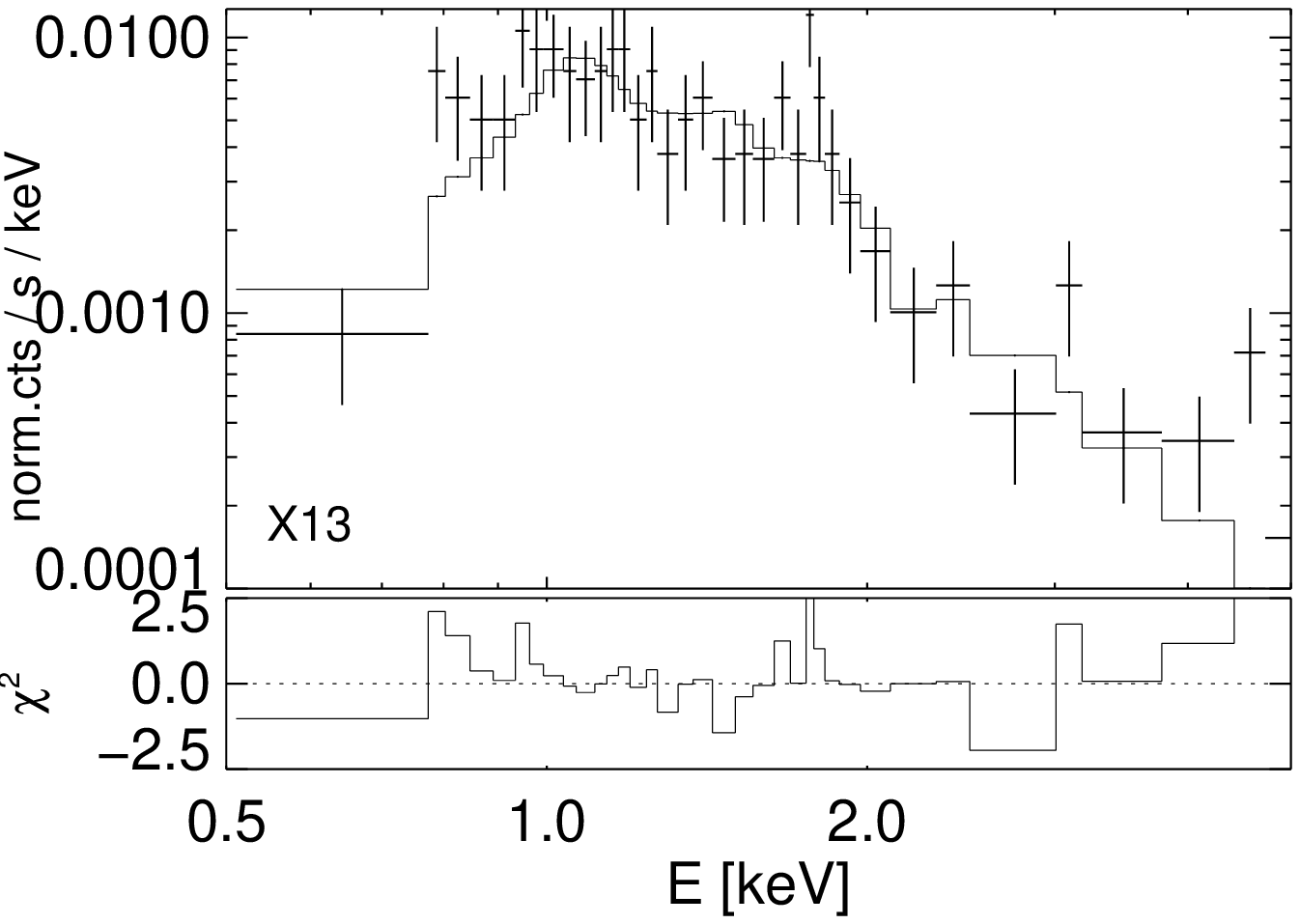}
}
}
\parbox{18cm}{
\parbox{6cm}{
\includegraphics[width=6cm]{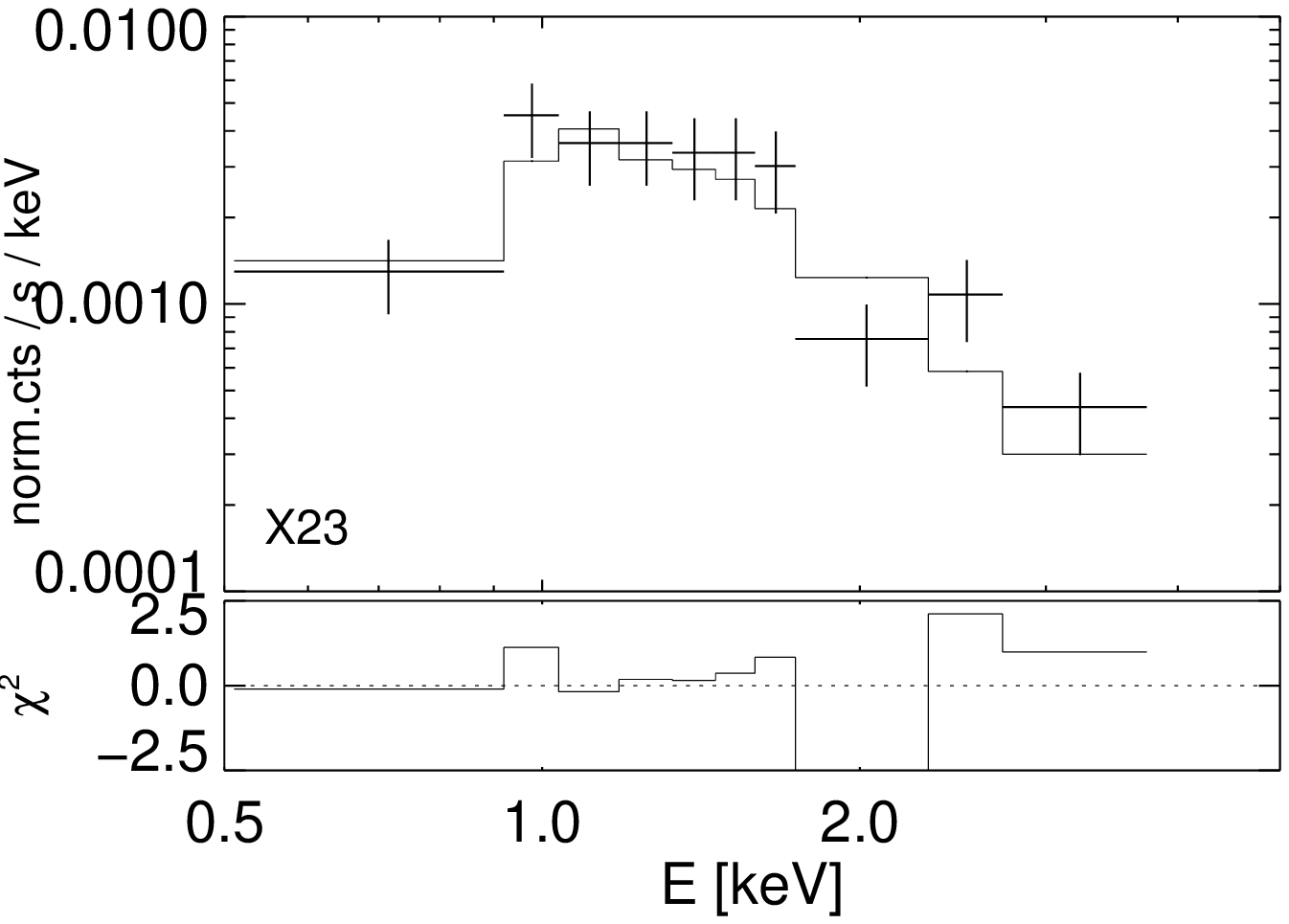}
}
\parbox{6cm}{
\includegraphics[width=6cm]{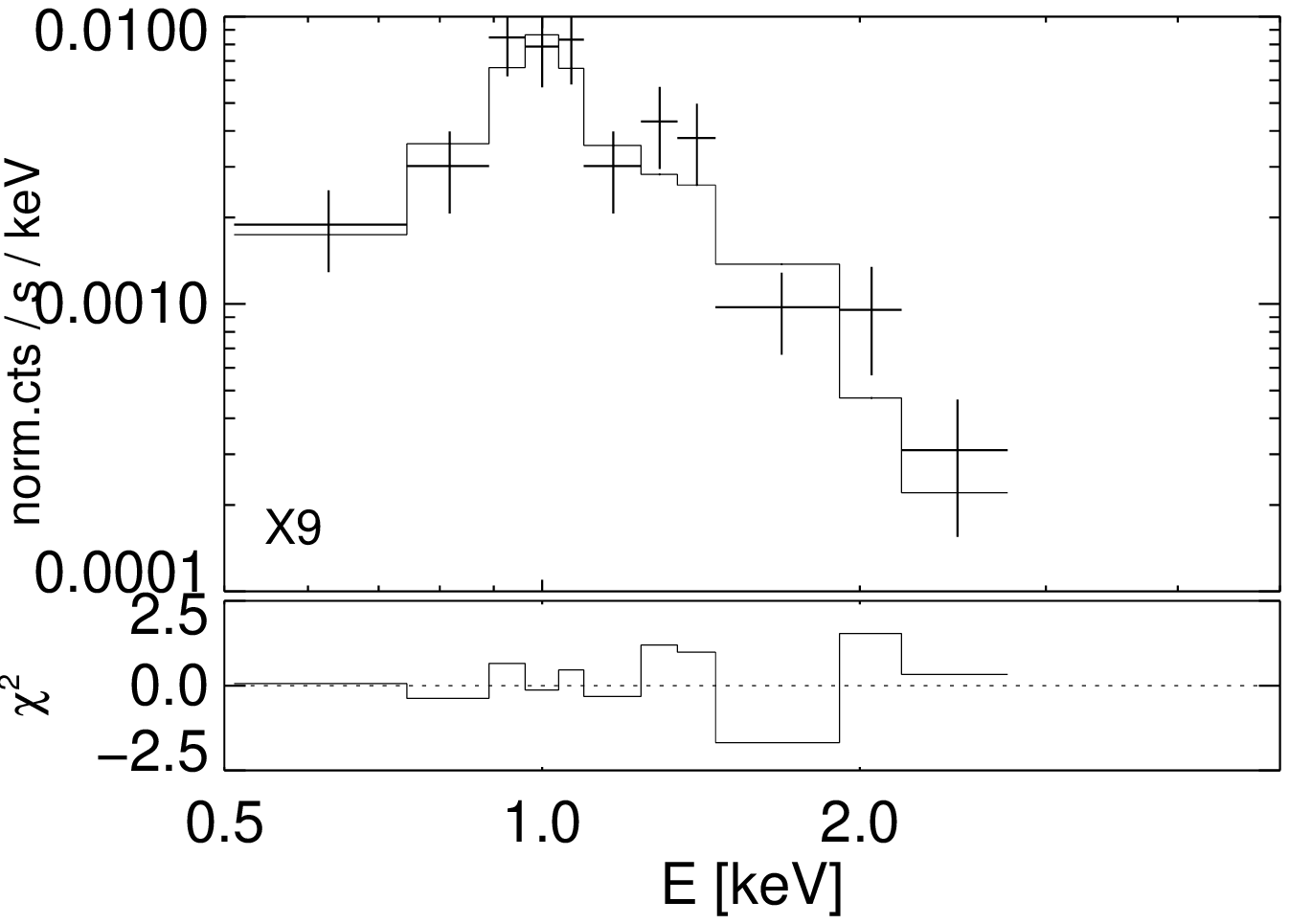}
}
}
\caption{X-ray spectra for the five brightest sources in NGC\,7129, best-fitting absorbed one-temperature
model and $\chi^2$ residuals.}
\label{fig:spectra}
\end{center}
\end{figure*}

There are only $5$ X-ray sources with more than $100$ counts in the $0.3-10$\,keV band 
(sources NGC\,7129-S3-X2,..-X12,...-X13,...-X23 and NGC\,7129-S2-X9) 
such that a detailed spectral analysis is not feasible for most objects. 
We have extracted and analysed the X-ray spectra for the five brightest sources mentioned above.
An individual response matrix and auxiliary response were extracted for each
of them using standard CIAO tools. Each spectrum was binned to a minimum of $5$ or more counts per bin
depending on the photon statistics. As the background of ACIS is very low ($<1$\,count in the source
extraction area) it can be neglected. We fitted each spectrum in the XSPEC\,12.4.0 environment 
with a one-temperature thermal model spectrum \citep{Raymond77.1} subject to photo-absorption ({\sc wabs * apec}). 
More complex spectral models are not considered due to the low statistics of these data. 
Fig.~\ref{fig:spectra} shows the data with the best fit spectral models and $\chi^2$ residuals. 
For NGC\,7129-S3-X12 and NGC\,7129-S3-X13, that are partially overlapping, we have used the photon extraction regions 
defined by \cite{Stelzer09.1}. All five spectra are compatible with $\log{N_{\rm H}}\,{\rm [cm^{-2}]} < 22$
and $kT$ in the range of $1...2$\,keV. 

\begin{figure}
\begin{center}
\epsfig{file=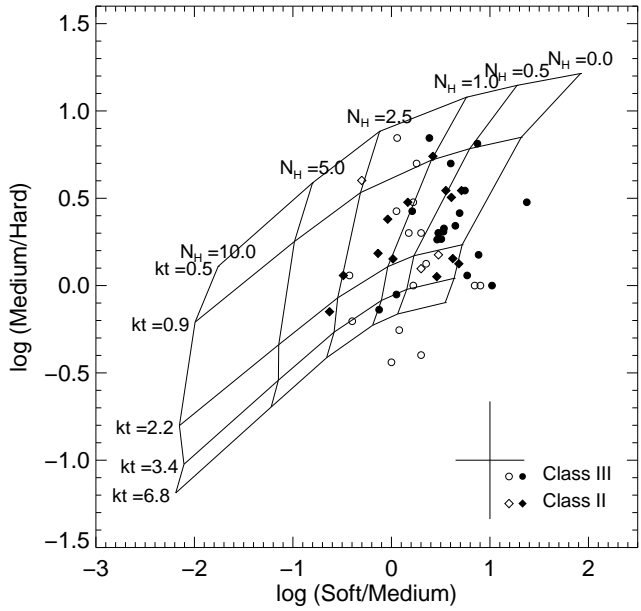,width=9cm}
\caption{X-ray hardness ratio diagram for the detected sources; see text 
in Sect.~\ref{sect:chandra} for a definition of the energy bands. The grid represents the predicted colors 
for isothermal Raymond-Smith models in a range of temperatures (in units of keV) and affected by different 
amounts of absorption (in units of $10^{22}\,{\rm cm^{-2}}$). 
The YSO classification from Sect.~\ref{sect:census} is anticipated here to show the trend for different
absorption between Class\,II and Class\,III stars. 
X-ray sources with more and less than $20$ net source counts, respectively, 
are shown with open and filled plotting symbols. A typical error bar is shown in the lower right.}
\label{fig:hrs}
\end{center}
\end{figure}

In cases of low photon statistics it is common practice to examine 
X-ray hardness ratios. 
We define three energy bands: soft  
($S = 0.2-1.7$\,keV), medium ($M = 1.7-2.8$\,keV) and hard ($H = 2.8-8.0$\,keV) from which we
construct two colors following \cite{Albacete07.1}, $S/M$ and $M/H$.
Fig.~\ref{fig:hrs} shows the hardness ratio diagram where a grid calculated from a one-temperature thermal model
with photo-absorption is overlaid. 
The majority of X-ray sources is compatible with a temperature
of $1...2$\,keV and moderate ($\leq 10^{22}\,{\rm cm^{-2}}$) absorption. 
For three of the five sources discussed in the previous paragraph, the $N_{\rm H}$ and $kT$ derived
from Fig.~\ref{fig:hrs} agree within the $90$\,\% confidence level with the values obtained
from the spectral fitting. For NGC\,7129-S3-X12 and NGC\,7129-S3-X13 the XSPEC fits yield a cooler and less absorbed spectrum
than the hardness ratios. These are the two stars with the best photon statistics and a $1-T$ approach
may not be appropriate. 

Column densities and temperatures 
inferred from X-ray hardness ratios can in principle be used to derive X-ray
luminosities of individual stars when the statistics are too poor for detailed spectral analysis. 
However, as seen above in the comparison to the spectral results for the
brightest stars, even the use of hardness ratios may be associated with large uncertainties.
Furthermore, not all data points lie in the region covered by physically reasonable models 
(see Fig.~\ref{fig:hrs}). For the calculation of X-ray luminosities of objects outside the grid 
we use the median temperature and absorbing column of the sample with $>20$ net
source counts, while for all objects on the grid we use their individual $N_{\rm H}$ and $kT$
estimates. Moreover, we anticipate in Fig.~\ref{fig:hrs} a trend of higher absorption for
Class\,II with respect to Class\,III sources. (The YSO classification scheme is described in 
Sect.~\ref{sect:census}.) Therefore, we compute separate medians for these two YSO groups.
We find for Class\,II a median of $N_{\rm H, C\,II} = 8.8 \cdot 10^{21}\,{\rm cm^{-2}}$ and for 
Class\,III of $N_{\rm H, C\,III} = 2.1 \cdot 10^{21}\,{\rm cm^{-2}}$. These differences can probably
be ascribed to additional absorbing material in the circumstellar environment of Class\,II stars. 
The median temperatures are $(kT)_{\rm C\,II} = 1.5$\,keV
and $(kT)_{\rm C\,III} = 1.9$\,keV for Class\,II and~III, respectively. A difference in $kT$ between these two groups
is not expected. We refrain from analysing its significance because the X-ray luminosity depends
very little on the temperature in this range. 

With the values of $N_{\rm H}$ and $kT$ described in the previous paragraph, 
a count-to-flux conversion factor is obtained for each X-ray source 
with PIMMS\footnote{The Portable Interactive Multi-Mission Simulator (PIMMS) is accessible at 
http://asc.harvard.edu/toolkit/pimms.jsp} from a one-temperature Raymond-Smith model. The corresponding 
X-ray luminosities are given in col.~9 of Table~\ref{tab:xraytab}. 
Evidently, these numbers are associated with large errors. 

As an alternative, we apply also a different method to derive the column density, by computing 
individual $N_{\rm H}$ estimates based on the $A_{\rm V}$ of each star using a
gas-to-dust conversion law of $N_{\rm H} = 1.8\,A_{\rm V}\,{\rm [mag]} \cdot 10^{21}\,{\rm cm^{-2}}$ 
\citep[e.g.][]{Predehl95.1}. 
Note that the $N_{\rm H}/A_{\rm V}$ ratio may vary for different star forming
environments as shown by \cite{Vuong03.1} but an assessment of its value in NGC\,7129 is
out of reach with the presently available data. 
The optical extinctions are obtained from dereddening of the objects in the $J$ vs. $J-H$ diagram,
and is, obviously, available only for X-ray sources with near-IR photometry. 
A comparison of the X-ray luminosity function derived with these two approaches is given 
in Sect.~\ref{subsect:results_xlf}. 
%
%

\section{Infrared catalog (Spitzer and 2\,MASS)}\label{sect:spitzer}

NGC\,7129 was observed by the {\em Spitzer} Space Telescope with IRAC in four
wavelength bands from $3.6$ to $8.0\,\mu$m and with MIPS at $24\,\mu$m. 
These observations are part of the {\em Spitzer} Young Stellar Cluster
Survey, initially presented by \cite{Megeath04.1} 
and further analysed by \cite{Gutermuth04.1}, \cite{Muzerolle04.2}, 
and \cite{Gutermuth09.1}.
Henceforth, we refer to the latter paper as G09. The Astronomical Observation Request (AOR)
number and dates of the {\em Spitzer} observations are 
AOR\,3655168 from 2003-12-24 and AOR\,3663616 from 2003-12-29.

Our analysis is based on a catalog comprising $JHK$ data from 2\,MASS
and IRAC and MIPS photometry kindly provided by R.Gutermuth prior to publication. 
The IRAC and MIPS fluxes have been measured by Gutermuth et~al. 
using the photometry tool PhotVis, which is based on DAOPHOT
routines ported to IDL. The magnitudes are corrected for aperture
losses and calibrated using large-aperture measurements of standard
stars. The {\em Spitzer} sources were visually inspected and
non-stellar contaminations rejected. 
To obtain near-IR counterparts to the {\em Spitzer} sources, G09 matched 
their positions with 2\,MASS using for the cross-identification radius 
a maximum of $1^{\prime\prime}$ in the case of IRAC and $1.3^{\prime\prime}$ in the case of MIPS.  
For more details on the IR
photometry, see \cite{Gutermuth04.1} and G09. 

The combined 2\,MASS and Spitzer catalog of NGC\,7129 contains $7139$ 
objects, of which $5389$ are detected in at least one {\em Spitzer} band and 
$287$ are detected in all four IRAC bands. The catalog includes the YSO
classification discussed in detail by G09;
see Sect.~\ref{subsect:census_g09}. 

%
%
\begin{figure*}[th]
\begin{center}
\parbox{18cm}{
\parbox{4.4cm}{
\epsfig{file=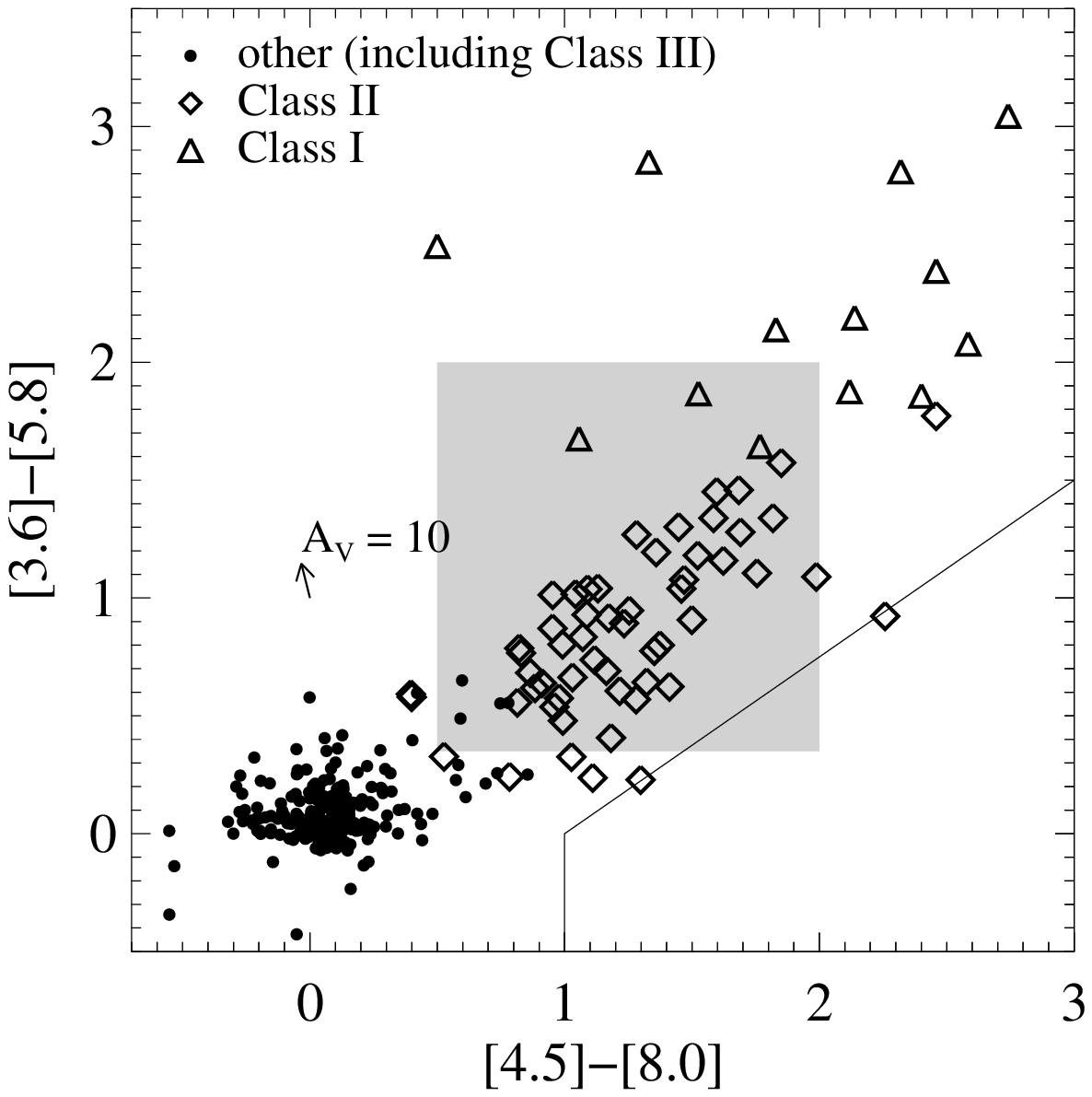,width=4.4cm}
}
\parbox{4.4cm}{
\epsfig{file=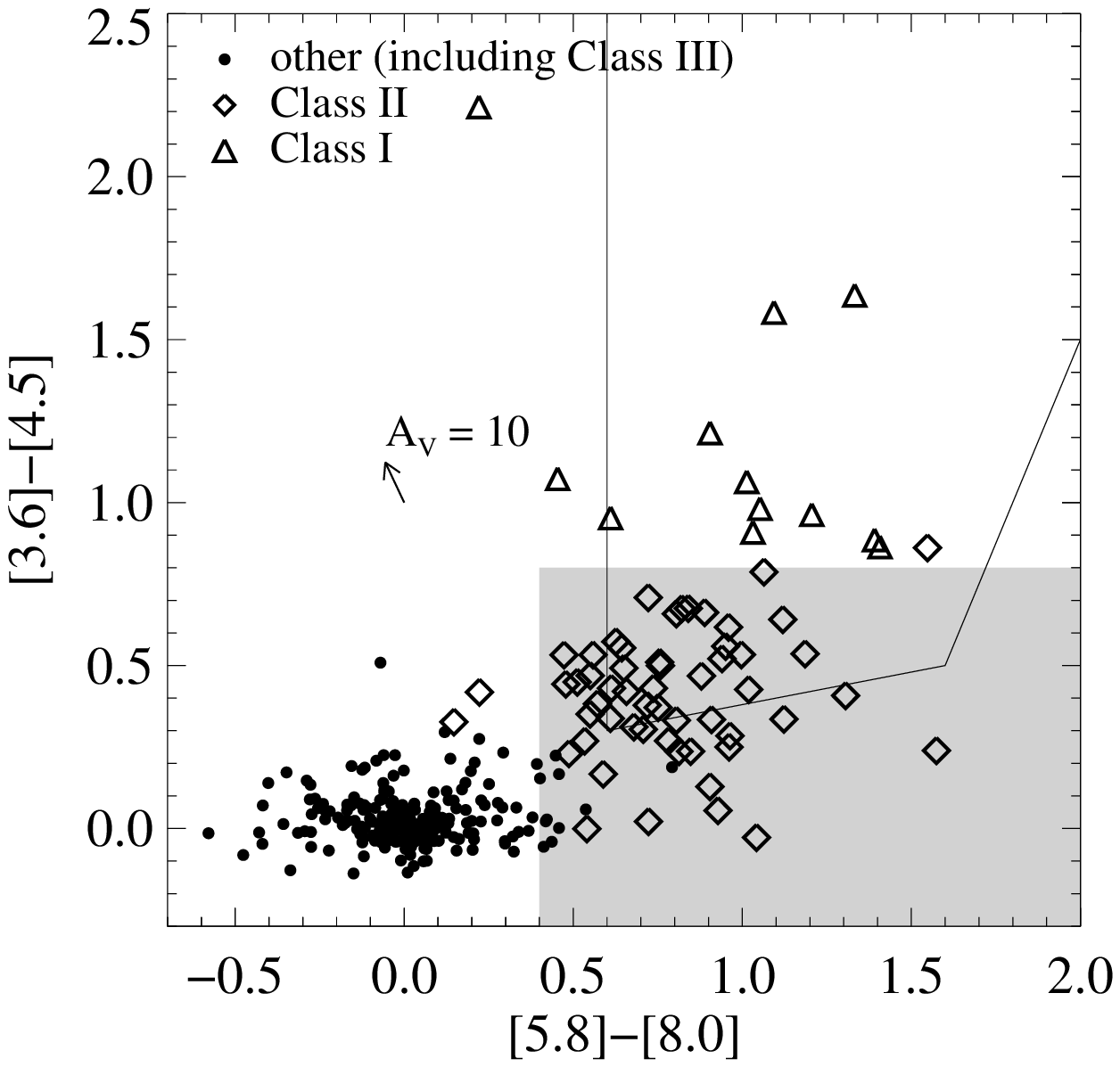,width=4.4cm}
}
\parbox{4.4cm}{
\epsfig{file=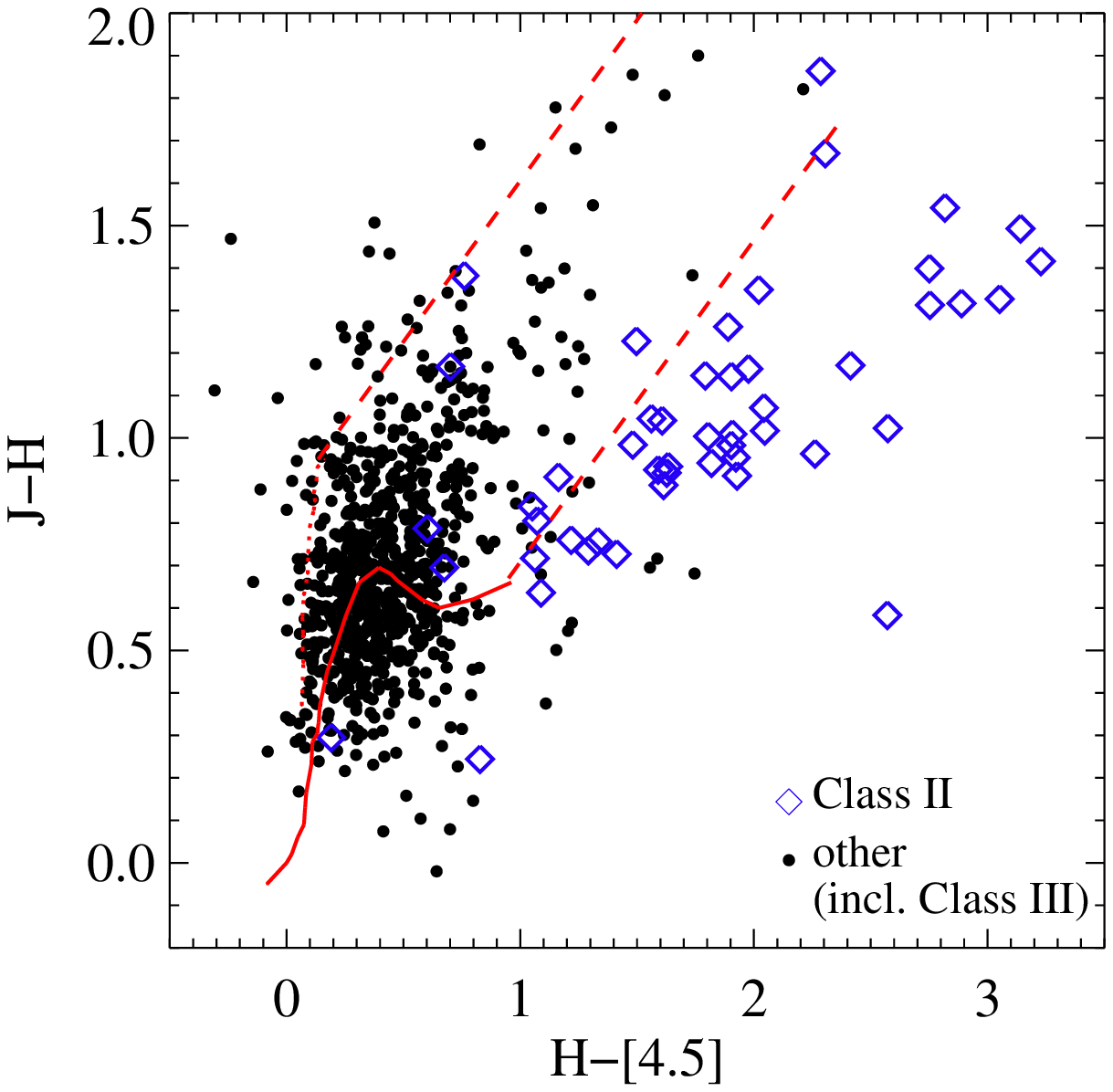,width=4.4cm}
}
\parbox{4.4cm}{
\epsfig{file=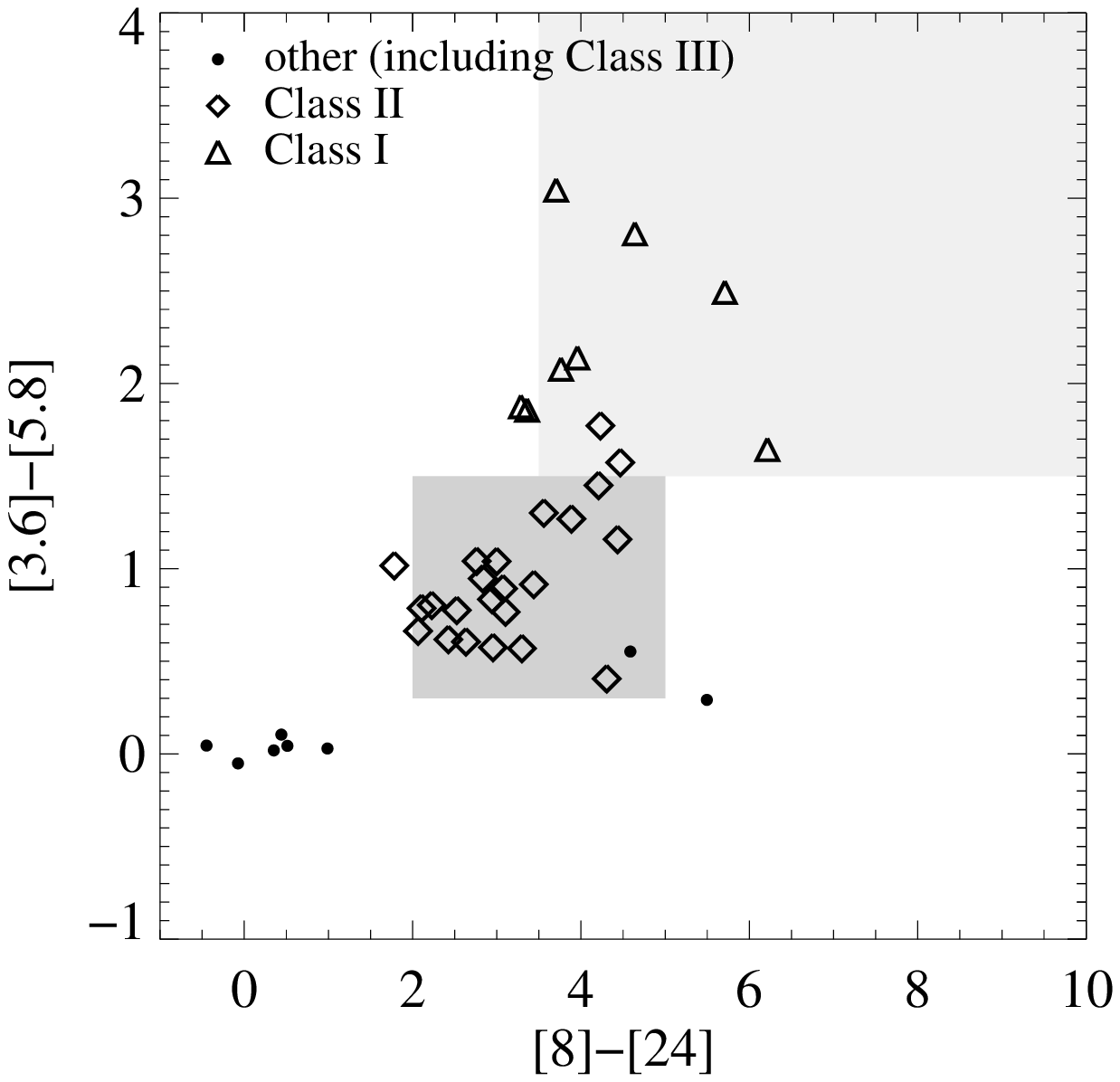,width=4.4cm}
}
}
\caption{Color-color diagrams for NGC\,7129 based on near-IR and mid-IR photometry. 
For these plots we used all {\em Spitzer} and 2\,MASS objects from the catalog described
in Sect.~\ref{sect:spitzer} that have photometry in the bands defining the displayed colors. 
The arrows denote reddening vectors with their length corresponding to $A_{\rm V} = 10$\,mag.  
The solid lines delineate the areas typical 
for star forming galaxies (below the line in the first panel) and for 
broad-line AGN (above the line in the second panel),  
while the shaded areas represent the typical location of Class\,II 
YSOs (dark grey) and of Class\,0/I YSOs (light grey in the rightmost diagram). 
Objects classified as Class\,II and Class\,0/I according to the criteria described in 
Sect.~\ref{subsect:census_ir} are highlighted with different plotting symbols: Class\,II (diamonds), Class\,0/I (triangles), and other IR sources including Class\,III (circles).}  
\label{fig:mir_ccds_all}
\end{center}
\end{figure*}

\section{Optical catalog}\label{sect:optical}

NGC\,7129 is poorly characterized in the optical.
We have compiled a list of studies that have identified pre-MS stars in the cluster. 
This list will be used to provide optical counterparts for the sample of YSOs that we define 
in Sect.~\ref{sect:census}. 

The bright B-type star SVS\,8, the two Herbig stars SVS\,7 
and SVS\,12, and evidence for some other emission line stars were reported by \cite{Herbig60.1}.
Several far-IR sources lack optical counterparts indicating that they are
deeply embedded in the reflection nebula. 
Most optial imaging surveys carried out in NGC\,7129 used narrow band filters to search for Herbig-Haro objects 
\cite[e.g.][]{Strom86.1, Eiroa92.2}. 
Three emission line stars are reported by \cite{Miranda93.1} together with four Herbig-Haro objects. 
Photometric measurements in the $VRI$ bands were presented by \cite{Magakian04.1} for
$22$ stars in the central region of the cluster, 
most of them newly discovered by means of slitless H$\alpha$ spectroscopy.

\section{YSO census in NGC\,7129}\label{sect:census}

Our census of the YSO population in NGC\,7129 is based on new X-ray and IR data
that yield complementary information. We consider as YSOs (1) those X-ray sources
that have IR properties compatible with young stars (Sect.~\ref{subsect:census_xdet}), 
and (2) those IR objects that
show evidence for emission in excess of the photospheric value irrespective of whether
they are detected in X-rays or not (Sect.~\ref{subsect:census_xuls}). The first group
comprises, in principle, all types of YSOs from Class\,0/I protostars to Class\,III objects, 
while the second group is biased against the identification of the diskless Class\,III objects. 
Our criteria for the classification of the YSOs on basis of IR photometry 
are described in Sect.~\ref{subsect:census_ir}
followed by the a description of the selected sample 
(Sects.~\ref{subsect:census_xdet} and~\ref{subsect:census_xuls}). We identify contaminating
foreground and background objects in Sect.~\ref{subsect:census_contam} and discuss the
completeness and spatial distribution of the sample in Sect.~\ref{subsect:census_spatial}. 
In Sect.~\ref{subsect:census_g09} our results are compared to a more complex selection 
process applied by G09.

\subsection{IR classification criteria for YSOs}\label{subsect:census_ir}

Color-color diagrams can be used to examine the evolutionary status of YSOs \citep[e.g.][]{Allen04.1}
and to discriminate contaminating background objects. 
Fig.~\ref{fig:mir_ccds_all} shows a selection of color-color 
diagrams based on IRAC photometry for the NGC\,7129 region. 
We determined the reddening vector for the IRAC bands with the extinction law of \cite{Indebetouw05.1}.  
Our first criterion to identify stars with circumstellar excess emission is based on the 
$[3.6]-[4.5]$ vs. $[5.8]-[8.0]$ diagram. 
We consider Class\,II candidates all objects with $1\,\sigma$ error bars contained in the region 
defined by \cite{Caramazza08.1} and shaded in Fig.~\ref{fig:mir_ccds_all} 
\cite[see e.g.][for previous similar definitions for Class\,II mid-IR colors given in the literature]{Megeath04.1, Hartmann05.1}.
We consider candidate Class\,0/I protostars all objects with $[3.6]-[4.5] > 0.8$. 
As outlined by \cite{White07.1}, the unambiguous distinction between 
Class\,II and Class\,0/I is not trivial and requires good SED coverage and information 
from spectroscopy. 
Detailed analyses of SEDs for young stars show that there is some overlap of the Class\,0/I and Class\,II 
populations as far as their IRAC colors are concerned \citep{Caramazza08.1}. In fact, 
it is evident from Fig.~\ref{fig:mir_ccds_all} that the Class\,0/I group may also contain reddened Class\,II stars.
Similarly, some protostars may be located below the chosen cutoff line. 
The unambiguous distinction of Class\,0/I and Class\,II for NGC\,7129 has to be postponed until spectroscopy 
is available. 

In a second step we examine the near-IR photometry. Generally, in near-IR color-color 
diagrams objects to the lower right
of the reddening band are considered to have circumstellar disks responsible for the red excess emission.
For the $J-H$ vs. $H - [4.5]$ diagram we interpolated the reddening law of \cite{Rieke85.1} 
to the $4.5\,\mu$m filter of IRAC. 
To exclude objects below the dwarf locus and likely from the background 
we impose as constraint $J-H > 0.65$.
We then added all objects above that line but below the reddening band in the $J-H$ vs. $H - [4.5]$ diagram 
to the list of Class\,II candidates. 
In principle, this diagram does not allow to distinguish
Class\,II from protostars. However, the fact that none of the IRAC selected
Class\,0/I candidates is in the $J-H$ vs. $H - [4.5]$ area shown in Fig.~\ref{fig:mir_ccds_all} makes
us confident that Class\,0/I make a very small contribution to the near-IR selected YSO sample.
In fact, only one of the $13$ Class\,0/I candidates selected from the IR catalog on basis of 
IRAC colors has $JH$ photometry,
and it has $H - [4.5] = 5.8$, thus lying outside the region plotted in the third panel 
of Fig.~\ref{fig:mir_ccds_all}.
(This object is outside the Chandra FOV and, therefore, not listed in our data tables.)
We conclude that most protostars are too faint for 2\,MASS. In fact, in the sample that we
use for scientific analysis (see Sect.~\ref{subsect:census_spatial}) there is none of the Class\,0/I sources
because we could not determine their mass. 

The sample from which our IR selected YSO candidates are drawn consists of the $287$ objects 
with photometry
in all four IRAC bands and the $811$ objects with photometry at $J$, $H$, and $[4.5]$. 
Our YSO classification results in $64$ Class\,II candidates and $13$ Class\,0/I candidates
across the field covered by IRAC. The remaining
IR sources are mostly not related to NGC\,7129 but they include also the Class\,III cluster members. 
In Sect.~\ref{subsect:census_xdet} we make use of the X-ray data to identify those latter ones.

We have not used MIPS data in our classification scheme because the majority of the objects 
have no $24\,\mu$m detection. However, as a cross-check we show in the rightmost diagram of 
Fig.~\ref{fig:mir_ccds_all} 
a color-color diagram with the reddest color available in the IR catalog, $[8]-[24]$. 
The Class\,II and the Class\,0/I regions defined by \cite{Muzerolle04.2} are highlighted with 
different grey-shades. The agreement between our individual classifications and the highlighted areas
is very good. There are only two Class\,III (or foreground) objects with very red $[8]-[24]$ color
such that the MIPS data is inconsistent with our classification from shorter wavelengths. 
These objects may have transition disks (Class\,II/III).

\subsection{YSO candidates among detected X-ray sources}\label{subsect:census_xdet}

The X-ray source list (Table~\ref{tab:xraytab}) 
was cross-correlated with the 2\,MASS point source catalog \citep{Cutri03.1}. 
We computed the median position offset between the X-ray sources and their 2\,MASS counterparts, 
and shifted the X-ray positions by this boresight value 
($\Delta \alpha = -0.22^{\prime\prime}$; $\Delta \delta = +0.07^{\prime\prime}$).
For our search radius of $1.5^{\prime\prime}$ the majority of the X-ray emitters ($47/59$) 
have 2\,MASS counterparts. Most ($43$) of these counterparts are found within 
$<0.5^{\prime\prime}$ of the X-ray position, and only one counterpart is at $>1^{\prime\prime}$
from the X-ray position.  
Then we searched for matches between the X-ray sources and the 2\,MASS positions
given in the combined 2\,MASS/Spitzer catalog introduced in Sect.~\ref{sect:spitzer}.
All but $9$ X-ray sources have 
a mid-IR counterpart in at least one {\em Spitzer} filter. 
For the cross-correlation of the X-ray sources with the 
optical positions from the catalogs described in Sect.~\ref{sect:optical} 
we allowed for a maximum matching radius of $2^{\prime\prime}$. This 
yielded $VRI$ photometry for nine X-ray sources. 
To this we add the optical photometry for the Herbig star SVS\,7  
from \cite{Hillenbrand92.1} and that for the B-type star SVS\,8  
from \cite{Racine68.1}.

Identifiers from published catalogs and the offset between X-ray and optical/IR sources 
are listed in Table~\ref{tab:counterparts_a}. 
A compilation of the optical and IR photometry for all X-ray sources in NGC\,7129
is given in Table~\ref{tab:counterparts_b}. 
We count $16$ Class\,II and $1$ Class\,0/I candidates among the X-ray detections.
The $30$ X-ray sources for which either 2\,MASS or IRAC photometry is complete and 
that are not classified Class\,0/I or~II are considered Class\,III candidates.
Another $12$ X-ray sources remain unclassified. 
In col.~13 of Table~\ref{tab:counterparts_b} we report the YSO classification for each
individual object. Class\,II sources with incomplete IRAC photometry and 
selected only on basis of near-IR data are labeled with
an asterisk. The names of the stars reported in this table go back to various literature sources. 
Some stars had been identified already in the near-IR survey by \cite{Strom76.1}. These stars
carry `SVS' numbers in col.~2 of Table~\ref{tab:counterparts_a}. 
Labels `HL85-N' and `HL85-S' refer to the northern and southern fields observed by \cite{Hartigan85.1}.
`MEG93' stands for \cite{Miranda93.1}, and `MMN' numbers are stars from \cite{Magakian04.1}.

\begin{figure*}
\begin{center}
\parbox{18cm}{
\parbox{6cm}{
\epsfig{file=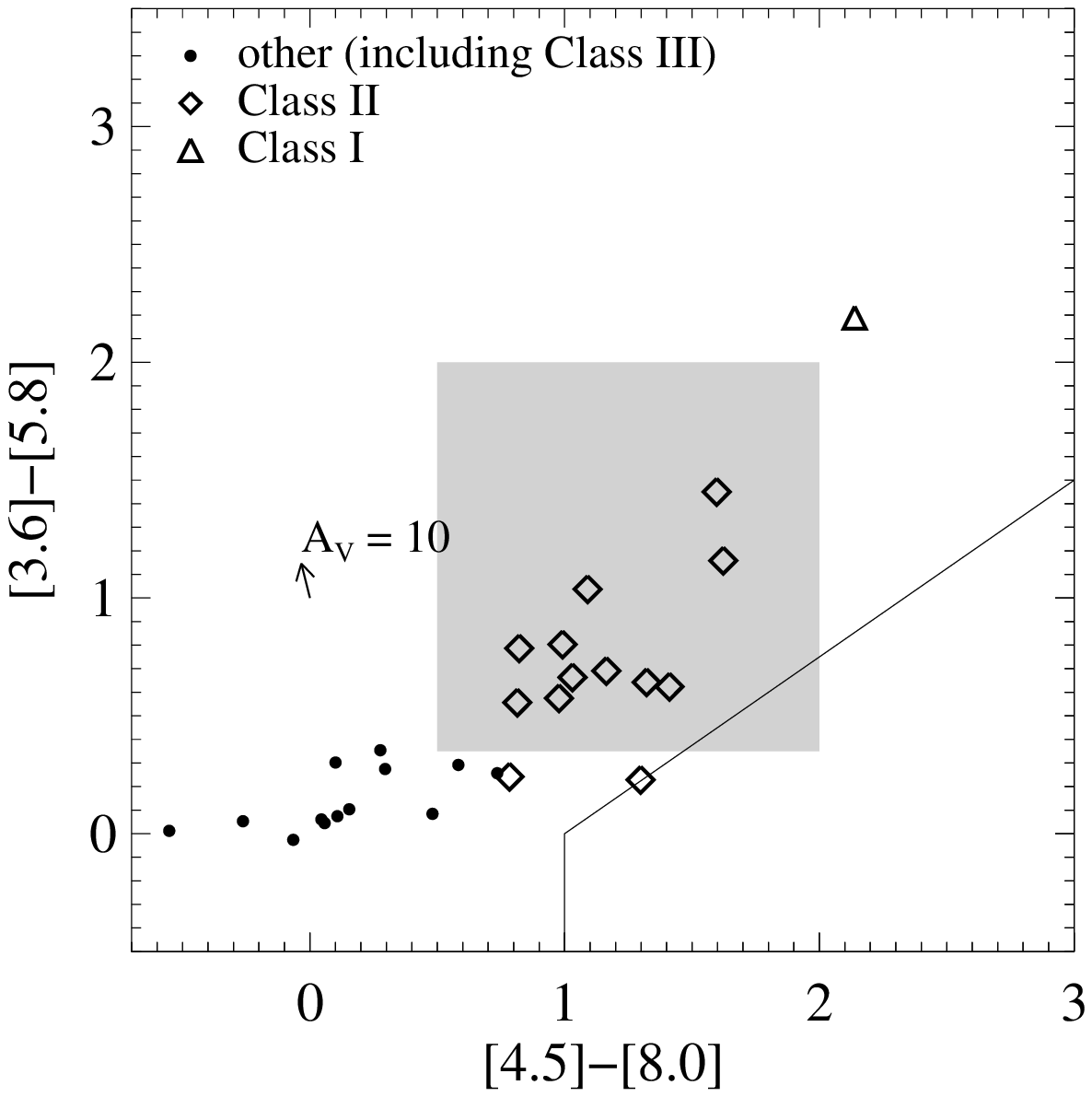,width=5.9cm}
}
\parbox{6cm}{
\epsfig{file=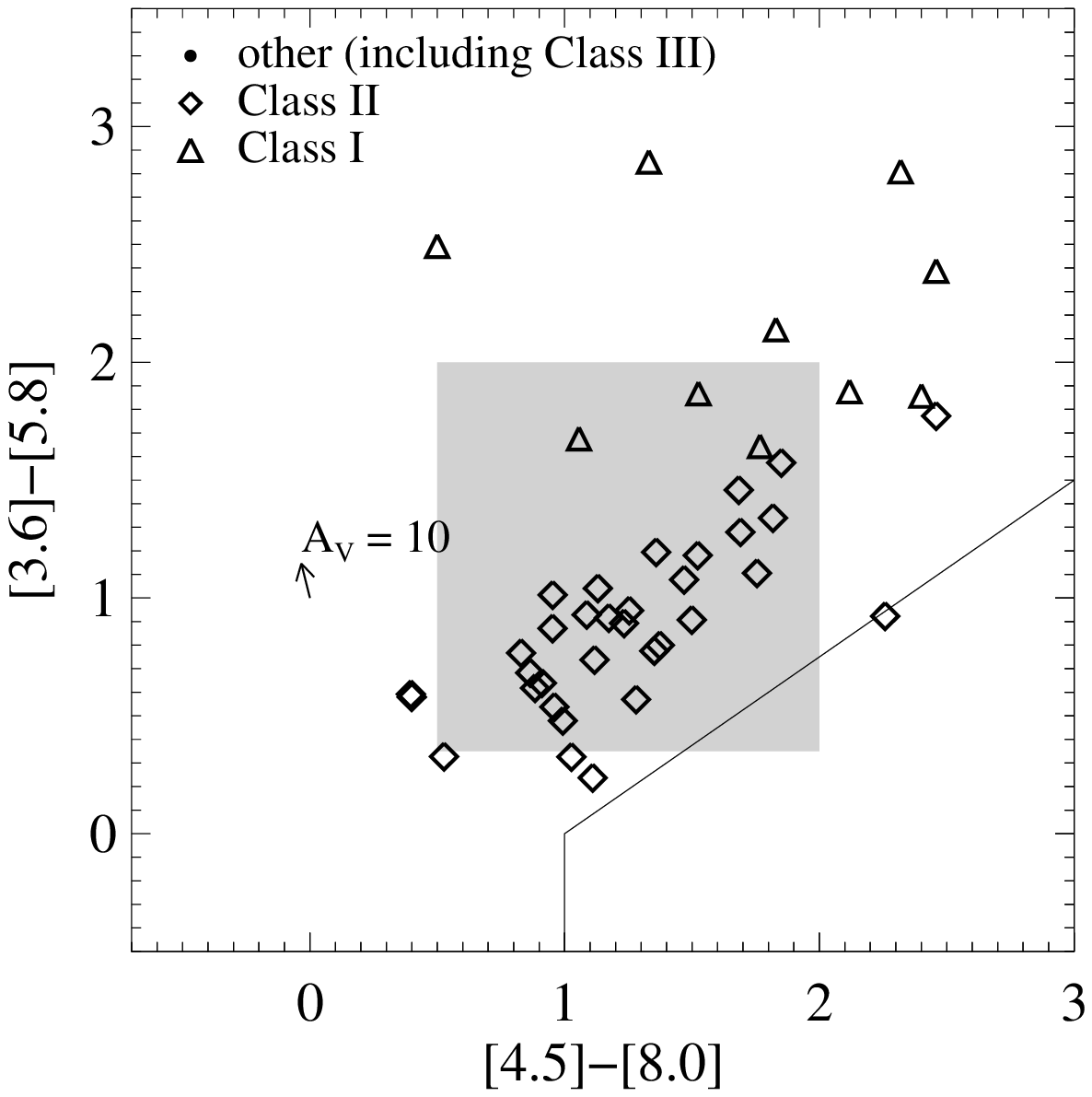,width=5.9cm}
}
}
\parbox{18cm}{
\parbox{6cm}{
\epsfig{file=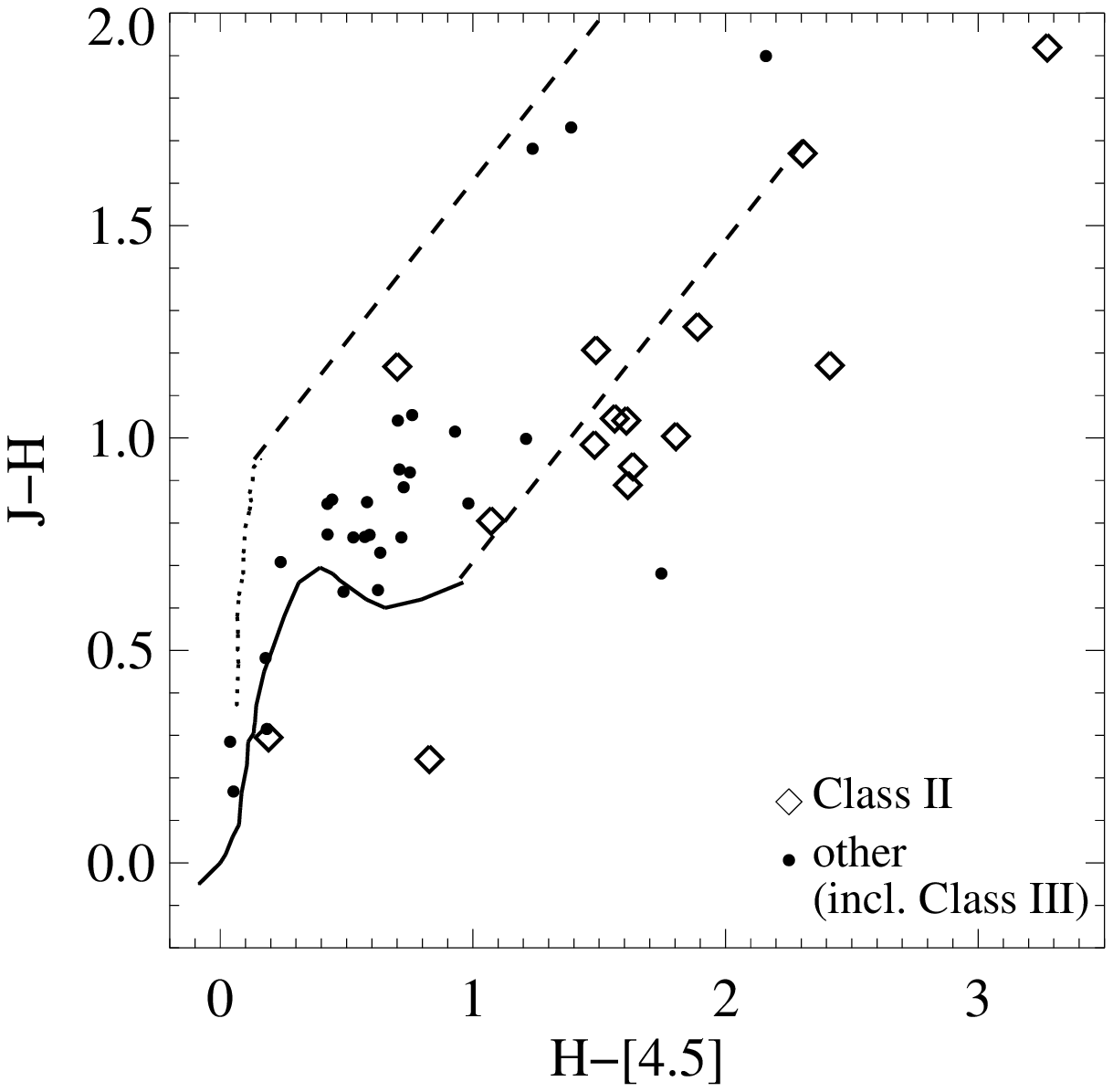,width=5.9cm}
}
\parbox{6cm}{
\epsfig{file=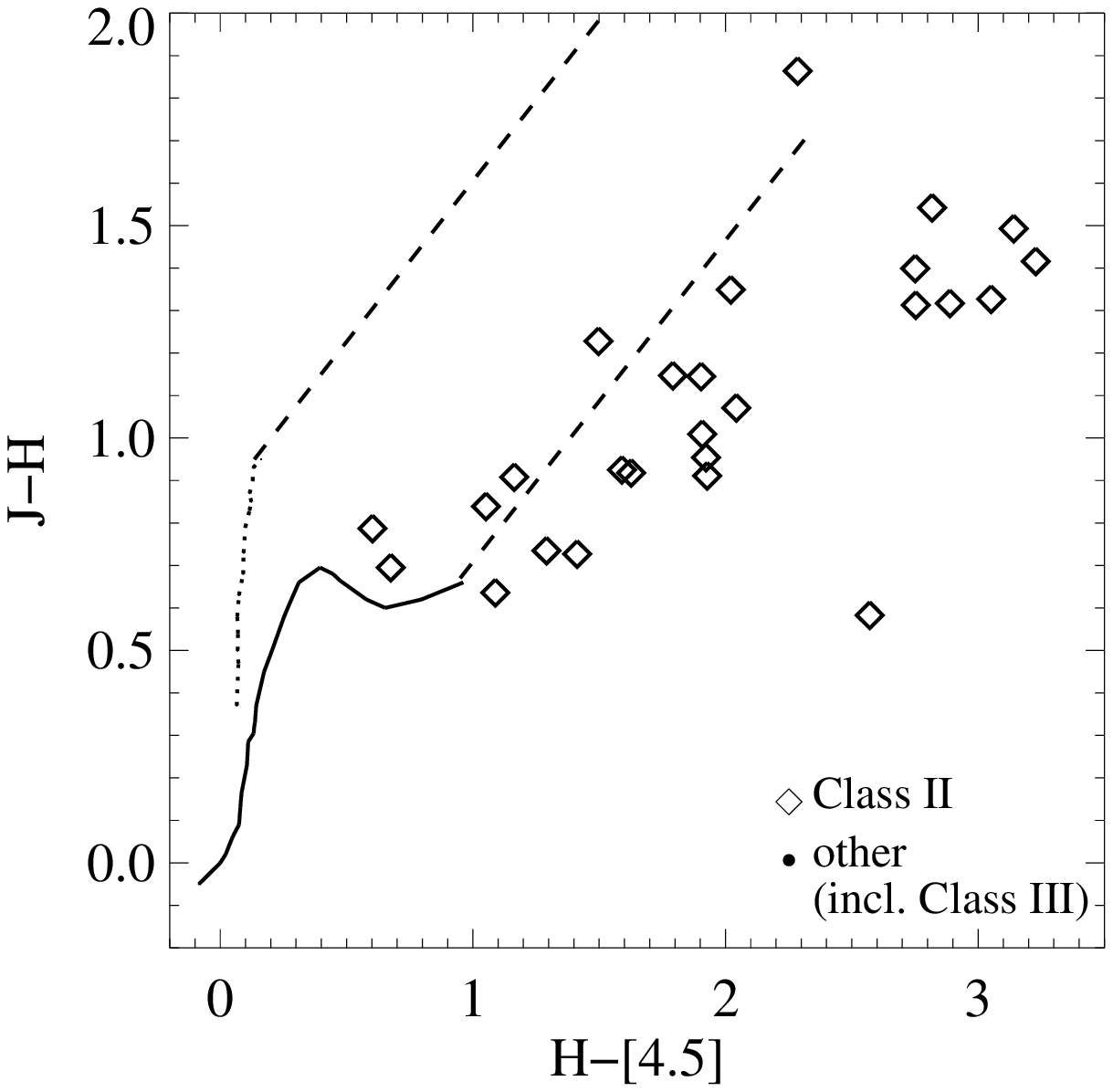,width=5.9cm}
}
}
\caption{Color-color diagrams for near-IR and mid-IR photometry used to classify the YSOs: 
left -- X-ray sources, right -- X-ray undetected YSO candidates selected with the criteria 
described in Sect.~\ref{subsect:census_ir}. In the top row, the solid line delineates the
area of star forming galaxies (equivalent to the first diagram in Fig.~\ref{fig:mir_ccds_all}). 
In the bottom row, the solid line represents the locus of main-sequence dwarf stars, 
the dotted line that of giants, and the dashed lines represent the reddening vector. 
} 
\label{fig:mir_ccds_sample}
\end{center}
\end{figure*}

\begin{figure*}
\begin{center}
\parbox{18cm}{
\parbox{6cm}{
\epsfig{file=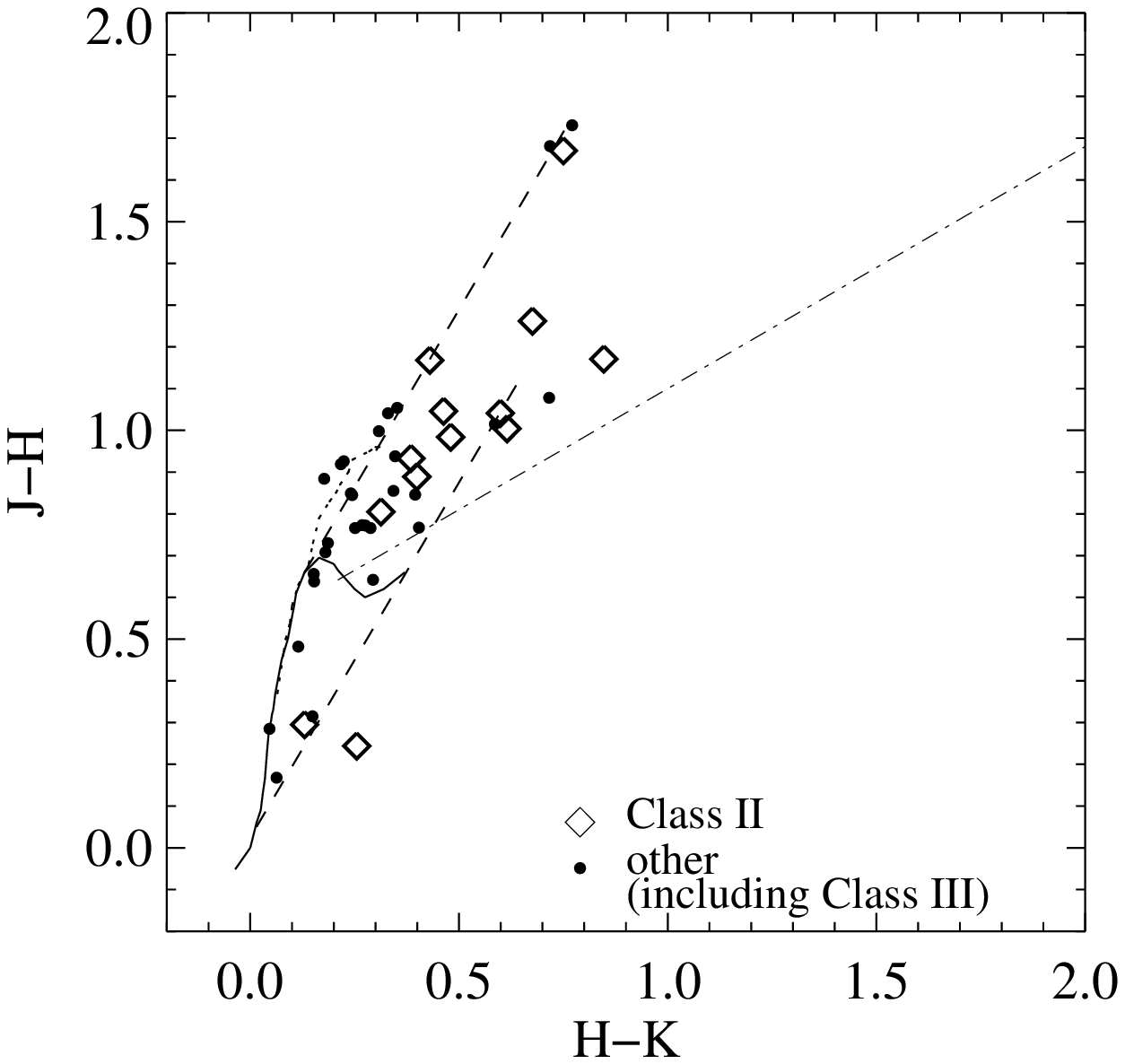,width=5.9cm}
}
\parbox{6cm}{
\epsfig{file=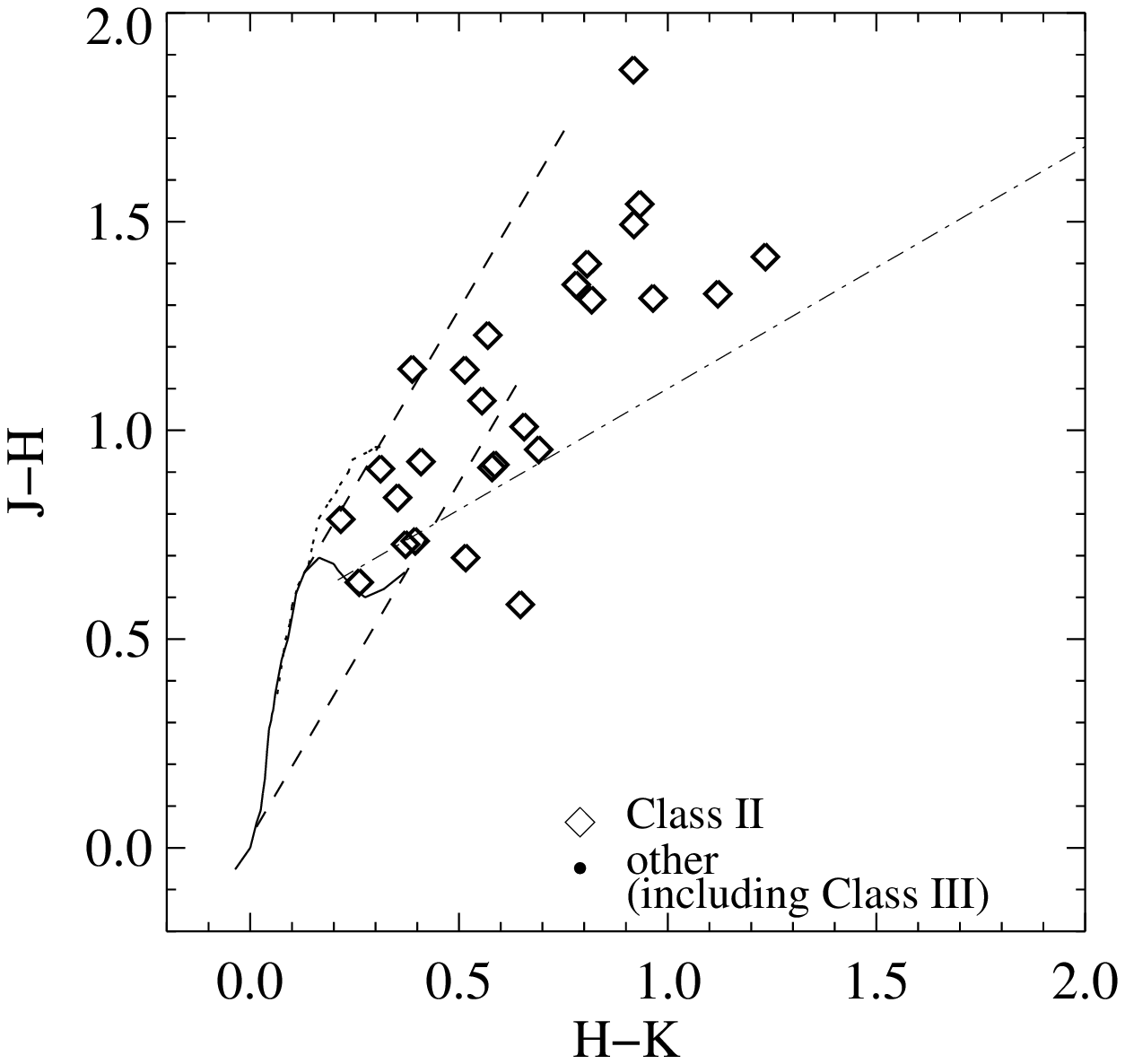,width=5.9cm}
}
}
\parbox{18cm}{
\parbox{6cm}{
\epsfig{file=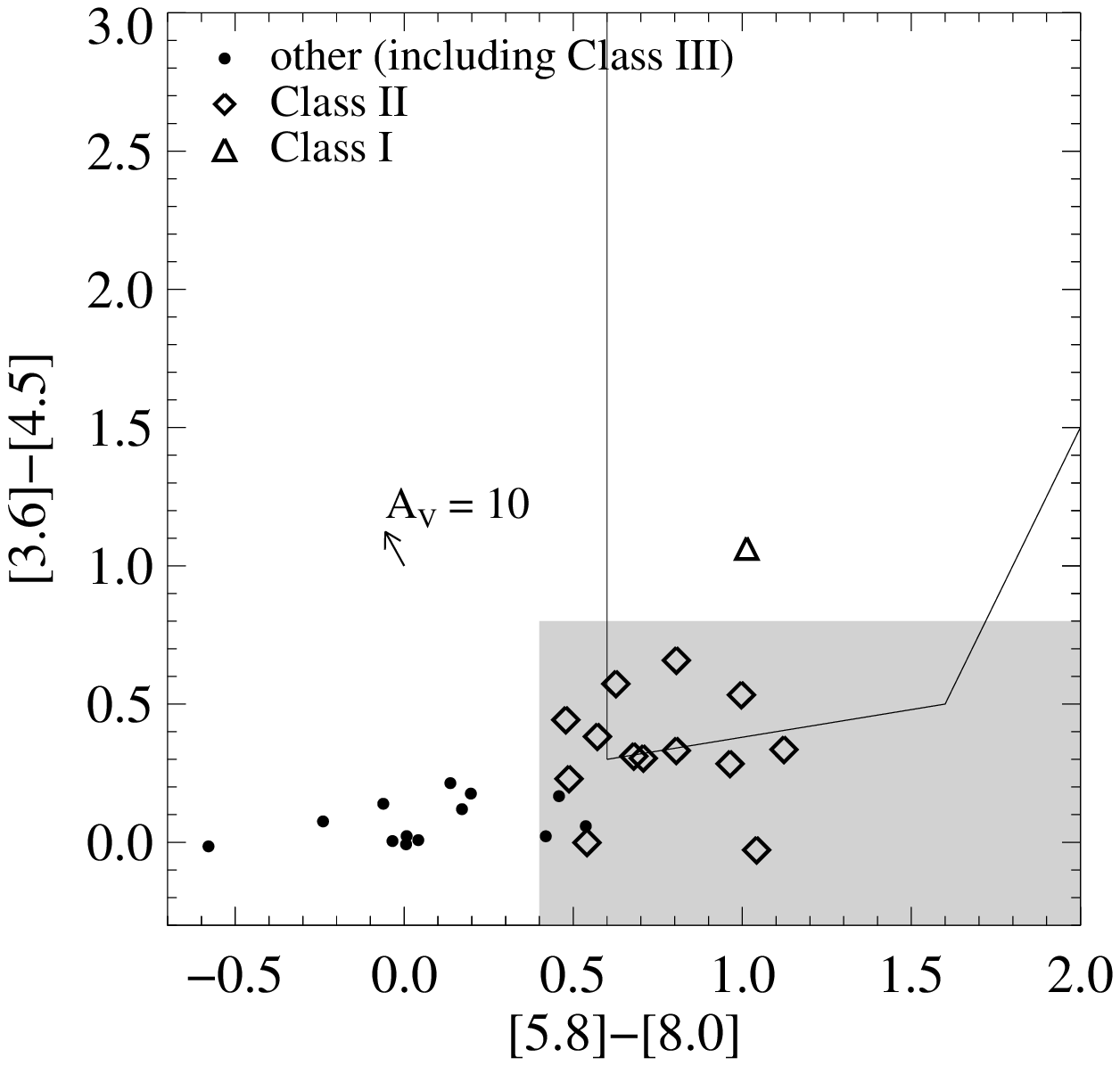,width=5.9cm}
}
\parbox{6cm}{
\epsfig{file=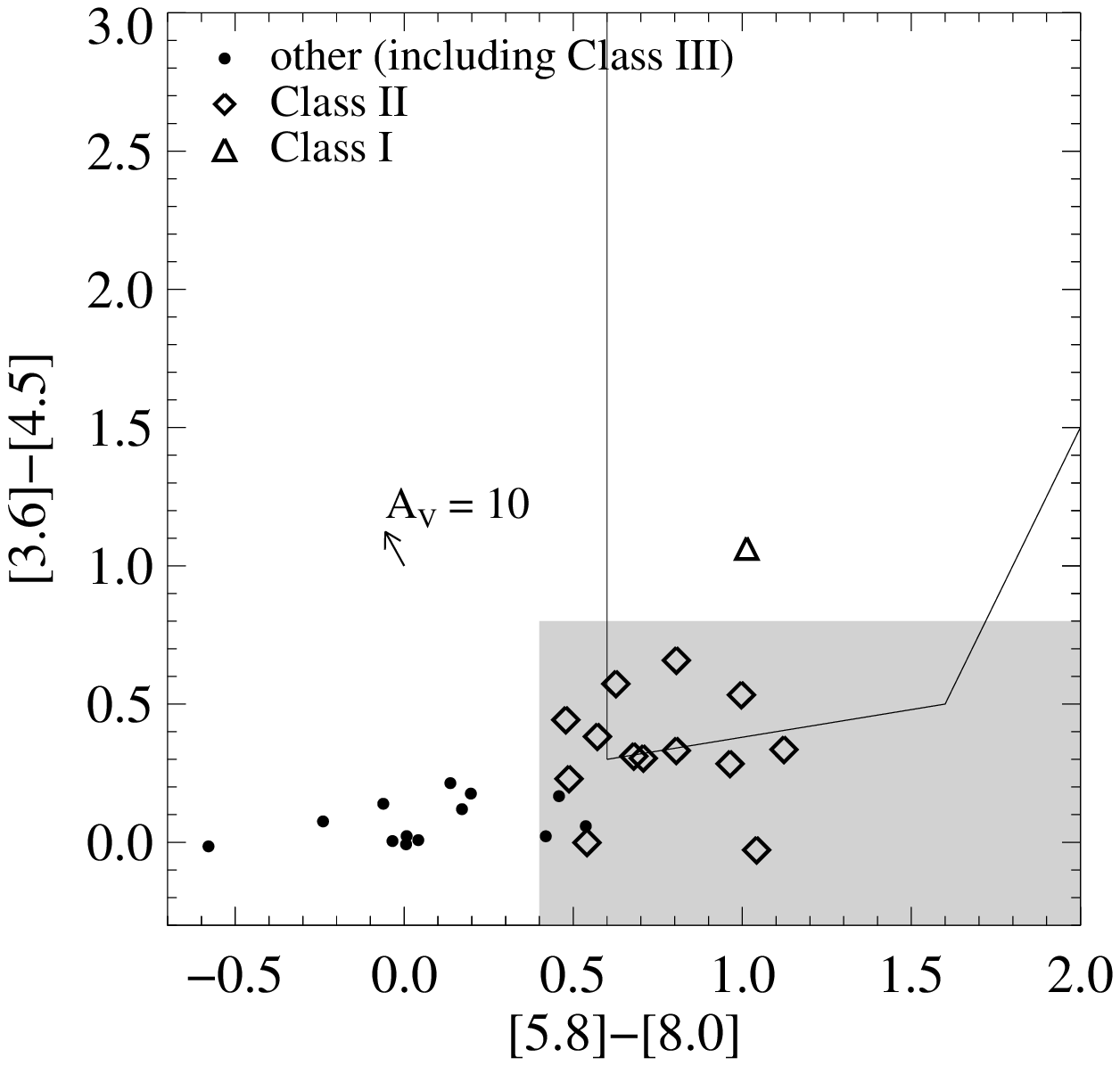,width=5.9cm}
}
}
\parbox{18cm}{
\parbox{6cm}{
\epsfig{file=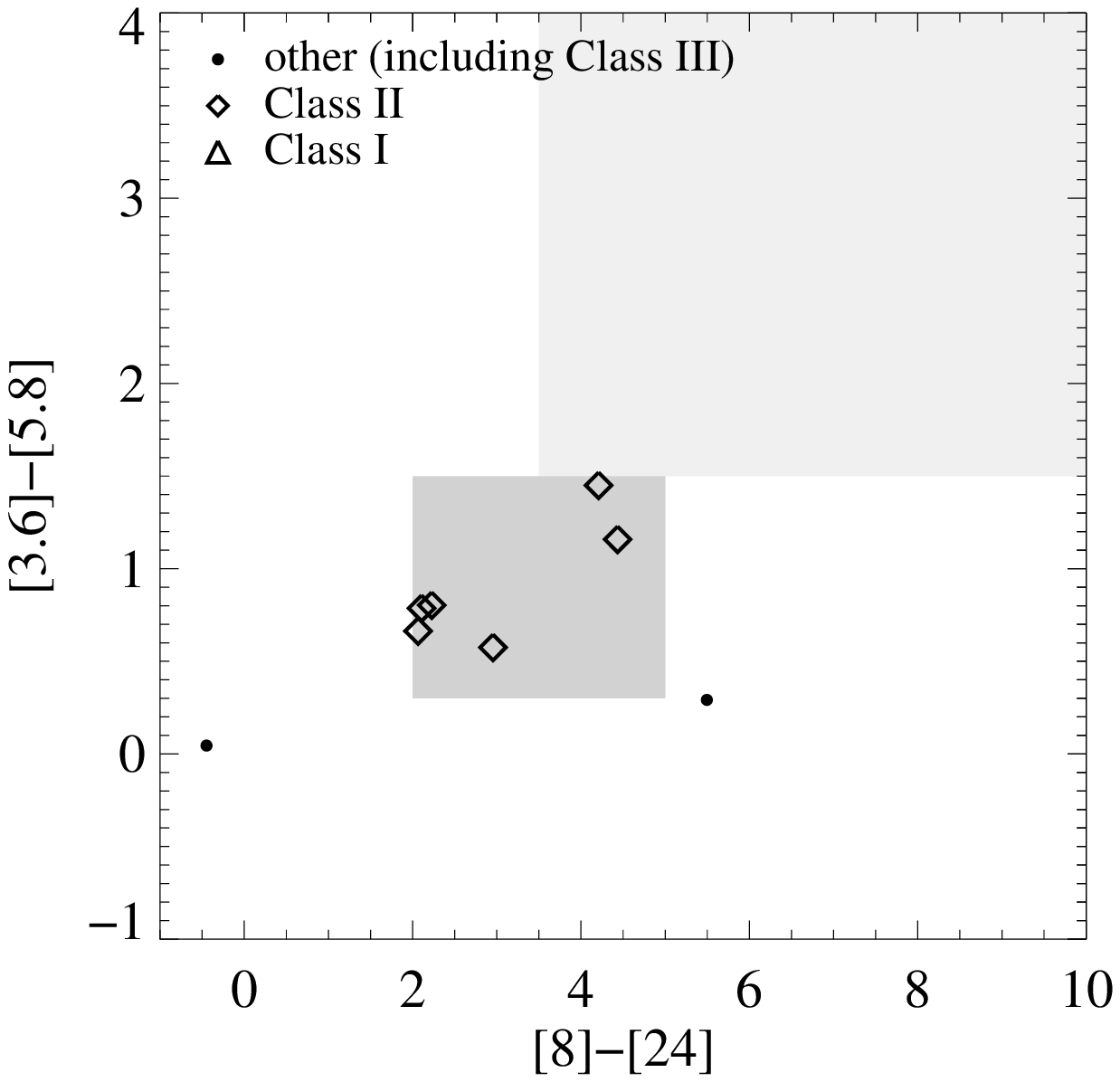,width=5.9cm}
}
\parbox{6cm}{
\epsfig{file=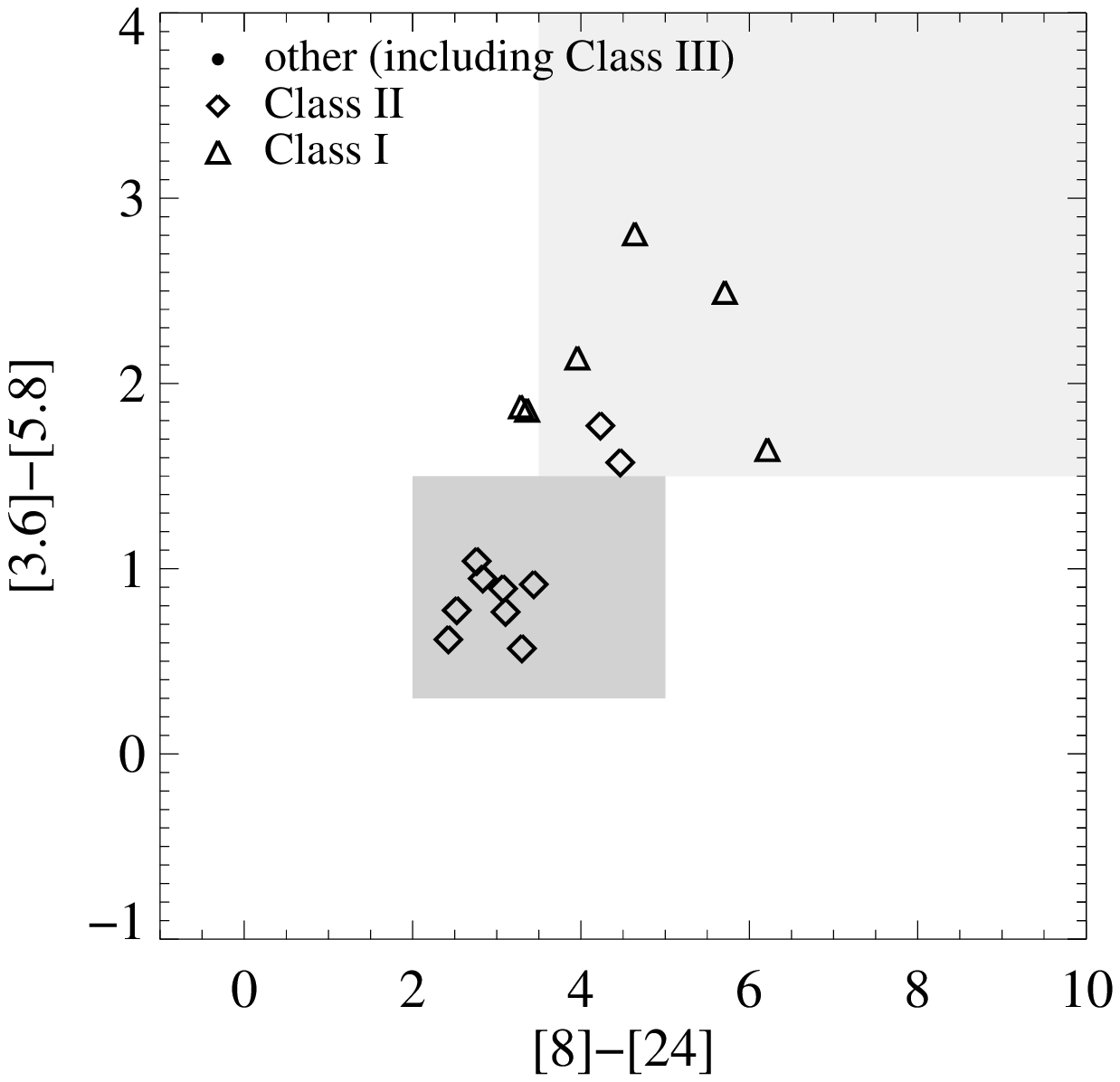,width=5.9cm}
}
}
\caption{Additional color-color diagrams for near-IR and mid-IR photometry: 
left -- X-ray sources, right -- X-ray undetected YSO candidates selected with the criteria 
described in Sect.~\ref{subsect:census_ir}.
The dash-dotted line in the $J-H$ vs. $H-K$ diagrams represents the locus of classical T\,Tauri stars
from \cite{Meyer97.1}. See Fig.~\ref{fig:mir_ccds_all} for the meaning of high-lighted regions.}
\label{fig:mips_ccds_sample}
\end{center}
\end{figure*}

Color-color diagrams of the X-ray sources are shown in the left column of 
Figs.~\ref{fig:mir_ccds_sample} and \ref{fig:mips_ccds_sample}. 
While Fig.~\ref{fig:mir_ccds_sample} show the
diagrams used for the YSO classification, Fig.~\ref{fig:mips_ccds_sample} provides a cross-check
of the resulting YSO status in colors not used for the selection process. It demonstrates that
our classification is consistent with the expectation from $JHK$ photometry (no Class\,III candidate
is in the area for T\,Tauri stars) and the expectation from MIPS photometry (only one Class\,III 
candidate is red at $[8]-[24]$). 

%
%

%
%

\subsection{X-ray undetected YSO candidates}\label{subsect:census_xuls}

As described in Sect.~\ref{subsect:census_ir} we have used the IR photometry to identify
YSOs with excess emission from circumstellar disks or envelopes in the {\em Spitzer} sample. 
There are $10$ Class\,0/I and $36$ Class\,II candidates within the {\em Chandra} FOV but 
not detected in X-rays. Their optical and IR photometry is summarized in Table~\ref{tab:counterparts_ul},
where objects are labeled by their running number in the catalog of R.Gutermuth.
Similar to the case of the X-ray detected ones, 
the majority of Class\,II is selected by their IRAC colors and only $5$ of them (labeled with an
asterisk in col.~14 of Table~\ref{tab:counterparts_ul}) are classified
solely on basis of near-IR photometry. 
Five stars have optical photometry in the catalogs described in Sect.~\ref{sect:optical},
and $VRI$ magnitudes for SVS\,12 
are extracted from \cite{Hillenbrand92.1}.  
The position of the X-ray undetected YSO candidates in the IR color-color 
diagrams are shown in the right column of Figs.~\ref{fig:mir_ccds_sample}
and~\ref{fig:mips_ccds_sample}. By definition this sample does not contain Class\,III stars.
Again, our classification scheme is supported by the MIPS data, 
as the Class\,0/I and Class\,II objects separate according to the empirical locations determined
by \cite{Muzerolle04.2}. 
  
We computed upper limits for the X-ray count rate of the undetected YSO candidates using the
algorithm of \cite{Kraft91.1}. For the conversion into X-ray flux 
we followed the procedure described in Sect.~\ref{sect:chandra} for the detected sources,
i.e. we estimate $L_{\rm x}$ in two ways: (A) 
We assumed a $1$-T RS-model with $kT$ and $N_{\rm H}$ corresponding to the median of the values measured
in the sample of X-ray detected Class\,II sources given in Sect.~\ref{sect:chandra}, 
and (B) we adopted for the $1$-T model the median $kT$ but individual $N_{\rm H}$ values for each star 
obtained from its $A_{\rm V}$. Then, for both approaches we computed 
a count-to-flux conversion factor with PIMMS. 
The X-ray parameters for the undetected YSO candidates are summarized in Table~\ref{tab:xraytab_ul}. 
We discuss possible effects on the X-ray luminosity upper limits resulting from the assumed column
density in Sects.~\ref{subsect:results_xlf} and~\ref{subsect:results_class1}.

\subsection{Contamination of the YSO sample}\label{subsect:census_contam}

The IR emitters 
comprise also extragalactic background objects, galactic background stars and foreground dwarf stars. 
Various authors have investigated the position of contaminating background AGN and galaxies
in the IRAC color-color diagrams, and we use their results. 
\cite{Stern05.1} have identified an area populated by broad-line AGNs enclosed in the solid 
line in the $[3.6]-[4.5]$ vs. $[5.8]-[8.0]$ diagram. 
A brightness cutoff can be used to discriminate AGN from YSO candidates on a statistical basis. 
Star forming galaxies appear very red at mid-IR
wavelengths and are usually found in the lower right area delineated by the solid lines in the
$[3.6]-[5.8]$ vs. $[4.5]-[8.0]$ diagram \citep{Gutermuth08.1}.
We elaborate on the problem of the sample contamination 
in Sect.~\ref{subsect:census_contam} after the YSO candidate sample of NGC\,7129 has been defined. 

In addition to the region marked by the solid line in the $[3.6]-[4.5]$ vs. $[5.8]-[8.0]$ diagram 
a brightness limit
of $[3.6] = 14.5$\,mag and $[4.5] = 14$\,mag \citep{Joergensen06.1, Hernandez07.1} has been used to
identify contaminating background AGN. 
Only two X-ray detected YSO candidates in the highlighted region 
and $4$ undetected YSO candidates are fainter than these magnitudes. 
They are flagged in Tables~\ref{tab:counterparts_b} and~\ref{tab:counterparts_ul}. 
None of these $6$ objects has a 2\,MASS counterpart, and we consider them likely AGNs. 
The $[3.6]-[5.8]$ vs. $[4.5]-[8.0]$ diagram demonstrates that none of our YSO candidates is in the area of the 
extragalactic PAH sources to the right and below the solid line defined by \cite{Gutermuth08.1}. 

The contribution of extragalactic objects to the X-ray emitters can also be estimated from
$\log{N} - \log{S}$ distributions taking into account the limiting flux of the observation. 
We use the expression given by \cite{Feigelson05.2} to make a rough 
estimate of the sensitivity limit in the {\em Chandra} observation. Using the median absorption,  
$\log{N_{\rm H}}\,{\rm [cm^{-2}]} = 21.4$, derived from the hardness ratios of the X-ray sources, we find  
$f_{\rm x, lim} \sim 7 \cdot 10^{-15}\,{\rm erg/cm^2/s}$. From the source counts in the {\em Chandra Deep 
Field North} \citep{Brandt01.1} we expect $\sim 17$ extragalactic objects in the combined ACIS-S2 and ACIS-S3 fields.
However, at large off-axis angles the sensitivity is significantly lower than the number given above, and  
in practice, most of the detected X-ray sources are found within or near the $1.7^\prime$ cluster core.
Thus, the true contamination is likely smaller. 
Indeed, most X-ray sources with large distance from the cluster center are not part of 
the mass limited sample that we define in Sect.~\ref{subsect:census_spatial}. They have no near-IR photometry
from which we could estimate the mass, i.e. they are faint, just as expected for extragalactic objects.  
For an area the size of the NGC\,7129 cluster core only $\sim 1$ extragalactic contaminant is predicted.
To summarize, both the considerations of the IR colors and magnitudes and the X-ray sensitivity limit indicate
that the fraction of AGN and background galaxies among our candidate NGC\,7129 members is probably small. 

The sample of X-ray selected Class\,III objects may also contain contaminating objects 
in the foreground of the cluster. The contribution of foreground field
 stars can be judged using the results from the {\em ROSAT}/NEXXUS survey: The typical
 X-ray luminosity for field stars is $\log{L_{\rm x}}\,{\rm [erg/s]} \sim 28$ for F/G dwarfs, 
$\sim 27.7$ for K dwarfs, and $\sim 27$ for M dwarfs \citep[][ their Fig.6]{Schmitt04.1}. 
 Comparing with the average X-ray luminosity in our sample ($\log{L_{\rm x}}\,{\rm [erg/s]} \sim 30$), 
we would thus be able to detect field F/G dwarfs for distances $<100$\,pc, K dwarfs 
for $<70$\,pc, and M dwarfs for $<30$\,pc. Based on the luminosity function in 
the solar neighbourhood, the space density of dwarf stars peaks in the M dwarf 
regime at $\sim 0.01$\,pc$^{-3}$ \citep{Reid97.1}, and is significantly lower for F/G/K 
stars \citep{Reid02.3}. In the core radius of $1.7^\prime$ for NGC\,7129, we therefore expect 
$<10^{-5}$ field stars with detectable X-ray emission. Although this estimate 
has been derived based on typical numbers, not taking into account the significant 
spread in X-ray luminosities at a given spectral type, the result is safely 
below 1 and thus negligible.

%
%
\begin{figure}
\begin{center}
\epsfig{file=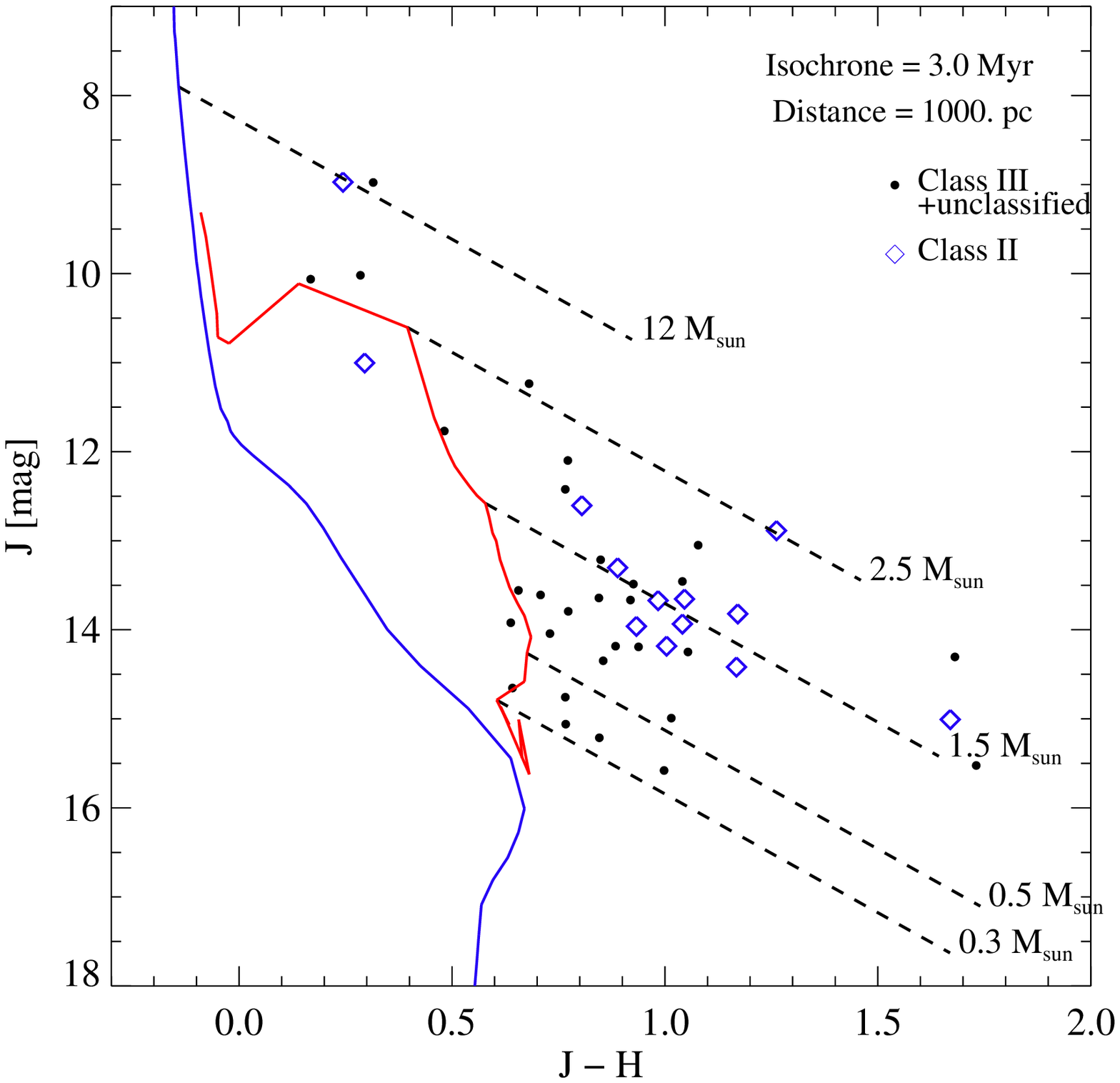,width=7.9cm}
\epsfig{file=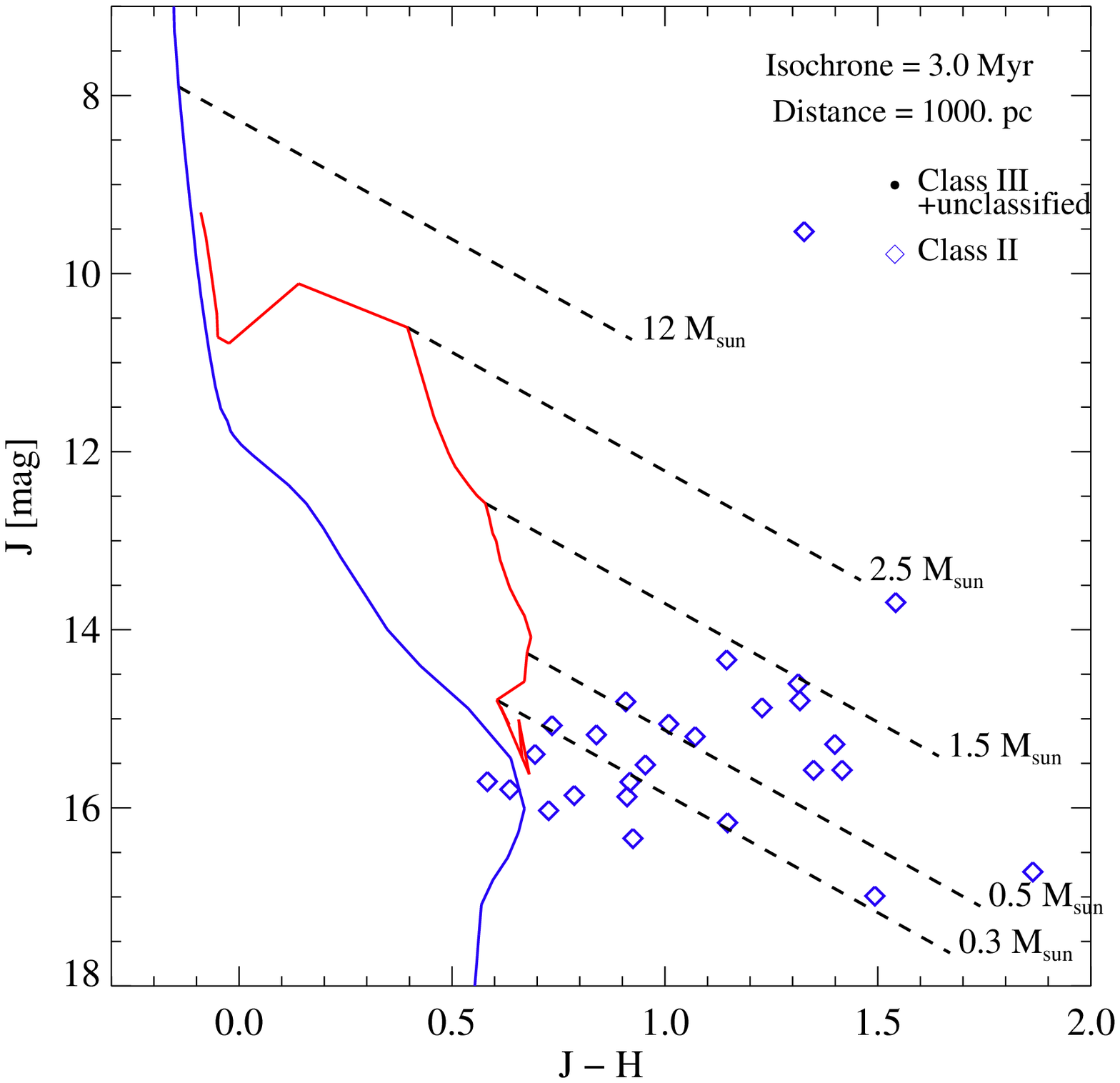,width=7.9cm}
\caption{$J$ vs. $J-H$ diagram for the 2\,MASS photometry of X-ray detected (top) and non-detected (bottom) 
cluster candidates in NGC\,7129. The two solid lines 
represent the $3$\,Myr isochrone for the pre-MS calculations from \protect\cite{Siess00.1} (right, red line) 
and the calculations from \protect\cite{Marigo08.1} (left, blue line), respectively. 
The dashed lines indicate the reddening
vector for a $0.3$, $0.5$, $1.5$, $2.5$, and $12\,M_\odot$ star following \cite{Rieke85.1}. 
Their length corresponds to a visual extinction $A_{\rm V}=10$\,mag.}
\label{fig:j_jh}
\end{center}
\end{figure}

\subsection{Completeness and spatial distribution}\label{subsect:census_spatial}

For a meaningful comparison of the YSO populations the possible differences in the completeness
of X-ray and IR selected samples must be taken into account. 
In this section we define several sub-samples for NGC\,7129 on basis of the stars
identified as NGC\,7129 candidate members in Sects.~\ref{subsect:census_ir} to~\ref{subsect:census_contam}. 
The selection criteria and the composition of these samples are summarized in Table~\ref{tab:samples}
and the individual stars of each sample are listed in Table~\ref{tab:samples_flag}. 
\begin{table*}
\begin{center}
\caption{Composition of different subsamples of NGC\,7129.}
\label{tab:samples}
\begin{tabular}{ccccccccc} \\ \hline
\multicolumn{3}{c}{------- Sample characteristics -------}         & \multicolumn{2}{c}{Class\,II} & \multicolumn{2}{c}{Class\,III} & \multicolumn{2}{c}{Total} \\ 
$M>0.5\,M_\odot$   & $A_{\rm V} < 5$\,mag & core   & $N$ & ($N_{\rm det}/N_{\rm u.l.}$) & $N$ & ($N_{\rm det}/N_{\rm u.l.}$) & $N$ & ($N_{\rm det}/N_{\rm u.l.}$) \\ \hline
$\surd$ &         &         & 26 & (13/13) & 25 & (25/0) & 51 & (38/13) \\
$\surd$ & $\surd$ &         & 13 & (9/4)   & 21 & (21/0) & 34 & (30/4)  \\
$\surd$ &         & $\surd$ & 16 & (7/9)   & 16 & (16/0) & 32 & (23/9) \\
$\surd$ & $\surd$ & $\surd$ &  7 & (5/2)   & 14 & (14/0) & 21 & (19/2) \\
\hline
\end{tabular}
\end{center}
\end{table*}

First, we estimate the mass distribution of the YSO candidates from a 
comparison of their IR photometry with the predictions from evolutionary calculations. 
We assume a cluster age of $3$\,Myr and an adequate combination of evolutionary models:
Low-mass stars
at an age of few Myrs are still in their pre-MS phase, and we use the model from \cite{Siess00.1}
converted to the observational plane using the calibration given by \cite{Kenyon95.1}. 
The Siess et al. calculations extend up to $7\,M_\odot$, i.e. they 
do not cover the full mass range of our sample. For the
high-mass stars we use the calculations from \cite{Marigo08.1}. At the high-mass end of 
the $3$\,Myr isochrones the Siess calculations are in good agreement with the Marigo model. 
Therefore, we use the Siess isochrone for $M \leq 2.7\,M_\odot$ and the Marigo isochrone
for higher masses. 

From the $J$ vs. $J-H$ diagram (Fig.~\ref{fig:j_jh}) it emerges that all X-ray detected stars in NGC\,7129 
with known $J$ and $H$ magnitudes have $M > 0.3\,M\odot$ according to the evolutionary calculations. 
In the Orion Nebula Cluster (ONC), at $1$\,Myr just slightly younger than NGC\,7129, stars with 
$\log{L_{\rm x}}\,{\rm [erg/s]} = 29.9$ corresponding to the sensitivity limit derived in 
Sect.~\ref{subsect:census_contam}, have a mass of $\sim 0.5\,{\rm M_\odot}$ \citep{Preibisch05.1}. 
Therefore, we take the X-ray sample to be complete down to this mass. 
About half of the 
YSOs that are not detected with {\em Chandra} are fainter in $J$ than predicted for a cluster member of 
$0.5\,M_\odot$. 
X-ray non-detected YSOs below 
the reddening track of an $M > 0.5\,M_\odot$ star according to the \cite{Siess00.1} model
are dropped from the sample for further analysis. 

Obviously, the sample defined this way comprises only stars with available photometry in the $J$ and $H$ bands.
This is not a serious limitation because {\em Spitzer} sources without known $JH$ magnitude are likely fainter
than our mass cutoff. The resulting sample consists of 
$38$ X-ray sources ($13$\,Class\,II and $25$\,Class\,III) 
and $13$ X-ray undetected Class\,II YSOs selected
from their {\em Spitzer} colors; see first row in Table~\ref{tab:samples}. 
By introducing this sensitivity cutoff the four AGN candidates discussed
in Sect.~\ref{subsect:census_contam} are also dropped. 
No Class\,0/I sources are in the mass-limited sample because none of them have $JH$ photometry. 

There is evidence for higher average extinction in the YSO sample that have X-ray upper limits with
respect to the sample of X-ray sources (see Fig.~\ref{fig:j_jh} and Fig.~\ref{fig:mips_ccds_sample} top panel).  
About half of the Class\,II stars are in the upper limit sample, while Class\,III stars are by definition present 
only in the X-ray detected sample. This might imply that the sensitivity limit is different for the two object groups. 
Class\,III sources with large extinction may, indeed, be absent from the sample because
high column density impedes the X-ray photons from being observed. 
To eliminate a possible bias we define a `lightly absorbed' sample limited by $A_{\rm V} < 5$\,mag.
There are very few X-ray undetected YSOs with low extinction above our mass cutoff (Fig.~\ref{fig:j_jh} bottom),
supporting our estimate that the X-ray observation is (nearly) complete down to $0.5\,M_\odot$. 
The mass-limited, lightly absorbed
sample consists of $13$\,Class\,II sources ($4$ of which not detected in X-rays) and $21$\,Class\,III sources;
second row in Table~\ref{tab:samples}.

\begin{figure*}
\begin{center}
\resizebox{18cm}{!}{\includegraphics{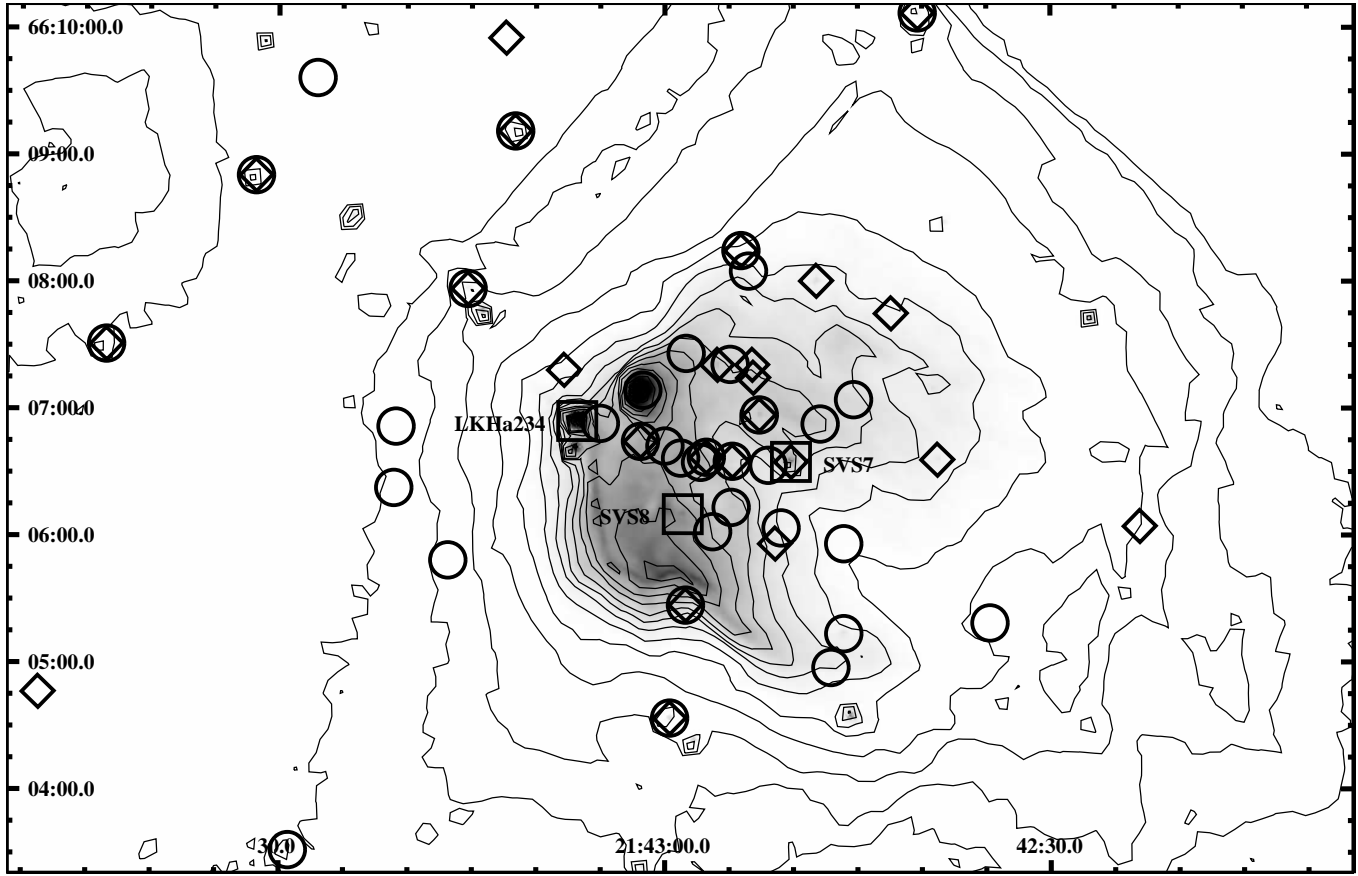}}
\caption{Contour plot for the IRAC $8\,\mu$m image of NGC\,7129 from Program\,\#\,6 with PI Fazio. Contour levels go
from $5$ to $1000$ MJy/sr in logarithmic steps. Squares show the 
position of the three hot stars. Circles are X-ray sources and diamonds are YSO candidates identified on basis
of near-IR/mid-IR emission irrespective of whether  
detected in X-rays. Only stars from the mass limited sample defined in Sect.~\ref{subsect:census_spatial} are shown. One 
X-ray detected and one non-detected member of this sample are outside the field shown in the figure.}
\label{fig:spatial}
\end{center}
\end{figure*}

We move now on to consider the spatial distribution of the YSO population in NGC\,7129. 
\cite{Gutermuth04.1} noticed that there is an accumulation of Class\,II YSOs around SVS\,12, 
at the outskirts of and pointing away from the molecular cloud. This is also the direction where
most X-ray sources, apart from those in the cluster core, are found.
A map with the cloud contours and our mass-limited sample is shown in Fig.~\ref{fig:spatial}.
To eliminate effects related to a possible difference in the spatial distribution of YSOs in different
evolutionary stages
we define a sample limited by the cluster core radius as described by \cite{Gutermuth04.1},
henceforth called the `core sample'. 
The mass limited core sample is composed of 
$23$ X-ray emitters, of which $7$ are Class\,II objects and $16$ are Class\,III. 
These numbers represent a fraction of $44$\,\% and $53$\,\% of the 
total sample of X-ray detections for Class\,II and Class\,III, respectively 
(cf. Table~\ref{tab:counterparts_b}). The mass-limited core sample, therefore, seems to
reflect roughly the spatial distribution of X-ray bright Class\,II and Class\,III in NGC\,7129. 
In addition to those $23$ X-ray sources, the mass-limited core sample includes     
$9$ Class\,II objects not detected with {\em Chandra} (see third row in Table~\ref{tab:samples}). 
Finally, for the reasons outlines above, we apply an extinction cutoff to this sample. This defines
the most restricted sample used in this work, the mass-limited, lightly absorbed core sample. 
It comprises $7$\,Class\,II (of which $2$ undetected in X-rays) and $14$\,Class\,III,
and is summarized in the last row of Table~\ref{tab:samples}. 
%
%

\subsection{Comparison to the classification scheme of Gutermuth et al.}\label{subsect:census_g09}

\cite{Gutermuth08.1} have devised a complex scheme for the classification of YSOs that was
subsequently revised and applied to a wide range of nebulous star forming environments (G09).
In this scheme the distinction between Class\,0/I and Class\,II sources is based on the $[4.5] - [5.8]$ color
rather than on the $[3.6] - [4.5]$ color, thus reducing effects of the reddening that are more severe
at shorter wavelengths \citep{Flaherty07.1}. In a second step, the near-IR photometry is used to deredden
the stars and to identify excess emission at wavelengths shorter than $[5.8]$. Finally, G09 use 
MIPS photometry to identify deeply embedded protostars without detectable emission in the IRAC bands. 
We refer to G09 for details of the procedure. 
Here, we present a brief comparison of the results for the case of NGC\,7129. 

Overall, we found our classification scheme to be consistent with the more exhaustive approach used by G09. 
Specifically, most of the objects classified as likely extragalactic are common to both studies. 
Among the $9$ Class\,0/I candidates of our list, $5$ are classified Class\,I by G09, one is Class\,0, one
is Class\,II and two are flagged as shock source. Better agreement is found for the Class\,II and Class\,III
samples that are more relevant to our study than the protostars. 
In particular, the mass-limited and the core sample defined in Sect.~\ref{subsect:census_spatial} are very clean. 
Our Class\,II core sample comprises only two objects that carry different flags in the scheme of G09
(one detected in X-rays and one not detected). 
Conversely, G09 assign Class\,II status to three X-ray undetected objects in the cluster core that are not identified as YSOs by our 
procedure. 
One of them has no 2\,MASS photometry, and would not have been selected into our mass-limited sample. 
Therefore, the Class\,II sample size is almost identical for the two selection approaches. 
In the framework of this paper an accurate YSO classification is relevant only for the evaluation 
of the disk fraction, and in practice, the differences are not significant 
(see Sect.~\ref{subsect:results_diskfraction}).

\section{Results}\label{sect:results}

\subsection{SEDs}\label{subsect:results_seds}

We examined the SEDs of all sources from Tables~\ref{tab:counterparts_b}
and~\ref{tab:counterparts_ul}. Due to the limited spectral range covered by the available photometry
of most sources we abstain from a systematic analysis and search only 
for outstanding objects. In particular, we are interested to identify objects with `holes' 
in the SED (i.e. excess emission is only visible at the longest wavelengths) and embedded objects 
or objects with edge-on disks (i.e. with strongly increasing flux levels towards longer wavelengths). 
This exercise was limited to sources for which the data cover at least the wavelength 
range from $1.25$ to $8\,\mu$m with a minimum of $6$ datapoints. These limits were imposed to guarantee a 
minimum amount of information about both the stellar photosphere and the circumstellar environment.

Three objects clearly stand out based on their SED characteristics: NGC\,7129-S3-X28 and NGC\,7129-S3-X9 from the X-ray 
detected sample, and  NGC\,7129-S3-U1211 from the non-detections in X-rays. These objects were selected for 
further analysis. We added object NGC\,7129-S3-U1012 (identified with the Herbig star LkH$\alpha$\,234), 
which is the brightest source in our sample at all IRAC wavelengths.

\begin{figure}
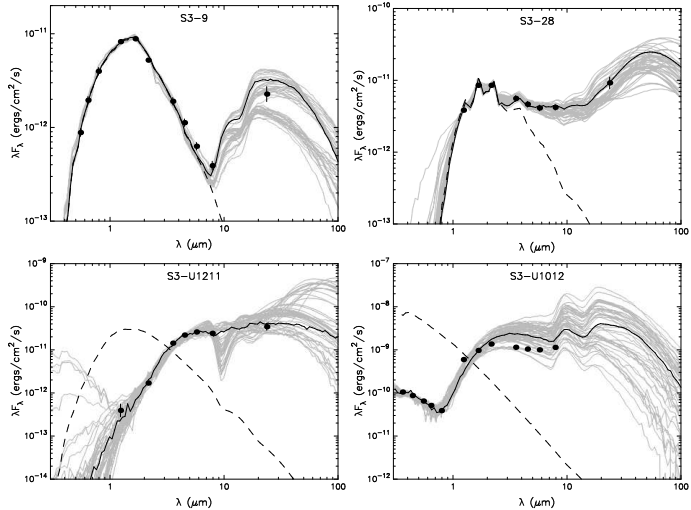

\begin{center}
\parbox{9cm}{
\parbox{4.5cm}{
\resizebox{4.5cm}{!}{\includegraphics{./f8a.ps}}
}
\parbox{4.5cm}{
\resizebox{4.5cm}{!}{\includegraphics{./f8b.ps}}
}
}
\parbox{9cm}{
\parbox{4.5cm}{
\resizebox{4.5cm}{!}{\includegraphics{./f8c.ps}}
}
\parbox{4.5cm}{
\resizebox{4.5cm}{!}{\includegraphics{./f8d.ps}}
}
}
\caption{Observed SEDs (datapoints) for four `peculiar' objects 
together with their $50$ best fitting model SEDs from the grid computed by 
\cite{Robitaille06.1, Robitaille07.1}. 
The best fitting model is shown as black solid line. The dashed line shows the photospheric 
spectrum without obscuration by disk or envelope, but with interstellar reddening.}
\label{fig:seds}
\end{center}
\end{figure}

For these four objects we compare the observed SED with the grid of model SEDs, pre-computed using 
a Monte Carlo radiation transfer code, provided by 
\cite{Robitaille06.1, Robitaille07.1}\footnote{The SED fitter is available under  
http://caravan.astro.wisc.edu/protostars/}. 
Interpreting SEDs using radiative transfer code is subject to
degeneracies, as discussed in detail in the literature \citep{Chiang01.1}.
Spatially resolved multi-wavelength observations can break the
degeneracies \citep{Watson07.1} but are only available for
relatively few object. Therefore, we do not attempt to derive a
unique model solution for the observed SED. Instead, the following
analysis is based on the $50$ best-fitting models according to $\chi^2$ for each object,
selected from a grid of $200 000$ pre-computed YSO models. For more information on
the specifications and limitations of the model grid see \cite{Robitaille06.1, Robitaille07.1}.
In Fig.~\ref{fig:seds} we show the observed SEDs  
and the $50$ best fitting models for each of the four stars. 
We have applied generic errors of $10$\,\% for all IRAC fluxes and $20$\,\% for all MIPS fluxes
to account for calibration uncertainties. (These errors are significantly higher than the nominal
errors in all cases.)
As can be seen, all four objects are adequately matched by a significant number of models. 

The multi-parameter fit produces partly ambiguous results; we limit 
the discussion to those parameters which are well-constrained by the procedure.
While the near-IR photometry mostly constrains the photospheric properties of the central 
objects, the fluxes at mid-IR wavelengths are strongly sensitive to the inner disk geometry, 
i.e. flaring, inclination, inner disk radius. In addition, the fluxes in our wavelength coverage are  
affected by the accretion rate. At $24\,\mu$m the envelope begins to contribute significantly, thus 
allowing us to constrain the evolutionary state of the object. The lack of longer wavelength data 
precludes putting limits on global parameters of disk and envelope 
\citep[see][for further information]{Wood02.2}.
In the following we briefly discuss the results for each of the four objects shown in Fig.~\ref{fig:seds}. 
The limitations of the model fit (e.g. finite resolution and parameter coverage of the grid)
sometimes do not allow a perfect fit of all datapoints. These type
of models do also not take into account variability or 3D structure in the disk. 

Source NGC\,7129-S3-X9 is identified with an H$\alpha$ emission line star 
\cite[$W_{\rm H\alpha} = 9.8$\,\AA; see][]{Magakian04.1}. Optical photometry in $VRI$ bands is available, 
which has allowed us to estimate the mass of the central source, which is around 1$\,M_{\odot}$ 
($0.4-2\,M_\odot$). 
NGC\,7129-S3-X9 does not show IR excess for $\lambda < 6\,\mu$m, 
but a strong excess level is seen at 24$\,\mu$m. This is the 
hallmark sign of a disk with an inner opacity hole, often called `transition disk' in the literature. 
Indeed, in its broadband shape, this SED resembles the ones measured for TW\,Hya \citep{Calvet02.1} 
and CoKu\,Tau/4 \citep{DAlessio05.1}. Possible explanations for opacity holes include the presence of a 
massive planet or a stellar companion. The best-fitting SED models give an inner disk radius of 10\,AU
($4-30$\,AU).  
This is $\sim 100$ times larger than the sublimation radius that usually limits the extent of the inner
dust disk.

The SED for object NGC\,7129-S3-X28 is significantly increasing between $8$ and $24\,\mu$m. The datapoints are 
well-reproduced by models for a disk with a high degree of flaring (i.e. the scale height 
increases strongly with increasing distance from the star) embedded in an envelope. The best estimate for the mass of the 
central object is $0.1-0.6\,M_\odot$ but optical photometry is required for a more reliable assessment. 
The high $24\,\mu$m flux indicates that this object is possibly in an early evolutionary state.
Indeed, in the $[3.6]-[5.8]$ vs. $[8]-[24]$ diagram it is just at the intersection between the
canonical areas for Class\,II and Class\,0/I objects.

Object NGC\,7129-S3-U1211 has been classified as Class\,0/I source in Sect.~\ref{subsect:census_xuls}. 
This is confirmed by the SED fit. 
This is one of the objects with the steepest SED slope in our sample. The flux levels increase steeply 
from near-IR wavelengths to $24\,\mu$m. 
For this object, most parameters are poorly constrained. 
As can be seen in the plot, the photospheric fluxes are strongly suppressed by circumstellar extinction. 
Optical and sub-millimeter measurements would 
permit to put more reliable limits on the physical properties of the source.

For NGC\,7129-S3-U1012 (alias SVS\,12) optical $UBVRI$ photometry was presented by \cite{Hillenbrand92.1}
in their study of Herbig stars. These data are combined with 2\,MASS and IRAC photometry in Fig.~\ref{fig:seds}. 
The $50$ best-matching 
SEDs constrain the mass of the central object to $5-15\,{\rm M_\odot}$.
Most of the best fitting models have a disk inclination $> 75^\circ$, i.e. close to edge-on. The disk thus
suppresses the observable photospheric flux responsible for the optical emission. It may even be 
conjectured that the non-detection of SVS\,12 in X-rays results from the absorption of its coronal
emission. Recall, however, that the presence of an X-ray production mechanism on Herbig stars is not obvious; 
we refer to \cite{Stelzer09.1} for more details on this subject.  
%
%

\subsection{Disk fraction}\label{subsect:results_diskfraction}

%
%
In the total sample of YSOs for which a mass could be estimated from the $J$ vs. $J-H$ diagram,
we count 
$26$ Class\,II objects and $25$ Class\,III objects (see first row of Table~\ref{tab:samples}). 
This corresponds to a disk fraction of 
$51 \pm 14$\,\%. The uncertainty represents the $95$\,\% confidence limit for a binominal distribution. 
Considering only the mass-limited core sample there are 
$16$ Class\,II and $16$ Class\,III objects. 
Therefore, the disk fraction for the core of NGC\,7129 is 
$50 \pm 18$\,\%. 
In the mass-limited lightly absorbed core sample, there are $7$ Class\,II and $14$\,Class\,III sources,
and we derive a disk fraction of $33^{+24}_{-19}$\,\%.
We recall, that this last sample is the least biased one but, obviously, also the least complete one in
terms of number of cluster members. 

\cite{Gutermuth04.1} have estimated the fraction of disk-bearing stars in the core of NGC\,7129 
using a sample selected from the $J-H$ vs. $H - [4.5]$ diagram with a brightness ($J<15.5$\,mag) 
and extinction ($A_{\rm V} < 6$\,mag) cutoff. With help of a control field they estimated 
that $20$\,\% in the sample without IR excess emission are background objects, and derive a
disk fraction of $54 \pm 14$\,\%. 
This number is statistically similar to the value obtained for our mass-limited core sample 
without extinction cutoff.
For the absorption limited case we obtain a smaller disk fraction than \cite{Gutermuth04.1}. 
The use of X-ray emission as a diagnostic for identifying the young
diskless population allows us to reduce the uncertainties due to background contamination. 
Note that, there are no significant differences in the derived disk fraction between our 
YSO classification method and the more sophisticated approach of G09. 
Using their YSO status together with our sample restrictions (mass-limit, extinction limit and confinement in
the core), we find a disk fraction of  
$39^{+25}_{-22}$\,\%. 

We compare the disk fraction in NGC\,7129 with literature values for other clusters and star forming 
regions in the same age range. Given the possibility of a mass dependence in the disk lifetimes, 
care has to be taken to compare similar mass regimes. The majority of the sources in NGC\,7129 
have masses between $0.5$ and $2.5$$\,M_{\odot}$ (see Fig.~\ref{fig:j_jh}). For consistency reasons, 
we limit ourselves to disk fractions estimated based on {\em Spitzer}/IRAC data.

In Cha\,I ($\sim 2$\,Myr), the disk fraction is roughly at 50\% with an error bar 
of $\pm 6$\% from 0.1 to 3$\,M_{\odot}$ \citep{Damjanov07.1, Luhman08.1}. This sample is dominated by stars 
with sub-solar masses; there are indications that the number is somewhat higher for $M>1\,M_{\odot}$. 
In the cluster IC\,348, at $2-3$\,Myr, the total fraction of objects harboring disks is 
$50\pm 6$\,\%. Here the stars with $M>1\,M_{\odot}$ have a significantly lower disk fraction of 
$\sim 20$\,\% \citep{Lada06.2}.
The cluster $\sigma$\,Ori with an estimated age of 3\,Myr has been found to have a disk fraction of 
$36\pm 4$\,\% for stars with $0.1<M<1\,M_{\odot}$, and $27\pm 7$\,\% for $1-2\,{\rm M_{\odot}}$ \citep{Hernandez07.1}. 
Finally, the disk fraction in the 5\,Myr old Upper\,Sco star forming region is $\sim 19$\% for 
$0.1-1\,M_{\odot}$, and close to zero for higher mass stars \citep{Carpenter06.1}.

As seen from this list, the disk fraction declines steeply with age between 1 and 5\,Myr 
\citep[see also ][]{Haisch01.3}. The value we derive for the disk fraction of 
the absorption limited core sample of NGC\,7129 with
$M>0.5\,M_\odot$ is clearly higher than in Upper\,Sco and lower than in Cha\,I and IC\,348.
Therefore, the age of NGC\,7129 is probably in between that of those regions, and similar to
$\sigma$\,Ori, i.e. $\sim 3$\,Myr.

\subsection{X-ray luminosity functions}\label{subsect:results_xlf}

Using the ASURV environment \citep{Feigelson85.1}, 
we have computed the XLF for the two core samples of NGC\,7129,
the mass-limited and the mass- and absorption-limited one. 
Three stars with $J < 10.0$\,mag are not considered
in this part of the analysis as they are hot stars according to Fig.~\ref{fig:j_jh} and
their X-ray emission may not originate in the corona but in winds. The two X-ray detected hot
stars are identified with the dominating B-type stars of the reflection nebula, SVS\,8 and SVS\,7,  
and the hot star not detected with our algorithm is the Herbig star SVS\,12. 
The X-ray emission of all three stars has been discussed in detail by \cite{Stelzer09.1}. 

\begin{figure} 
\begin{center}
\resizebox{8.5cm}{!}{\includegraphics{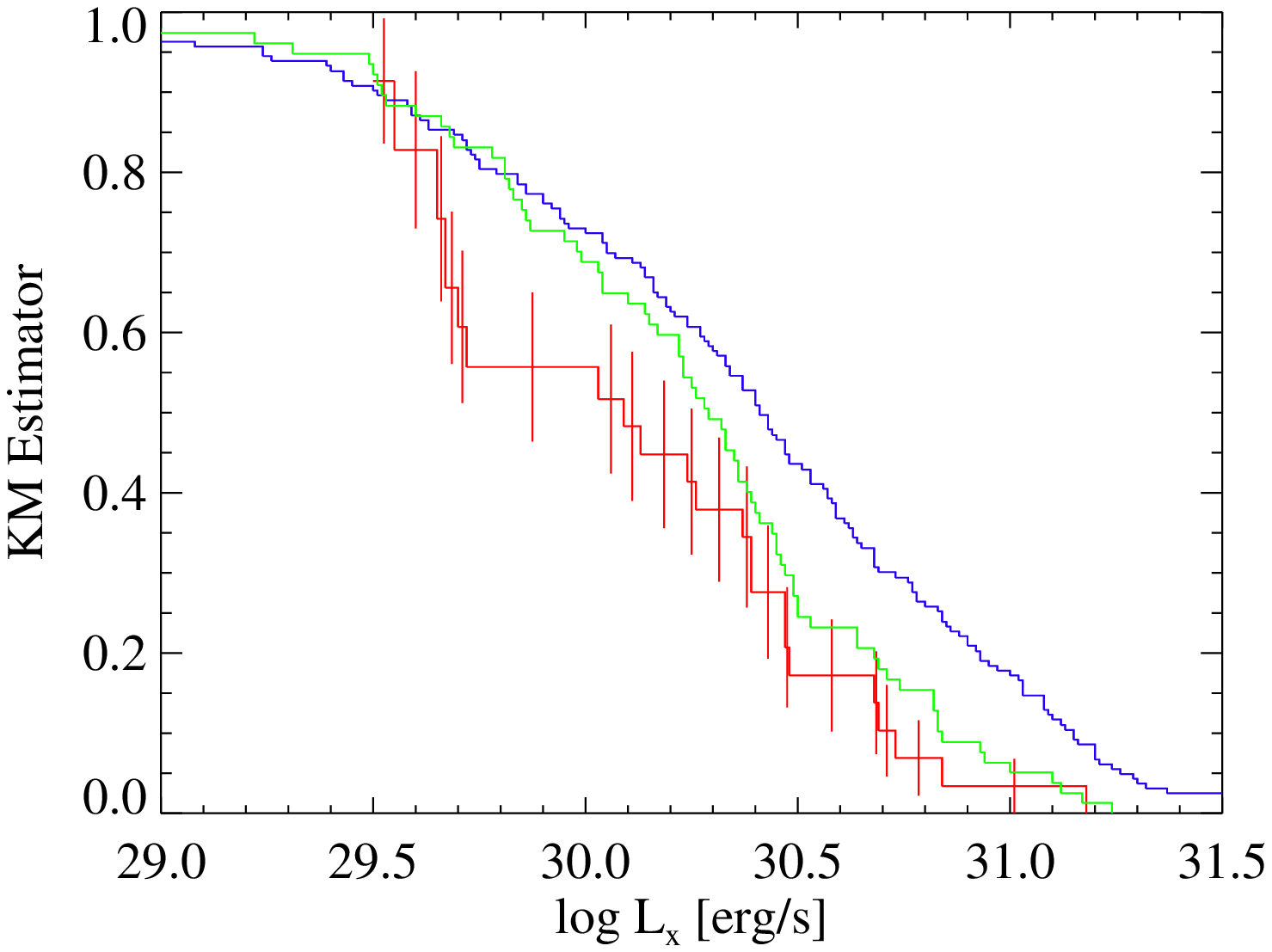}}
\resizebox{8.5cm}{!}{\includegraphics{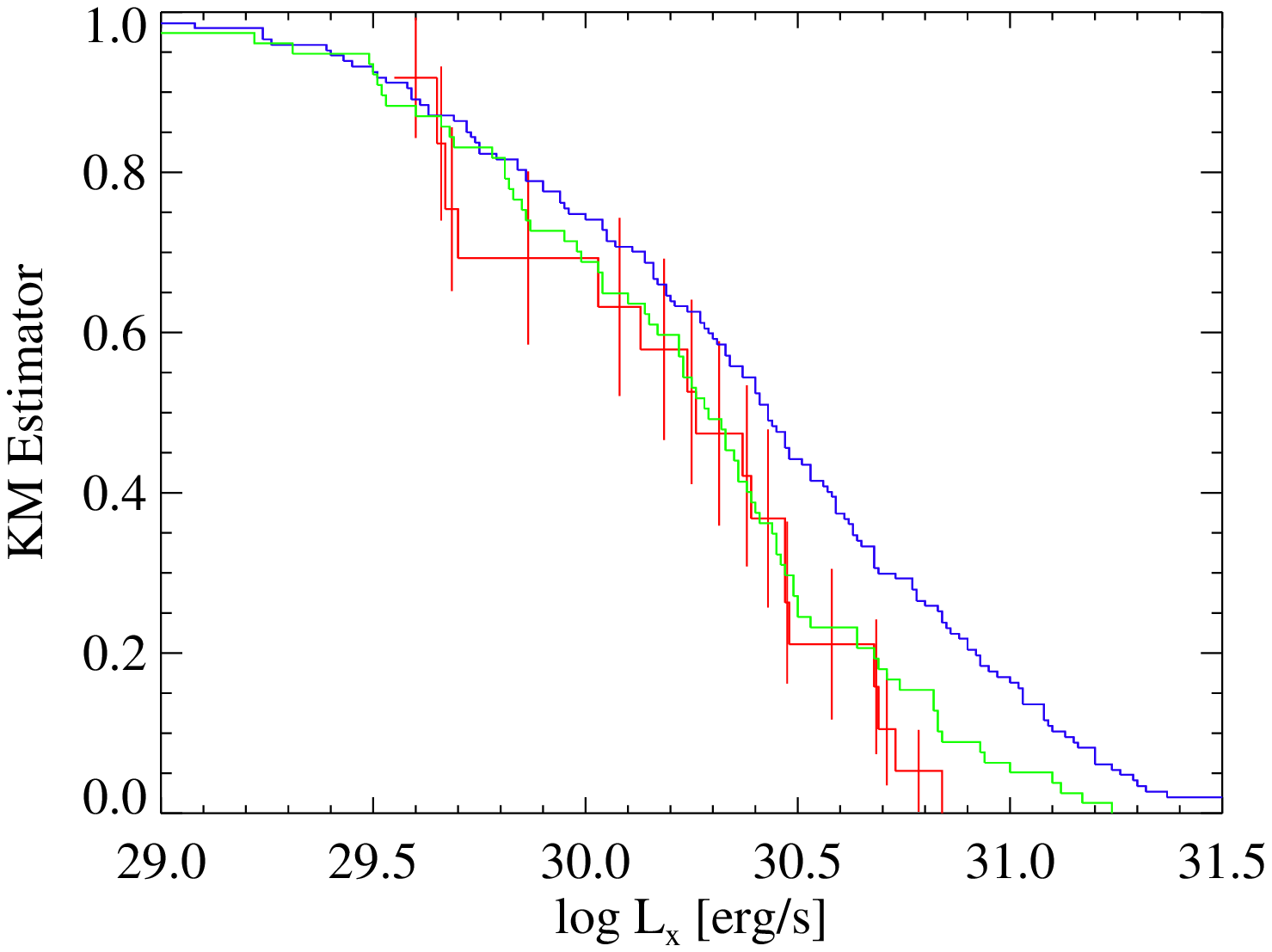}}
\caption{X-ray luminosity functions for the core sample of NGC\,7129 (red, with error bars) 
compared to the distributions for the ONC (blue, rightmost line) and NGC\,2264 (green). 
The ONC and NGC\,2264 comprise the mass range $0.5 > M > 2.5\,M_\odot$. 
The NGC\,7129 distributions are composed from the X-ray 
luminosities derived with the $N_{\rm H}$ values obtained from the X-ray hardness ratios
(see Sect.~\ref{sect:chandra}).
In the top panel we show the mass-limited core sample, 
while in the bottom panel an extinction cutoff was introduced ($A_{\rm V} < 5$\,mag). 
} 
\label{fig:xlf}
\end{center}
\end{figure}

The XLF for the mass-limited core sample of NGC\,7129 with X-ray luminosities 
using the $N_{\rm H}$ obtained from the hardness ratios is shown in the top panel of Fig.~\ref{fig:xlf}.  
Without the three hot stars this sample consists of 
$29$ stars composed of $15$ Class\,III and $14$ Class\,II sources. Eight of the latter ones 
have X-ray upper limits. 
The median X-ray luminosity is 
$(\log{L_{\rm x}})_{\rm med}\,{\rm [erg/s]} = 30.1$. 
For comparison we also show the XLF for two other samples of young stars, the 
ONC and NGC\,2264. The data for the ONC were 
extracted from the {\em Chandra Orion Ultradeep Project} (COUP) \citep[see][]{Getman05.1}.
After eliminating the probable non-members identified by \cite{Getman05.2} from the COUP source list,
we have limited the XLF of the ONC to stars with masses in the range $0.5...2.5\,{\rm M_\odot}$.
As described in Sect.~\ref{subsect:census_contam} this is the approximate mass range covered by the NGC\,7129 
sample. Similarly, the XLF of NGC\,2264 shown in green in 
Fig.~\ref{fig:xlf}, obtained with data from \cite{Flaccomio06.1}, 
is restricted to stars with spectroscopically determined masses between $0.5$ and $2.5\,M_\odot$.
In this mass range, the median X-ray luminosity for the ONC is $(\log{L_{\rm x}})_{\rm med}\,{\rm [erg/s]} = 30.4$ 
and for NGC\,2264 it is $(\log{L_{\rm x}})_{\rm med}\,{\rm [erg/s]} = 30.3$. 
The precise choice of the upper mass bound for the two comparison clusters is not a concern. We have lowered it 
to $2.0\,{\rm M_\odot}$ and found no dramatic change of the median X-ray luminosities. 

In the lower panel of Fig.~\ref{fig:xlf} we show the mass-limited lightly absorbed core sample
of NGC\,7129 together with the ONC and NGC\,2264 samples to which the same restrictions have been applied (i.e. mass range
of $0.5...2.5\,M_\odot$ and $A_{\rm V} < 5$\,mag). The median X-ray luminosity for this NGC\,7129 sample is
higher than for the one without limitations on the absorption ($(\log{L_{\rm x}})_{\rm med}\,{\rm [erg/s]} = 30.3$)
because it has fewer upper limits. For the ONC less than $10$\,\% of the sample is removed due to the absorption
cutoff and there is no change in the median $L_{\rm x}$. In NGC\,2264 all stars have $A_{\rm V} < 5$\,mag. 

As mentioned in Sect.~\ref{sect:chandra} the column densities (derived from the hardness ratios)
that are used to compute the X-ray luminosities are associated with large uncertainties, and as a cross-check
we have applied an independent method to derive $N_{\rm H}$ values from the optical extinction extracted
from near-IR photometry. 
The XLF for the mass-limited core sample and that for the mass-limited lightly absorbed core sample of NGC\,7129 
computed with those latter $N_{\rm H}$ estimates  
both have a median of $(\log{L_{\rm x}})_{\rm med}\,{\rm [erg/s]} = 30.2$.
This shows that the XLF are relatively stable despite the considerable uncertainties in the computation
of the luminosities.  
 
Various uncertainties are related to the calculation of the XLFs, such that the similarity to that
of the other clusters is remarkable. 
We can speculate on the origin of the remaining differences. A possible contribution is from the uncertain
mass estimates by means of photometric observations. 
In addition, the uncertain distance estimate for NGC\,7129
may play a role. To bring its XLF up by a factor $\sim 2$ the distance would have to be 
$1.4$\,kpc instead of the canonical value of $1$\,kpc. While this does not seem impossible, 
an investigation of the consequences (that include a higher mass limit for our observations) is
clearly unfeasible with the presently available data. 
Spectroscopic observations of NGC\,7129 will shed light on this issue. 
%
%

\subsection{X-ray emission from protostars}\label{subsect:results_class1}

None of the Class\,0/I candidates is detected in X-rays. The only X-ray detected object classified
Class\,0/I according to our IR selection does not survive our contamination tests and is flagged as
AGN in Table~\ref{tab:counterparts_b}.
We have derived upper limits on the order of $\log{L_{\rm x}}\,{\rm [erg/s]} \sim 29.5...30.0$
for the undetected Class\,0/I candidates. 
However, the extinction of these embedded objects is likely to be much higher than the assumed
median of the X-ray detected Class\,II sources, and
consequently the X-ray luminosities are probably higher than estimated. For $A_{\rm V} = 15$\,mag,
a moderate absorption for a protostar, the gas column density is expected to be 
$\log{N_{\rm H}}\,{\rm [cm^{-2}]} \sim 22.4$, and the upper limits 
to the X-ray luminosities higher by a factor four with
respect to the values given in Table~\ref{tab:xraytab_ul}, i.e. $\log{L_{\rm x}}\,{\rm [erg/s]} \sim 30.1...30.6$.  
This is in the middle of the range of X-ray luminosities for Class\,0/I sources 
seen in the ONC \citep{Prisinzano08.1}. Therefore, the non-detection of most Class\,0/I sources in 
NGC\,7129 is not surprising.
There is also no X-ray source at the position of the Class\,0 object FIRS-2.

\section{Summary}\label{sect:conclusions}

We present a combined X-ray and near/mid-IR census of the young
star cluster associated with the reflection nebula NGC\,7129. 
YSOs are characterised by excess emission in the IR 
due to the presence of disks and elevated levels of X-ray emission
due to enhanced coronal activity compared with field stars. Thus, the
two wavelength regimes provide complementary information which, in
principle, allow us to obtain a complete census of the cluster
population.

In a $22$\,ksec long {\em Chandra} observation pointed at the Herbig star
SVS\,12 we find in total $59$ X-ray sources, 
from which $47$ are matched with 2\,MASS photometry. $50$ X-ray
sources have a counterpart seen in at least one IRAC or MIPS band in
the {\em Spitzer} catalog kindly provided to us by R.Gutermuth
(see also G09).

Based on literature recommendations, 
we define a set of criteria to distinguish between Class\,I/II objects (with disks and/or envelope)
and diskless Class\,III objects from 
their color in IRAC and 2\,MASS photometry. 
For Class\,III candidates we require, besides their detection in X-rays,
that either 2\,MASS or IRAC photometry is complete. 
After likely extragalactic sources have been removed, 
there are $0$ Class\,0/I, $15$ Class\,II and $30$ Class\,III candidates among the X-ray
detections.
 In addition, we have identified $8$\,Class\,0/I candidates and $34$\,Class\,II candidates 
without X-ray detection in the {\em Spitzer}/2\,MASS catalog.

Based on a comparison with evolutionary tracks, we estimate that this
sample is complete down to $0.5\,M_{\odot}$. The mass-limited census
($M>0.5\,M_{\odot}$) is composed of 26 Class\,II and 25 Class\,III
objects, from which 16 Class\,II and 16 Class\,III are densely
clustered within $1.7^\prime$ around the center of the cloud. 
The sample may not be complete for strongly absorbed cluster stars.
Therefore, we defined an absorption limited subsample, that comprises
$7$ Class\,II and $14$ Class\,III sources.
Spectroscopy is required to further assess the completeness and properties of this
sample.

From our census, we estimate the disk fraction in NGC\,7129 to be 
$\sim 33$\,\%. This is lower than the disk fraction of Cha\,I and IC\,348 ($2-3$\,Myr),
higher than the disk fraction of Upper Sco ($5$\,Myr), and comparable with the disk fraction of 
$\sigma$\,Ori ($3$\,Myr). Therefore, the age of 
NGC\,7129 is likely to be $\sim 3$\,Myr. 
The same age is suggested by the XLF that we find to be similar to that of the $3$\,Myr
old NGC\,2264 cluster but fainter than that of the ONC ($1$\,Myr).

For both methods used to derive the luminosities, the XLF 
of the mass-limited core sample in NGC\,7129 are somewhat lower than those of NGC\,2264 and the ONC. Better agreement
with NGC\,2264 is obtained for the lightly absorbed mass-limited core sample, indicating that this sample is likely
representative for the cluster.

Based on the broadband shape of their SED, we identify four peculiar
sources. Further analysis using the grid of model SEDs by 
\cite{Robitaille06.1} reveals that one object is a clear candidate for a
`transition' disk with an inner opacity hole (radius $10$\,AU). For the
well-known HAeBe star LkH$\alpha$\,234 we find evidence for an
edge-on disk which blocks the optical light from the central object.

The presently available data set is subject to 
various uncertainties related with the cluster properties (such as mass distribution or distance) 
and with the X-ray data itself (such as the spectral shape of the sources). 
This underlines the need to characterize the cluster members with optical photometric and spectroscopic
observations.

\clearpage

\begin{table*}\tiny\begin{center}
\caption{X-ray parameters for all detected sources.}
\label{tab:xraytab}
\begin{tabular}{lccrrrrrr}\hline
ID     &  \multicolumn{2}{c}{X-ray position}        & Offax      & Expo & Rate               & signif & S/N & $\log{L_{\rm x}}^{(a)}$  \\
       &  $\alpha_{\rm 2000}$ & $\delta_{\rm 2000}$ & [$\prime$] & [s]  & [$10^{-3}$\,cts/s] &        &     & [erg/s]            \\ \hline
NGC7129-S3-X1  & 21 42 59.62 & +66 04 33.9 & $  2.4$ & $  22462$ & $ 3.3 \pm 0.4$ & $   30.0$ & $  110.8$ & $ 30.9$ \\
NGC7129-S3-X2  & 21 42 47.09 & +66 04 57.8 & $  2.8$ & $  22193$ & $ 5.8 \pm 0.5$ & $   49.4$ & $  188.4$ & $ 30.7$ \\
NGC7129-S3-X3  & 21 42 46.09 & +66 05 13.8 & $  2.7$ & $  22224$ & $ 2.2 \pm 0.3$ & $   20.1$ & $   71.8$ & $ 30.4$ \\
NGC7129-S3-X4  & 21 42 34.73 & +66 05 18.8 & $  3.7$ & $  21969$ & $ 3.3 \pm 0.4$ & $   25.5$ & $   69.8$ & $ 30.5$ \\
NGC7129-S3-X6  & 21 42 58.39 & +66 05 27.3 & $  1.7$ & $  22587$ & $ 1.5 \pm 0.3$ & $   14.9$ & $   53.2$ & $ 30.3$ \\
NGC7129-S3-X7  & 21 43 18.10 & +66 05 35.1 & $  1.6$ & $  22499$ & $ 0.8 \pm 0.2$ & $    8.5$ & $   33.9$ & $ 30.5$ \\
NGC7129-S3-X8  & 21 42 49.89 & +66 05 42.7 & $  2.1$ & $  22397$ & $ 0.3 \pm 0.1$ & $    3.2$ & $   11.7$ & $ 29.5$ \\
NGC7129-S3-X9  & 21 43 16.86 & +66 05 48.6 & $  1.4$ & $  22516$ & $ 0.6 \pm 0.2$ & $    6.4$ & $   28.6$ & $ 30.4$ \\
NGC7129-S3-X10 & 21 42 56.28 & +66 06 02.0 & $  1.4$ & $  22548$ & $ 3.6 \pm 0.4$ & $   32.9$ & $  126.8$ & $ 30.5$ \\
NGC7129-S3-X11 & 21 42 50.96 & +66 06 03.8 & $  1.9$ & $  22489$ & $ 2.2 \pm 0.3$ & $   19.4$ & $   83.7$ & $ 30.4$ \\
NGC7129-S3-X12 & 21 42 58.82 & +66 06 10.2 & $  1.1$ & $  22595$ & $24.5 \pm 1.0$ & $   72.4$ & $  883.9$ & $  ...$ \\
NGC7129-S3-X13 & 21 42 58.60 & +66 06 10.3 & $  1.2$ & $  22586$ & $25.7 \pm 1.1$ & $  105.7$ & $  942.4$ & $ 31.4$ \\
NGC7129-S3-X14 & 21 43 21.07 & +66 06 22.8 & $  1.4$ & $  22389$ & $ 1.2 \pm 0.2$ & $   12.5$ & $   58.3$ & $ 30.1$ \\
NGC7129-S3-X15 & 21 42 57.22 & +66 06 34.8 & $  1.1$ & $  22567$ & $ 0.4 \pm 0.1$ & $    4.6$ & $   15.1$ & $ 29.6$ \\
NGC7129-S3-X16 & 21 42 50.19 & +66 06 35.0 & $  1.8$ & $  22480$ & $ 3.8 \pm 0.4$ & $   33.0$ & $  136.2$ & $ 30.6$ \\
NGC7129-S3-X17 & 21 42 58.77 & +66 06 36.8 & $  0.9$ & $  22576$ & $ 1.1 \pm 0.2$ & $   11.2$ & $   45.5$ & $ 30.2$ \\
NGC7129-S3-X18 & 21 42 56.77 & +66 06 37.2 & $  1.1$ & $  22565$ & $ 1.7 \pm 0.3$ & $   16.5$ & $   62.1$ & $ 30.7$ \\
NGC7129-S3-X19 & 21 42 59.99 & +66 06 42.5 & $  0.8$ & $  22599$ & $ 0.3 \pm 0.1$ & $    3.6$ & $   15.5$ & $ 29.6$ \\
NGC7129-S3-X20 & 21 43 01.89 & +66 06 44.7 & $  0.6$ & $  21330$ & $ 2.7 \pm 0.4$ & $   25.4$ & $   99.4$ & $ 30.5$ \\
NGC7129-S3-X21 & 21 43 20.90 & +66 06 51.8 & $  1.3$ & $  22331$ & $ 0.8 \pm 0.2$ & $    8.1$ & $   37.2$ & $ 30.2$ \\
NGC7129-S3-X22 & 21 43 05.01 & +66 06 53.2 & $  0.3$ & $  22598$ & $ 2.1 \pm 0.3$ & $   20.0$ & $   83.9$ & $ 30.4$ \\
NGC7129-S3-X23 & 21 42 47.92 & +66 06 52.9 & $  2.0$ & $  22425$ & $ 5.1 \pm 0.5$ & $   41.0$ & $  197.4$ & $ 30.7$ \\
NGC7129-S3-X24 & 21 42 59.78 & +66 06 55.7 & $  0.8$ & $  22571$ & $ 1.0 \pm 0.2$ & $   10.8$ & $   37.3$ & $  ...$ \\
NGC7129-S3-X25 & 21 42 52.65 & +66 06 57.2 & $  1.5$ & $  22493$ & $ 0.8 \pm 0.2$ & $    7.9$ & $   27.7$ & $ 30.1$ \\
NGC7129-S3-X26 & 21 42 54.92 & +66 07 21.1 & $  1.4$ & $  22465$ & $ 1.2 \pm 0.2$ & $   11.0$ & $   37.7$ & $ 30.1$ \\
NGC7129-S3-X27 & 21 43 05.18 & +66 07 33.9 & $  0.8$ & $  22475$ & $ 2.0 \pm 0.3$ & $   19.7$ & $  116.4$ & $ 31.1$ \\
NGC7129-S3-X28 & 21 43 15.31 & +66 07 56.9 & $  1.3$ & $  22291$ & $ 2.8 \pm 0.4$ & $   27.0$ & $  108.6$ & $ 31.0$ \\
NGC7129-S3-X29 & 21 42 53.49 & +66 08 05.2 & $  1.9$ & $  22331$ & $ 2.8 \pm 0.4$ & $   24.3$ & $  115.6$ & $ 30.5$ \\
NGC7129-S3-X30 & 21 43 11.62 & +66 09 11.4 & $  2.4$ & $  21997$ & $ 4.0 \pm 0.4$ & $   34.1$ & $  201.6$ & $ 30.5$ \\
NGC7129-S3-X31 & 21 42 46.09 & +66 05 56.3 & $  2.4$ & $  20019$ & $ 1.2 \pm 0.2$ & $   11.0$ & $   36.6$ & $ 30.8$ \\
NGC7129-S3-X32 & 21 42 54.84 & +66 06 13.5 & $  1.5$ & $  22540$ & $ 1.0 \pm 0.2$ & $    6.7$ & $   36.4$ & $ 30.0$ \\
NGC7129-S3-X35 & 21 42 51.96 & +66 06 33.4 & $  1.6$ & $  22502$ & $ 0.5 \pm 0.1$ & $    4.6$ & $   18.0$ & $ 29.7$ \\
NGC7129-S3-X36 & 21 42 54.75 & +66 06 35.3 & $  1.3$ & $  22535$ & $ 1.2 \pm 0.2$ & $   12.4$ & $   50.0$ & $ 30.7$ \\
NGC7129-S3-X37 & 21 42 55.72 & +66 06 45.0 & $  1.2$ & $  22541$ & $ 0.2 \pm 0.1$ & $    3.2$ & $    9.7$ & $ 29.4$ \\
NGC7129-S3-X38 & 21 43 07.15 & +66 06 54.0 & $  0.1$ & $  22595$ & $ 0.7 \pm 0.2$ & $    6.8$ & $   34.9$ & $  ...$ \\
NGC7129-S3-X39 & 21 42 46.91 & +66 06 57.4 & $  2.1$ & $  22383$ & $ 0.6 \pm 0.2$ & $    5.7$ & $   24.6$ & $ 29.8$ \\
NGC7129-S3-X40 & 21 42 45.32 & +66 07 04.4 & $  2.3$ & $  22327$ & $ 1.0 \pm 0.2$ & $   10.6$ & $   39.9$ & $ 30.4$ \\
NGC7129-S3-X41 & 21 42 38.91 & +66 07 08.7 & $  2.9$ & $  21036$ & $ 0.7 \pm 0.2$ & $    5.0$ & $   22.3$ & $  ...$ \\
NGC7129-S3-X42 & 21 43 19.39 & +66 07 21.4 & $  1.3$ & $  22266$ & $ 0.6 \pm 0.2$ & $    5.9$ & $   26.6$ & $ 29.8$ \\
NGC7129-S3-X43 & 21 42 58.34 & +66 07 26.3 & $  1.1$ & $  21325$ & $ 0.5 \pm 0.2$ & $    4.7$ & $   17.7$ & $ 29.7$ \\
NGC7129-S3-X45 & 21 42 54.08 & +66 08 14.9 & $  2.0$ & $  21951$ & $ 0.6 \pm 0.2$ & $    7.0$ & $   22.7$ & $ 31.2$ \\
NGC7129-S3-X47 & 21 42 58.26 & +66 05 40.0 & $  1.5$ & $  22571$ & $ 0.3 \pm 0.1$ & $    3.3$ & $   10.6$ & $  ...$ \\
NGC7129-S3-X48 & 21 42 55.87 & +66 05 42.8 & $  1.7$ & $  22541$ & $ 0.4 \pm 0.1$ & $    3.6$ & $   13.4$ & $ 29.6$ \\
NGC7129-S3-X50 & 21 43 03.01 & +66 06 55.9 & $  0.5$ & $  21651$ & $ 0.4 \pm 0.1$ & $    4.0$ & $   16.3$ & $  ...$ \\
NGC7129-S3-X51 & 21 43 01.71 & +66 07 08.9 & $  0.7$ & $  21646$ & $ 0.4 \pm 0.1$ & $    4.6$ & $   19.0$ & $ 29.7$ \\
NGC7129-S3-X52 & 21 42 40.34 & +66 10 07.2 & $  4.3$ & $   8466$ & $ 9.3 \pm 1.0$ & $   16.7$ & $  127.7$ & $ 31.3$ \\
NGC7129-S3-X53 & 21 42 24.73 & +66 01 41.0 & $  6.8$ & $  20570$ & $ 4.1 \pm 0.4$ & $   12.5$ & $   22.2$ & $ 31.5$ \\
NGC7129-S3-X56 & 21 43 29.28 & +66 03 31.5 & $  4.0$ & $  21103$ & $ 0.5 \pm 0.2$ & $    3.9$ & $   14.9$ & $ 29.8$ \\
NGC7129-S3-X58 & 21 42 19.32 & +66 07 26.6 & $  4.9$ & $  21272$ & $ 0.6 \pm 0.2$ & $    3.7$ & $    7.7$ & $  ...$ \\
NGC7129-S2-X1  & 21 43 31.80 & +66 08 50.5 & $  3.1$ & $  22254$ & $ 1.9 \pm 0.3$ & $   11.3$ & $  104.5$ & $ 30.7$ \\
NGC7129-S2-X2  & 21 43 24.89 & +66 07 34.1 & $  1.9$ & $  10708$ & $ 1.7 \pm 0.4$ & $    9.4$ & $   68.4$ & $  ...$ \\
NGC7129-S2-X3  & 21 43 20.37 & +66 08 20.2 & $  2.0$ & $  10346$ & $ 0.8 \pm 0.3$ & $    4.5$ & $   30.3$ & $  ...$ \\
NGC7129-S2-X4  & 21 43 27.02 & +66 09 36.5 & $  3.4$ & $  22154$ & $ 2.1 \pm 0.3$ & $   12.7$ & $  116.0$ & $ 30.2$ \\
NGC7129-S2-X5  & 21 43 43.43 & +66 07 30.7 & $  3.7$ & $  22071$ & $ 1.9 \pm 0.3$ & $   15.7$ & $   68.2$ & $ 30.7$ \\
NGC7129-S2-X7  & 21 43 53.11 & +66 08 09.8 & $  4.8$ & $  20210$ & $ 0.7 \pm 0.2$ & $    3.9$ & $   15.1$ & $  ...$ \\
NGC7129-S2-X8  & 21 43 33.96 & +66 10 43.4 & $  4.7$ & $  19757$ & $ 1.4 \pm 0.3$ & $    6.8$ & $   37.1$ & $ 30.5$ \\
NGC7129-S2-X9  & 21 43 36.29 & +66 11 32.9 & $  5.5$ & $  19218$ & $ 5.7 \pm 0.5$ & $   21.8$ & $   99.0$ & $ 30.8$ \\
NGC7129-S2-X10 & 21 44 29.55 & +66 07 06.8 & $  8.3$ & $  19639$ & $ 1.2 \pm 0.3$ & $    5.6$ & $   17.1$ & $  ...$ \\
NGC7129-S2-X14 & 21 44 26.43 & +66 10 29.1 & $  8.7$ & $  18704$ & $ 1.0 \pm 0.2$ & $    4.2$ & $   11.0$ & $  ...$ \\
\hline
\multicolumn{9}{l}{(a) - The X-ray luminosities were computed using the $kT$ and $N_{\rm H}$ obtained from the hardness ratios as described in} \\
\multicolumn{9}{l}{Sect.~\ref{sect:chandra}. $L_{\rm x}$ is given only for objects selected as pre-MS candidates with the criteria described in Sect.~\ref{sect:census}.} \\
 \end{tabular}\end{center}\end{table*}

\begin{sidewaystable*}\tiny\begin{center}
\caption{Identification and offset between X-ray and optical/IR position for counterparts of X-ray sources.}
\label{tab:counterparts_a}
\begin{tabular}{llrrrrrrrrrrrrrcc}\hline
NGC7129-...  &  Name & $\Delta_{\rm O-X}$   & $\Delta_{\rm 2M-X}$  \\
             &       & [${\prime\prime}$]   & [${\prime\prime}$]           \\
\hline
...-S3-X1 & HL85-S\,25 & $0.46$ & $0.10$ \\
...-S3-X2 &            & $   -$ & $0.11$ \\
...-S3-X3 & HL85-S\,II & $1.82$ & $0.23$ \\
...-S3-X4 &            & $   -$ & $0.22$ \\
...-S3-X6 &            & $   -$ & $0.10$ \\
...-S3-X7 &            & $   -$ & $0.11$ \\
...-S3-X8 &            & $   -$ & $0.11$ \\
...-S3-X9 & MMN-19     & $1.15$ & $0.09$ \\
...-S3-X10 &           & $   -$ & $0.10$ \\
...-S3-X11 &           & $   -$ & $0.22$ \\
...-S3-X12 &           & $   -$ & $   -$ \\
...-S3-X13 & SVS\,8    & $1.27$ & $0.07$ \\
...-S3-X14 &           & $   -$ & $1.44$ \\
...-S3-X15 &           & $   -$ & $0.17$ \\
...-S3-X16 & SVS\,7    & $1.30$ & $0.09$ \\
...-S3-X17 &           & $   -$ & $0.18$ \\
...-S3-X18 &           & $   -$ & $0.10$ \\
...-S3-X19 &           & $   -$ & $0.13$ \\
...-S3-X20 &           & $   -$ & $0.05$ \\
...-S3-X21 &           & $   -$ & $0.08$ \\
...-S3-X22 &           & $   -$ & $0.22$ \\
...-S3-X23 &           & $   -$ & $0.01$ \\
...-S3-X24 &           & $   -$ & $   -$ \\
...-S3-X25 & MMN-6     & $1.09$ & $0.10$ \\
...-S3-X26 &           & $   -$ & $0.12$ \\
...-S3-X27 &           & $   -$ & $0.17$ \\
...-S3-X28 &           & $   -$ & $0.16$ \\
...-S3-X29 & MMN-9     & $1.94$ & $0.18$ \\
...-S3-X30 & MMN-17    & $1.30$ & $0.12$ \\
...-S3-X31 &           & $   -$ & $0.13$ \\
...-S3-X32 & MMN-10    & $1.94$ & $0.82$ \\
...-S3-X35 &           & $   -$ & $0.27$ \\
...-S3-X36 & MEG93\,Star\,2 & $1.51$ & $0.16$ \\
...-S3-X37 &           & $   -$ & $0.30$ \\
...-S3-X38 &           & $   -$ & $   -$ \\
...-S3-X39 &           & $   -$ & $0.13$ \\
...-S3-X40 &           & $   -$ & $0.13$ \\
...-S3-X41 &           & $   -$ & $   -$ \\
...-S3-X42 &           & $   -$ & $0.11$ \\
...-S3-X43 &           & $   -$ & $0.09$ \\
...-S3-X45 &           & $   -$ & $0.15$ \\
...-S3-X47 &           & $   -$ & $0.29$ \\
...-S3-X48 &           & $   -$ & $0.29$ \\
...-S3-X50 &           & $   -$ & $   -$ \\
...-S3-X51 & SVS\,13   & $1.49$ & $0.05$ \\
...-S3-X52 & SVS\,2    & $1.40$ & $0.31$ \\
...-S3-X53 &           & $   -$ & $   -$ \\
...-S3-X56 &           & $   -$ & $0.63$ \\
...-S3-X58 &           & $   -$ & $   -$ \\
...-S2-X1 & HL85-N\,27 & $1.42$ & $0.18$ \\
...-S2-X2 &            & $   -$ & $   -$ \\
...-S2-X3 &            & $   -$ & $0.54$ \\
...-S2-X4 &            & $   -$ & $0.38$ \\
...-S2-X5 & MMN-22     & $1.32$ & $0.14$ \\
...-S2-X7 &            & $   -$ & $   -$ \\
...-S2-X8 &            & $   -$ & $   -$ \\
...-S2-X9 & SVS\,15    & $1.96$ & $0.17$ \\
...-S2-X10 &           & $   -$ & $   -$ \\
...-S2-X14 &           & $   -$ & $   -$ \\
\hline \end{tabular}
\end{center}\end{sidewaystable*}

\begin{sidewaystable*}\tiny\begin{center}
\caption{Photometry and YSO classification for optical and IR counterparts of X-ray sources. Asterisks in column `YSO Class' identify objects selected on basis of near-IR data.}
\label{tab:counterparts_b}
\begin{tabular}{lrrrrrrrrrrrcc}\hline
NGC7129-...  & $V$   & $R$   & $I$   & $J$   & $H$   & $K$   & $[3.6]$ & $[4.5]$ &  $[5.8]$ & $[8.0]$ & $[24]$   & YSO   & AGN? \\
             & [mag] & [mag] & [mag] & [mag] & [mag] & [mag] & [mag]   & [mag]   & [mag]    & [mag]   & [mag]    & Class &      \\ \hline
...-S3-X1 & $  18.32$ & $  16.23$ & $  15.11$ & $  12.89 \pm    0.03$ & $  11.62 \pm    0.03$ & $  10.95 \pm    0.02$ & $  10.18 \pm    0.00$ & $   9.73 \pm    0.00$ & $   9.39 \pm    0.00$ & $   8.91 \pm    0.02
$ & $   6.81 \pm    0.21$ &   II & \\
...-S3-X2 & $    ...$ & $    ...$ & $    ...$ & $  12.10 \pm    0.02$ & $  11.33 \pm    0.03$ & $  11.05 \pm    0.02$ & $  10.79 \pm    0.01$ & $  10.74 \pm    0.01$ & $  10.54 \pm    0.09$ & $  10.00 \pm    0.28$ & $
                ...$ &   III & \\
...-S3-X3 & $  10.64$ & $  10.39$ & $  10.24$ & $  10.06 \pm    0.02$ & $   9.90 \pm    0.02$ & $   9.83 \pm    0.02$ & $   9.85 \pm    0.00$ & $   9.84 \pm    0.00$ & $   9.87 \pm    0.04$ & $   9.91 \pm    0.16$ & $                ...$ &   III & \\
...-S3-X4 & $    ...$ & $    ...$ & $    ...$ & $  13.92 \pm    0.03$ & $  13.28 \pm    0.04$ & $  13.13 \pm    0.04$ & $  12.87 \pm    0.01$ & $  12.80 \pm    0.01$ & $  12.82 \pm    0.05$ & $  13.06 \pm    0.19$ & $                ...$ &   III & \\
...-S3-X6 & $    ...$ & $    ...$ & $    ...$ & $  13.96 \pm    0.03$ & $  13.03 \pm    0.03$ & $  12.64 \pm    0.03$ & $  11.79 \pm    0.03$ & $  11.39 \pm    0.03$ & $                ...$ & $                ...$ & $                ...$ &   II$^*$ & \\
...-S3-X7 & $    ...$ & $    ...$ & $    ...$ & $  14.76 \pm    0.04$ & $  13.99 \pm    0.05$ & $  13.70 \pm    0.06$ & $  13.39 \pm    0.01$ & $  13.27 \pm    0.01$ & $  13.29 \pm    0.17$ & $  13.12 \pm    0.28$ & $                ...$ &   III & \\
...-S3-X8 & $    ...$ & $    ...$ & $    ...$ & $  14.65 \pm    0.05$ & $  14.01 \pm    0.06$ & $  13.72 \pm    0.06$ & $  13.53 \pm    0.03$ & $  13.39 \pm    0.04$ & $                ...$ & $                ...$ & $                ...$ &   III & \\
...-S3-X9 & $  18.42$ & $  17.12$ & $  16.00$ & $  14.18 \pm    0.03$ & $  13.30 \pm    0.04$ & $  13.12 \pm    0.04$ & $  12.74 \pm    0.01$ & $  12.57 \pm    0.01$ & $  12.45 \pm    0.09$ & $  11.99 \pm    0.12$ & $   6.50 \pm    0.21$ &   III & \\
...-S3-X10 & $    ...$ & $    ...$ & $    ...$ & $  12.42 \pm    0.02$ & $  11.66 \pm    0.03$ & $  11.41 \pm    0.03$ & $  11.28 \pm    0.06$ & $  11.13 \pm    0.09$ & $  10.78 \pm    0.39$ & $                ...$ & $                ...$ &   III & \\
...-S3-X11 & $    ...$ & $    ...$ & $    ...$ & $  13.64 \pm    0.03$ & $  12.80 \pm    0.04$ & $  12.55 \pm    0.03$ & $  12.42 \pm    0.03$ & $  12.37 \pm    0.03$ & $  12.24 \pm    0.43$ & $                ...$ & $                ...$ &   III & \\
...-S3-X12 & $    ...$ & $    ...$ & $    ...$ & $                ...$ & $                ...$ & $                ...$ & $                ...$ & $                ...$ & $                ...$ & $                ...$ & $                ...$ & ... & \\
...-S3-X13 & $    ...$ & $    ...$ & $    ...$ & $   8.98 \pm    0.02$ & $   8.66 \pm    0.02$ & $   8.51 \pm    0.02$ & $   8.50 \pm    0.02$ & $   8.48 \pm    0.03$ & $   8.42 \pm    0.09$ & $   8.00 \pm    0.28$ & $                ...$ &   III & \\
...-S3-X14 & $    ...$ & $    ...$ & $    ...$ & $  13.56 \pm    0.03$ & $  12.90 \pm    0.04$ & $  12.75 \pm    0.04$ & $  12.44 \pm    0.00$ & $                ...$ & $  12.32 \pm    0.03$ & $  12.42 \pm    0.10$ & $                ...$ &   III & \\
...-S3-X15 & $    ...$ & $    ...$ & $    ...$ & $  13.67 \pm    0.05$ & $  12.75 \pm    0.07$ & $  12.53 \pm    0.03$ & $                ...$ & $  12.00 \pm    0.04$ & $                ...$ & $                ...$ & $                ...$ &   III & \\
...-S3-X16 & $    ...$ & $    ...$ & $    ...$ & $   8.97 \pm    0.02$ & $   8.73 \pm    0.02$ & $   8.47 \pm    0.02$ & $   8.13 \pm    0.00$ & $   7.90 \pm    0.00$ & $   7.57 \pm    0.01$ & $   7.09 \pm    0.05$ & $                ...$ &   II & \\
...-S3-X17 & $    ...$ & $    ...$ & $    ...$ & $  13.49 \pm    0.03$ & $  12.56 \pm    0.03$ & $  12.34 \pm    0.03$ & $  11.98 \pm    0.03$ & $  11.85 \pm    0.04$ & $                ...$ & $                ...$ & $                ...$ &   III & \\
...-S3-X18 & $    ...$ & $    ...$ & $    ...$ & $  13.67 \pm    0.03$ & $  12.69 \pm    0.03$ & $  12.21 \pm    0.05$ & $  11.47 \pm    0.05$ & $  11.21 \pm    0.04$ & $  10.44 \pm    0.26$ & $                ...$ & $                ...$ &   II$^*$ & \\
...-S3-X19 & $    ...$ & $    ...$ & $    ...$ & $  14.35 \pm    0.05$ & $  13.49 \pm    0.04$ & $  13.15 \pm    0.04$ & $  12.73 \pm    0.07$ & $  13.05 \pm    0.11$ & $                ...$ & $                ...$ & $                ...$ &   III & \\
...-S3-X20 & $    ...$ & $    ...$ & $    ...$ & $  13.30 \pm    0.03$ & $  12.41 \pm    0.03$ & $  12.02 \pm    0.02$ & $  11.12 \pm    0.03$ & $  10.80 \pm    0.02$ & $  11.15 \pm    0.41$ & $                ...$ & $                ...$ &   II$^*$ & \\
...-S3-X21 & $    ...$ & $    ...$ & $    ...$ & $  15.52 \pm    0.07$ & $  13.79 \pm    0.05$ & $  13.02 \pm    0.04$ & $  12.54 \pm    0.01$ & $  12.40 \pm    0.01$ & $  12.24 \pm    0.02$ & $  12.30 \pm    0.05$ & $                ...$ &   III & \\
...-S3-X22 & $    ...$ & $    ...$ & $    ...$ & $  13.05 \pm    0.04$ & $  11.97 \pm    0.05$ & $  11.26 \pm    0.07$ & $                ...$ & $                ...$ & $                ...$ & $                ...$ & $                ...$ &   III & \\
...-S3-X23 & $    ...$ & $    ...$ & $    ...$ & $  13.46 \pm    0.03$ & $  12.42 \pm    0.03$ & $  12.09 \pm    0.03$ & $  11.69 \pm    0.02$ & $  11.71 \pm    0.01$ & $  10.89 \pm    0.12$ & $                ...$ & $                ...$ &   III & \\
...-S3-X24 & $    ...$ & $    ...$ & $    ...$ & $                ...$ & $                ...$ & $                ...$ & $                ...$ & $                ...$ & $                ...$ & $                ...$ & $                ...$ & ... & \\
...-S3-X25 & $  18.30$ & $  17.04$ & $  15.78$ & $  13.82 \pm    0.03$ & $  12.65 \pm    0.03$ & $  11.80 \pm    0.03$ & $  10.81 \pm    0.01$ & $  10.24 \pm    0.00$ & $   9.77 \pm    0.04$ & $   9.15 \pm    0.06$ & $                ...$ &   II & \\
...-S3-X26 & $    ...$ & $    ...$ & $    ...$ & $  14.25 \pm    0.05$ & $  13.19 \pm    0.06$ & $  12.84 \pm    0.04$ & $  12.50 \pm    0.03$ & $  12.44 \pm    0.03$ & $  11.96 \pm    0.28$ & $                ...$ & $                ...$ &   III & \\
...-S3-X27 & $    ...$ & $    ...$ & $    ...$ & $  >17.89  $ & $  >15.97$ & $  14.21 \pm    0.08$ & $  13.03 \pm    0.01$ & $  12.69 \pm    0.01$ & $  12.41 \pm    0.25$ & $  11.28 \pm    0.44$ & $                ...$ & II & \\
...-S3-X28 & $    ...$ & $    ...$ & $    ...$ & $  15.01 \pm    0.05$ & $  13.34 \pm    0.04$ & $  12.59 \pm    0.03$ & $  11.57 \pm    0.01$ & $  11.03 \pm    0.01$ & $  10.41 \pm    0.01$ & $   9.41 \pm    0.01$ & $   4.98 \pm    0.02$ &   II & \\
...-S3-X29 & $  17.66$ & $  16.22$ & $  14.97$ & $  13.21 \pm    0.03$ & $  12.36 \pm    0.03$ & $  12.12 \pm    0.03$ & $  11.85 \pm    0.01$ & $  11.78 \pm    0.02$ & $  11.22 \pm    0.08$ & $                ...$ & $                ...$ &   III & \\
...-S3-X30 & $  16.31$ & $  15.23$ & $  14.14$ & $  12.60 \pm    0.02$ & $  11.80 \pm    0.03$ & $  11.49 \pm    0.03$ & $  11.04 \pm    0.00$ & $  10.73 \pm    0.00$ & $  10.38 \pm    0.01$ & $   9.70 \pm    0.01$ & $   7.63 \pm    0.05$ &   II & \\
...-S3-X31 & $    ...$ & $    ...$ & $    ...$ & $  13.79 \pm    0.03$ & $  13.02 \pm    0.03$ & $  12.75 \pm    0.03$ & $  12.58 \pm    0.01$ & $  12.60 \pm    0.01$ & $  12.57 \pm    0.09$ & $  13.15 \pm    0.90$ & $                ...$ &   III & \\
...-S3-X32 & $  17.78$ & $  17.51$ & $  16.19$ & $  14.19 \pm    0.03$ & $  13.25 \pm    0.03$ & $  12.91 \pm    0.04$ & $  12.58 \pm    0.05$ & $                ...$ & $                ...$ & $                ...$ & $                ...$ &   III & \\
...-S3-X35 & $    ...$ & $    ...$ & $    ...$ & $  14.04 \pm    0.04$ & $  13.31 \pm    0.04$ & $  13.13 \pm    0.05$ & $  12.63 \pm    0.04$ & $  12.68 \pm    0.02$ & $                ...$ & $                ...$ & $                ...$ &   III & \\
...-S3-X36 & $    ...$ & $    ...$ & $    ...$ & $  14.18 \pm    0.05$ & $  13.18 \pm    0.06$ & $  12.56 \pm    0.04$ & $  11.66 \pm    0.02$ & $  11.37 \pm    0.02$ & $  11.02 \pm    0.19$ & $  10.05 \pm    0.39$ & $                ...$ &   II & \\
...-S3-X37 & $    ...$ & $    ...$ & $    ...$ & $  15.58 \pm    0.08$ & $  14.58 \pm    0.06$ & $  14.27 \pm    0.08$ & $  13.62 \pm    0.07$ & $  13.37 \pm    0.06$ & $                ...$ & $                ...$ & $                ...$ &   III & \\
...-S3-X38 & $    ...$ & $    ...$ & $    ...$ & $                ...$ & $                ...$ & $                ...$ & $                ...$ & $                ...$ & $                ...$ & $                ...$ & $                ...$ & ... & \\
...-S3-X39 & $    ...$ & $    ...$ & $    ...$ & $  13.60 \pm    0.04$ & $  >12.40$ & $  >11.97$ & $  11.24 \pm    0.01$ & $  10.91 \pm    0.01$ & $  10.55 \pm    0.09$ & $   9.74 \pm    0.22$ & $                ...$ & II & \\
...-S3-X40 & $    ...$ & $    ...$ & $    ...$ & $  14.99 \pm    0.05$ & $  13.98 \pm    0.05$ & $  13.39 \pm    0.05$ & $  13.13 \pm    0.05$ & $  13.05 \pm    0.05$ & $  12.35 \pm    0.38$ & $                ...$ & $                ...$ &   III & \\
...-S3-X41 & $    ...$ & $    ...$ & $    ...$ & $                ...$ & $                ...$ & $                ...$ & $                ...$ & $                ...$ & $                ...$ & $                ...$ & $                ...$ & ... & \\
...-S3-X42 & $    ...$ & $    ...$ & $    ...$ & $  15.06 \pm    0.05$ & $  14.29 \pm    0.07$ & $  13.89 \pm    0.08$ & $  13.90 \pm    0.05$ & $  13.72 \pm    0.05$ & $  13.62 \pm    0.06$ & $  13.43 \pm    0.16$ & $                ...$ &   III & \\
...-S3-X43 & $    ...$ & $    ...$ & $    ...$ & $  14.31 \pm    0.03$ & $  12.62 \pm    0.03$ & $  11.91 \pm    0.03$ & $  11.44 \pm    0.05$ & $  11.39 \pm    0.03$ & $  10.93 \pm    0.44$ & $                ...$ & $                ...$ &   III & \\
...-S3-X45 & $    ...$ & $    ...$ & $    ...$ & $  14.42 \pm    0.04$ & $  13.25 \pm    0.04$ & $  12.82 \pm    0.03$ & $  12.52 \pm    0.02$ & $  12.55 \pm    0.01$ & $  12.29 \pm    0.21$ & $  11.25 \pm    0.50$ & $                ...$ &   II & \\
...-S3-X47 & $    ...$ & $    ...$ & $    ...$ & $  >14.17$ & $  >13.38$ & $  13.20 \pm    0.05$ & $                ...$ & $                ...$ & $                ...$ & $                ...$ & $                ...$ & ... & \\
...-S3-X48 & $    ...$ & $    ...$ & $    ...$ & $  15.21 \pm    0.06$ & $  14.37 \pm    0.07$ & $  13.97 \pm    0.09$ & $  13.43 \pm    0.27$ & $  13.38 \pm    0.38$ & $                ...$ & $                ...$ & $                ...$ &   III & \\
...-S3-X50 & $    ...$ & $    ...$ & $    ...$ & $                ...$ & $                ...$ & $                ...$ & $  12.79 \pm    0.21$ & $  12.59 \pm    0.20$ & $                ...$ & $                ...$ & $                ...$ & ... & \\
...-S3-X51 & $    ...$ & $    ...$ & $    ...$ & $  11.24 \pm    0.04$ & $  10.56 \pm    0.04$ & $  >10.25$ & $   9.19 \pm    0.07$ & $   8.81 \pm    0.05$ & $                ...$ & $                ...$ & $                ...$ &   III & \\
...-S3-X52 & $    ...$ & $    ...$ & $    ...$ & $  11.00 \pm    0.02$ & $  10.71 \pm    0.03$ & $  10.58 \pm    0.02$ & $  10.52 \pm    0.00$ & $  10.52 \pm    0.00$ & $  10.27 \pm    0.02$ & $   9.73 \pm    0.07$ & $                ...$ &   II & \\
...-S3-X53 & $    ...$ & $    ...$ & $    ...$ & $                ...$ & $                ...$ & $                ...$ & $  14.88 \pm    0.02$ & $  14.22 \pm    0.02$ & $  13.43 \pm    0.05$ & $  12.62 \pm    0.05$ & $   8.41 \pm    0.07$ &   II & $\surd$ \\
...-S3-X56 & $    ...$ & $    ...$ & $    ...$ & $  10.02 \pm    0.02$ & $   9.73 \pm    0.03$ & $   9.69 \pm    0.02$ & $   9.69 \pm    0.00$ & $   9.69 \pm    0.00$ & $   9.64 \pm    0.00$ & $   9.64 \pm    0.01$ & $  10.09 \pm    0.18$ &   III & \\
...-S3-X58 & $    ...$ & $    ...$ & $    ...$ & $                ...$ & $                ...$ & $                ...$ & $                ...$ & $                ...$ & $                ...$ & $                ...$ & $                ...$ & ... & \\
...-S2-X1 & $  18.12$ & $  17.04$ & $  15.82$ & $  13. \pm    0.03$ & $  12.89 \pm    0.03$ & $  12.30 \pm    0.03$ & $  11.59 \pm    0.00$ & $  11.29 \pm    0.00$ & $  11.02 \pm    0.01$ & $  10.31 \pm    0.01$ & $   7.35 \pm    0.05$ &   II & \\
...-S2-X2 & $    ...$ & $    ...$ & $    ...$ & $                ...$ & $                ...$ & $                ...$ & $  17.12 \pm    0.23$ & $  16.15 \pm    0.12$ & $                ...$ & $                ...$ & $                ...$ & ... & \\
...-S2-X3 & $    ...$ & $    ...$ & $    ...$ & $  >16.66$ & $  14.76 \pm    0.08$ & $  13.60 \pm    0.05$ & $  12.81 \pm    0.01$ & $  12.60 \pm    0.01$ & $  12.46 \pm    0.02$ & $  12.32 \pm    0.06$ & $                ...$ & ... & \\
...-S2-X4 & $    ...$ & $    ...$ & $    ...$ & $  13.61 \pm    0.03$ & $  12.90 \pm    0.03$ & $  12.72 \pm    0.03$ & $  12.67 \pm    0.00$ & $  12.66 \pm    0.01$ & $  12.60 \pm    0.02$ & $  12.55 \pm    0.05$ & $                ...$ &   III & \\
...-S2-X5 & $  18.25$ & $  16.82$ & $  15.56$ & $  13.65 \pm    0.03$ & $  12.61 \pm    0.03$ & $  12.15 \pm    0.03$ & $  11.43 \pm    0.00$ & $  11.05 \pm    0.00$ & $  10.63 \pm    0.01$ & $  10.06 \pm    0.01$ & $   7.83 \pm    0.05$ &   II & \\
...-S2-X7 & $    ...$ & $    ...$ & $    ...$ & $                ...$ & $                ...$ & $                ...$ & $                ...$ & $                ...$ & $                ...$ & $                ...$ & $                ...$ & ... & \\
...-S2-X8 & $    ...$ & $    ...$ & $    ...$ & $                ...$ & $                ...$ & $                ...$ & $  16.53 \pm    0.06$ & $  15.46 \pm    0.04$ & $  14.34 \pm    0.11$ & $  13.33 \pm    0.10$ & $                ...$ &   0/I & $\surd$ \\
...-S2-X9 & $    ...$ & $    ...$ & $    ...$ & $  11.77 \pm    0.02$ & $  11.29 \pm    0.03$ & $  11.17 \pm    0.02$ & $  11.13 \pm    0.00$ & $  11.11 \pm    0.00$ & $  11.07 \pm    0.01$ & $  11.06 \pm    0.$ & $                ...$ &   III & \\
...-S2-X10 & $    ...$ & $    ...$ & $    ...$ & $                ...$ & $                ...$ & $                ...$ & $                ...$ & $                ...$ & $                ...$ & $                ...$ & $                ...$ & ... & \\
...-S2-X14 & $    ...$ & $    ...$ & $    ...$ & $                ...$ & $                ...$ & $                ...$ & $                ...$ & $                ...$ & $                ...$ & $                ...$ & $                ...$ & ... & \\
\hline \end{tabular}
\end{center}\end{sidewaystable*}

\begin{table*}\begin{center}
\caption{X-ray upper limits and 2\,MASS counterparts for Class 1 sources not detected in X-rays.}
\label{tab:xraytab_ul}
\begin{tabular}{lccrrrrrr}\hline
ID     &  \multicolumn{2}{c}{2\,MASS position}        & Offax      & Expo & Rate               & $\log{L_{\rm x}}$  \\
       &  $\alpha_{\rm 2000}$ & $\delta_{\rm 2000}$ & [$\prime$] & [s]  & [$10^{-3}$\,cts/s] & [erg/s]            \\ \hline
NGC7129-S3-U419  & 21 43 01.78 & +66 03 24.4 & $  3.5$ & $ 21865$ & $ <  0.2$ & $ <  29.7$ \\
NGC7129-S3-U546  & 21 42 57.75 & +66 04 23.5 & $  2.7$ & $ 22270$ & $ <  0.3$ & $ <  29.8$ \\
NGC7129-S3-U968  & 21 43 06.96 & +66 06 41.7 & $  0.2$ & $ 22640$ & $ <  0.1$ & $ <  29.5$ \\
NGC7129-S3-U1059 & 21 43 24.90 & +66 07 04.7 & $  1.7$ & $ 11355$ & $ <  0.3$ & $ <  29.8$ \\
NGC7129-S3-U1169 & 21 43 14.83 & +66 07 37.5 & $  1.0$ & $ 22349$ & $ <  0.2$ & $ <  29.7$ \\
NGC7129-S3-U1178 & 21 41 55.30 & +66 07 41.5 & $  7.4$ & $ 10077$ & $ <  1.4$ & $ <  30.5$ \\
NGC7129-S3-U1211 & 21 43 14.17 & +66 07 46.5 & $  1.1$ & $ 22328$ & $ <  0.2$ & $ <  29.7$ \\
NGC7129-S3-U1242 & 21 43 05.92 & +66 07 58.5 & $  1.1$ & $ 22367$ & $ <  0.1$ & $ <  29.5$ \\
NGC7129-S3-U1521 & 21 42 42.86 & +66 09 24.0 & $  3.6$ & $ 21139$ & $ <  0.2$ & $ <  29.7$ \\
NGC7129-S2-U1350 & 21 43 24.13 & +66 08 31.5 & $  2.3$ & $ 22573$ & $ <  0.3$ & $ <  29.8$ \\
\hline
NGC7129-S3-U270  & 21 42 59.82 & +66 01 54.9 & $  5.0$ & $ 20825$ & $ <  0.1$ & $ <  29.5$ \\
NGC7129-S3-U500  & 21 43 02.01 & +66 04 02.7 & $  2.9$ & $ 22274$ & $ <  0.1$ & $ <  29.5$ \\
NGC7129-S3-U550  & 21 43 11.17 & +66 04 25.6 & $  2.5$ & $ 22485$ & $ <  0.2$ & $ <  29.7$ \\
NGC7129-S3-U722  & 21 42 54.63 & +66 05 20.3 & $  2.0$ & $ 22520$ & $ <  0.3$ & $ <  29.8$ \\
NGC7129-S3-U815  & 21 42 51.43 & +66 05 56.4 & $  1.9$ & $ 22499$ & $ <  0.2$ & $ <  29.7$ \\
NGC7129-S3-U821  & 21 43 29.32 & +66 05 55.7 & $  2.4$ & $ 20607$ & $ <  0.2$ & $ <  29.5$ \\
NGC7129-S3-U822  & 21 43 04.38 & +66 05 56.4 & $  1.0$ & $ 22647$ & $ <  0.1$ & $ <  29.5$ \\
NGC7129-S3-U840  & 21 42 23.08 & +66 06 04.5 & $  4.6$ & $ 21643$ & $ <  0.6$ & $ <  30.1$ \\
NGC7129-S3-U849  & 21 43 12.30 & +66 06 05.6 & $  0.9$ & $ 22600$ & $ <  0.3$ & $ <  29.8$ \\
NGC7129-S3-U939  & 21 42 38.81 & +66 06 36.0 & $  2.9$ & $ 22113$ & $ <  0.2$ & $ <  29.7$ \\
NGC7129-S3-U1012 & 21 43 06.81 & +66 06 54.3 & $  0.1$ & $ 22600$ & $ <  0.6$ & $ <  30.1$ \\
NGC7129-S3-U1026 & 21 43 12.42 & +66 06 55.9 & $  0.5$ & $ 22507$ & $ <  0.2$ & $ <  29.7$ \\
NGC7129-S3-U1085 & 21 42 53.13 & +66 07 14.8 & $  1.5$ & $ 22465$ & $ <  0.3$ & $ <  29.8$ \\
NGC7129-S3-U1103 & 21 43 07.84 & +66 07 18.5 & $  0.5$ & $ 22477$ & $ <  0.1$ & $ <  29.5$ \\
NGC7129-S3-U1107 & 21 42 55.92 & +66 07 20.9 & $  1.3$ & $ 22470$ & $ <  0.5$ & $ <  30.1$ \\
NGC7129-S3-U1109 & 21 42 53.21 & +66 07 20.9 & $  1.6$ & $ 22454$ & $ <  0.5$ & $ <  30.1$ \\
NGC7129-S3-U1194 & 21 42 42.42 & +66 07 45.2 & $  2.7$ & $ 22107$ & $ <  0.2$ & $ <  29.7$ \\
NGC7129-S3-U1246 & 21 42 48.23 & +66 08 00.6 & $  2.3$ & $ 22214$ & $ <  0.4$ & $ <  30.0$ \\
NGC7129-S3-U1294 & 21 43 02.89 & +66 08 14.2 & $  1.5$ & $ 22316$ & $ <  0.1$ & $ <  29.5$ \\
NGC7129-S3-U1367 & 21 42 17.67 & +66 08 40.3 & $  5.4$ & $ 19816$ & $ <  0.5$ & $ <  30.0$ \\
NGC7129-S3-U1433 & 21 43 14.40 & +66 08 58.7 & $  2.2$ & $ 15537$ & $ <  0.2$ & $ <  29.7$ \\
NGC7129-S3-U1504 & 21 42 53.47 & +66 09 19.7 & $  2.9$ & $ 21725$ & $ <  0.2$ & $ <  29.7$ \\
NGC7129-S3-U1522 & 21 42 41.92 & +66 09 24.5 & $  3.7$ & $ 21067$ & $ <  0.1$ & $ <  29.5$ \\
NGC7129-S3-U1542 & 21 43 05.17 & +66 09 29.4 & $  2.6$ & $ 21816$ & $ <  0.1$ & $ <  29.5$ \\
NGC7129-S3-U1611 & 21 43 02.62 & +66 09 50.7 & $  3.0$ & $ 21560$ & $ <  0.3$ & $ <  29.8$ \\
NGC7129-S3-U1612 & 21 42 40.55 & +66 09 51.8 & $  4.1$ & $ 20281$ & $ <  0.5$ & $ <  30.0$ \\
NGC7129-S3-U1780 & 21 42 59.47 & +66 10 35.8 & $  3.8$ & $ 20598$ & $ <  0.3$ & $ <  29.9$ \\
NGC7129-S2-U613  & 21 43 48.70 & +66 04 46.2 & $  4.6$ & $ 21164$ & $ <  0.2$ & $ <  29.7$ \\
NGC7129-S2-U804  & 21 43 33.12 & +66 05 49.0 & $  2.8$ & $  9668$ & $ <  0.3$ & $ <  29.9$ \\
NGC7129-S2-U820  & 21 44 05.38 & +66 05 53.3 & $  5.9$ & $ 19261$ & $ <  0.6$ & $ <  30.1$ \\
NGC7129-S2-U1083 & 21 43 29.92 & +66 07 09.2 & $  2.3$ & $ 20393$ & $ <  0.2$ & $ <  29.7$ \\
NGC7129-S2-U1313 & 21 43 26.64 & +66 08 20.5 & $  2.4$ & $ 22559$ & $ <  0.1$ & $ <  29.5$ \\
NGC7129-S2-U1640 & 21 43 12.35 & +66 09 55.5 & $  3.1$ & $ 10420$ & $ <  0.3$ & $ <  29.8$ \\
NGC7129-S2-U1660 & 21 43 47.04 & +66 10 01.2 & $  5.1$ & $ 21507$ & $ <  0.4$ & $ <  29.9$ \\
NGC7129-S2-U1713 & 21 43 49.35 & +66 10 12.9 & $  5.4$ & $ 17909$ & $ <  0.4$ & $ <  29.9$ \\
NGC7129-S2-U2219 & 21 43 38.59 & +66 12 30.6 & $  6.5$ & $ 14592$ & $ <  0. 30.1$ \\
\hline
\multicolumn{7}{l}{(a) - The upper limits to the X-ray luminosity were computed using the median of $kT$ and $N_{\rm H}$} \\
\multicolumn{7}{l}{from the sample of X-ray detected Class\,II sources. These numbers may be underestimated if} \\
\multicolumn{7}{l}{the extinction is much higher, as is expected for the Class\,I sources (see Sect.~\ref{subsect:results_class1}).} \\
\end{tabular}\end{center}\end{table*}

\begin{sidewaystable*}\tiny\begin{center}
\caption{Candidate YSOs without X-ray emission selected on basis of their IR photometry. Asterisks in column `YSO Class' identify objects selected on basis of near-IR data.}
\label{tab:counterparts_ul}
\begin{tabular}{llrrrrrrrrrrrrcc}\hline
NGC7129-...  &  Name & $\Delta_{\rm O-S}$ & $V$   & $R$   & $I$   & $J$   & $H$   & $K$   & $[3.6]$ & $[4.5]$ & $[5.8]$ & $[8.0]$ & $[24]$ & YSO   & AGN? \\
      &       & [${\prime\prime}$] & [mag] & [mag] & [mag] & [mag] & [mag] & [mag] & [mag]   & [mag]   & [mag]   & [mag]   & [mag]  & Class &      \\ \hline
...-S3-U419  & & $   -$ &  $    ...$ & $    ...$ & $    ...$ & $                ...$ & $                ...$ & $                ...$ & $  13.02 \pm    0.02$ & $  10.41 \pm    0.01$ & $  10.17 \pm    0.01$ & $   9.08 \pm    0.01$ & $                ...$ & I & \\
...-S3-U546  & & $   -$ &  $    ...$ & $    ...$ & $    ...$ & $                ...$ & $  14.48 \pm    0.10$ & $  13.66 \pm    0.07$ & $  11.80 \pm    0.01$ & $  10.94 \pm    0.00$ & $   9.95 \pm    0.01$ & $   8.54 \pm    0.01$ & $   5.18 \pm    0.05$ & I & \\
...-S3-U968  & & $   -$ &  $    ...$ & $    ...$ & $    ...$ & $                ...$ & $  13.23 \pm    0.06$ & $  10.89 \pm    0.03$ & $   7.54 \pm    0.01$ & $   6.59 \pm    0.00$ & $   5.67 \pm    0.00$ & $   5.06 \pm    0.02$ & $                ...$ & I & \\
...-S3-U1059 & & $   -$ &  $    ...$ & $    ...$ & $    ...$ & $                ...$ & $                ...$ & $                ...$ & $  14.23 \pm    0.01$ & $  13.33 \pm    0.01$ & $  12.59 \pm    0.02$ & $  11.56 \pm    0.02$ & $   5.34 \pm    0.03$ & I & \\
...-S3-U1169 & & $   -$ &  $    ...$ & $    ...$ & $    ...$ & $                ...$ & $                ...$ & $  14.77 \pm    0.14$ & $  12.93 \pm    0.04$ & $  10.72 \pm    0.02$ & $  10.44 \pm    0.02$ & $  10.22 \pm    0.05$ & $   4.51 \pm    0.07$ & I & \\
...-S3-U1178 & & $   -$ &  $    ...$ & $    ...$ & $    ...$ & $                ...$ & $                ...$ & $                ...$ & $  15.86 \pm    0.05$ & $  14.90 \pm    0.03$ & $  13.98 \pm    0.08$ & $  12.78 \pm    0.05$ & $   9.49 \pm    0.15$ & I & $\surd$ \\
...-S3-U1211 & & $   -$ &  $    ...$ & $    ...$ & $    ...$ & $  17.48 \pm    0.37$ & $                ...$ & $  14.34 \pm    0.10$ & $  10.55 \pm    0.00$ & $   9.34 \pm    0.00$ & $   8.41 \pm    0.00$ & $   7.51 \pm    0.00$ & $   3.55 \pm    0.03$ & I & \\
...-S3-U1242 & & $   -$ &  $    ...$ & $    ...$ & $    ...$ & $                ...$ & $                ...$ & $                ...$ & $  13.78 \pm    0.01$ & $  12.70 \pm    0.01$ & $  12.10 \pm    0.06$ & $  11.65 \pm    0.11$ & $                ...$ & I & \\
...-S3-U1521 & & $   -$ &  $    ...$ & $    ...$ & $    ...$ & $                ...$ & $                ...$ & $                ...$ & $  16.23 \pm    0.05$ & $  15.25 \pm    0.04$ & $  13.84 \pm    0.09$ & $  12.79 \pm    0.17$ & $                ...$ & I & $\surd$ \\
...-S2-U1350 & & $   -$ &  $    ...$ & $    ...$ & $    ...$ & $                ...$ & $                ...$ & $                ...$ & $  12.86 \pm    0.01$ & $  11.28 \pm    0.01$ & $  10.05 \pm    0.01$ & $   8.96 \pm    0.01$ & $   4.32 \pm    0.00$ & I & \\
\hline 
...-S3-U270 & & $   -$ & $    ...$ & $    ...$ & $    ...$ & $                ...$ & $                ...$ & $                ...$ & $  16.83 \pm    0.06$ & $  16.59 \pm    0.08$ & $  15.91 \pm    0.26$ & $  14.33 \pm    0.14$ & $                ...$ & II & \\
...-S3-U500 & & $   -$ & $    ...$ & $    ...$ & $    ...$ & $                ...$ & $                ...$ & $                ...$ & $  13.64 \pm    0.01$ & $  12.98 \pm    0.01$ & $  12.18 \pm    0.04$ & $  11.29 \pm    0.07$ & $                ...$ & II & \\
...-S3-U550 & & $   -$ & $    ...$ & $    ...$ & $    ...$ & $                ...$ & $                ...$ & $                ...$ & $  13.91 \pm    0.01$ & $  13.42 \pm    0.01$ & $  12.98 \pm    0.08$ & $  12.33 \pm    0.18$ & $                ...$ & II & \\
...-S3-U722 & & $   -$ & $    ...$ & $    ...$ & $    ...$ & $  15.52 \pm    0.08$ & $  14.56 \pm    0.08$ & $  13.87 \pm    0.08$ & $  12.51 \pm    0.14$ & $  12.64 \pm    0.14$ & $                ...$ & $                ...$ & $                ...$ & II$^*$ & \\
...-S3-U815 & MMN-5 & $1.72$ & $  20.35$ & $  18.77$ & $  17.28$ & $  15.20 \pm    0.06$ & $  14.13 \pm    0.06$ & $  13.57 \pm    0.05$ & $  12.62 \pm    0.04$ & $  12.09 \pm    0.03$ & $  11.52 \pm    0.26$ & $  10.33 \pm    0.49$ & $                ...$ & II & \\
...-S3-U821 &  & $   -$ & $    ...$ & $    ...$ & $    ...$ & $  15.88 \pm    0.10$ & $  14.96 \pm    0.09$ & $  14.39 \pm    0.10$ & $  13.47 \pm    0.01$ & $  13.04 \pm    0.01$ & $  12.52 \pm    0.02$ & $  11.78 \pm    0.02$ & $   8.95 \pm    0.15$ & II & \\
...-S3-U822 &  & $   -$ & $    ...$ & $    ...$ & $    ...$ & $  15.70 \pm    0.10$ & $  15.12 \pm    0.12$ & $  14.47 \pm    0.12$ & $                ...$ & $  12.55 \pm    0.20$ & $                ...$ & $                ...$ & $                ...$ & II$^*$ & \\
...-S3-U840 & MMN-1 & $1.42$ & $  20.11$ & $  19.02$ & $  17.52$ & $  15.06 \pm    0.05$ & $  14.05 \pm    0.05$ & $  13.39 \pm    0.05$ & $  12.51 \   0.01$ & $  12.14 \pm    0.01$ & $  11.78 \pm    0.06$ & $  11.02 \pm    0.14$ & $                ...$ & II & \\
...-S3-U849 &  & $   -$ & $    ...$ & $    ...$ & $    ...$ & $  15.07 \pm    0.05$ & $  14.34 \pm    0.06$ & $  13.94 \pm    0.07$ & $  13.36 \pm    0.02$ & $  13.05 \pm    0.02$ & $  12.48 \pm    0.18$ & $                ...$ & $                ...$ & II$^*$ & \\
...-S3-U939 & HL85-S\,14 & $1.79$ & $  18.67$ & $  17.66$ & $  16.45$ & $  14.80 \pm    0.05$ & $  13.48 \pm    0.04$ & $  12.51 \pm    0.03$ & $  11.01 \pm    0.01$ & $  10.59 \pm    0.00$ & $  10.42 \pm    0.04$ & $  10.19 \pm    0.16$ & $                ...$ & II$^*$ & \\
...-S3-U1012 & SVS\,12   & $1.83$ & $    ...$ & $    ...$ & $    ...$ & $   9.53 \pm    0.02$ & $   8.20 \pm    0.02$ & $   7.08 \pm    0.02$ & $   5.79 \pm    0.00$ & $   5.15 \pm    0.00$ & $   4.45 \pm    0.00$ & $   3.33 \pm    0.03$ & $                ...$ & II & \\
...-S3-U1026 &  & $   -$ & $    ...$ & $    ...$ & $    ...$ & $                ...$ & $                ...$ & $                ...$ & $  14.16 \pm    0.07$ & $  13.73 \pm    0.07$ & $  13.25 \pm    0.27$ & $  12.23 \pm    0.$ & $                ...$ & II & \\
...-S3-U1085 & MMN-7 & $1.33$ & $  20.58$ & $  19.66$ & $  18.24$ & $  15.58 \pm    0.08$ & $  14.16 \pm    0.05$ & $  12.93 \pm    0.04$ & $  11.61 \pm    0.01$ & $  10.93 \pm    0.01$ & $  10.41 \pm    0.07$ & $   9.57 \pm    0.11$ & $                ...$ & II & \\
...-S3-U1103 &  & $   -$ & $    ...$ & $    ...$ & $    ...$ & $  16.72 \pm    0.21$ & $  14.85 \pm    0.09$ & $  13.94 \pm    0.08$ & $  13.04 \pm    0.06$ & $  12.57 \pm    0.06$ & $  11.76 \pm    0.21$ & $  10.88 \pm    0.40$ & $                ...$ & II & \\
...-S3-U1107 &  & $   -$ & $    ...$ & $    ...$ & $    ...$ & $  14.60 \pm    0.04$ & $  13.29 \pm    0.04$ & $  12.47 \pm    0.03$ & $  11.16 \pm    0.02$ & $  10.54 \pm    0.01$ & $   9.98 \pm    0.08$ & $   9.02 \pm    0.17$ & $                ...$ & II & \\
...-S3-U1109 &  & $   -$ & $    ...$ & $    ...$ & $    ...$ & $  14.34 \pm    0.04$ & $  13.19 \pm    0.05$ & $  12.68 \pm    0.05$ & $  11.62 \pm    0.02$ & $  11.29 \pm    0.01$ & $  10.82 \pm    0.12$ & $   9.92 \pm    0.29$ & $                ...$ & II & \\
...-S3-U1194 &  & $   -$ & $    ...$ & $    ...$ & $    ...$ & $  15.58 \pm    0.07$ & $  14.23 \pm    0.05$ & $  13.45 \pm    0.05$ & $  12.56 \pm    0.02$ & $  12.21 \pm    0.01$ & $  12.23 \pm    0.18$ & $  11.68 \pm    0.56$ & $                ...$ & II & \\
...-S3-U1246 &  & $   -$ & $    ...$ & $    ...$ & $    ...$ & $  15.29 \pm    0.06$ & $  13.89 \pm    0.04$ & $  13.08 \pm    0.04$ & $  11.69 \pm    0.01$ & $  11.14 \pm    0.00$ & $  10.62 \pm    0.03$ & $   9.67 \pm    0.06$ & $                ...$ & II & \\
...-S3-U1294 &  & $   -$ & $    ...$ & $    ...$ & $    ...$ & $  16.34 \pm    0.14$ & $  15.42 \pm    0.13$ & $  15.01 \pm    0.16$ & $  14.15 \pm    0.01$ & $  13.83 \pm    0.01$ & $  13.58 \pm    0.12$ & $  13.43 \pm    0.48$ & $                ...$ & II$^*$ & \\
...-S3-U1367 &  & $   -$ & $    ...$ & $    ...$ & $    ...$ & $  16.17 \pm    0.12$ & $  15.02 \pm    0.11$ & $  14.63 \pm    0.10$ & $  13.68 \pm    0.01$ & $  13.23 \pm    0.01$ & $  12.91 \pm    0.04$ & $  12.40 \pm    0.04$ & $   9.29 \pm    0.13$ & II & \\
...-S3-U1433 &  & $   -$ & $    ...$ & $    ...$ & $    ...$ & $  16.03 \pm    0.10$ & $  15.30 \pm    0.12$ & $  14.93 \pm    0.16$ & $  14.23 \pm    0.01$ & $  13.89 \pm    0.01$ & $  13.59 \pm    0.05$ & $  12.98 \pm    0.05$ & $                ...$ & II & \\
...-S3-U1504 & MMN-8 & $1.57$ & $  21.06$ & $  19.99$ & $  19.01$ & $  16.99 \pm    0.25$ & $  15.50 \pm    0.14$ & $  14.58 \pm    0.12$ & $  12.91 \pm    0.01$ & $  12.36 \pm    0.01$ & $  11.87 \pm    0.02$ & $  11.23 \pm    0.04$ & $   8.47 \pm    0.20$ & II & \\
...-S3-U1522 &  & $   -$ &  $    ...$ & $    ...$ & $    ...$ & $  15.18 \pm    0.05$ & $  14.34 \pm    0.06$ & $  13.99 \pm    0.07$ & $  13.56 \pm    0.01$ & $  13.29 \pm    0.01$ & $  13.08 \pm    0.04$ & $  12.30 \pm    0.11$ & $                ...$ & II & \\
...-S3-U1542 &  & $   -$ &  $    ...$ & $    ...$ & $    ...$ & $  15.79 \pm    0.09$ & $  15.16 \pm    0.12$ & $  14.90 \pm    0.15$ & $  14.23 \pm    0.01$ & $  14.07 \pm    0.01$ & $  13.70 \pm    0.04$ & $  13.11 \pm    0.07$ & $                ...$ & II & \\
...-S3-U1611 &  & $   -$ &  $    ...$ & $    ...$ & $    ...$ & $  15.71 \pm    0.09$ & $  14.79 \pm    0.08$ & $  14.20 \pm    0.09$ & $  13.54 \pm    0.01$ & $  13.16 \pm    0.01$ & $  12.65 \pm    0.02$ & $  11.93 \pm    0.03$ & $   8.86 \pm    0.12$ & II & \\
...-S3-U1612 &  & $   -$ &  $    ...$ & $    ...$ & $    ...$ & $                ...$ & $                ...$ & $  14.86 \pm    0.15$ & $  13.12 \   0.01$ & $  12.33 \pm    0.01$ & $  11.54 \pm    0.02$ & $  10.48 \pm    0.$ & $   6.01 \pm    0.07$ & II & \\
...-S3-U1780 &  & $   -$ &  $    ...$ & $    ...$ & $    ...$ & $                ...$ & $  16.05 \pm    0.23$ & $  14.61 \pm    0.12$ & $  13.54 \pm    0.01$ & $  13.00 \pm    0.01$ & $  12.52 \pm    0.02$ & $  12.05 \pm    0.04$ & $                ...$ & II & \\
...-S2-U613 &  & $   -$ & $    ...$ & $    ...$ & $    ...$ & $  14.88 \pm    0.05$ & $  13.65 \pm    0.05$ & $  13.08 \pm    0.04$ & $  12.40 \pm    0.00$ & $  12.15 \pm    0.01$ & $  11.83 \pm    0.01$ & $  10.87 \pm    0.01$ & $   7.57 \pm    0.03$ & II & \\
...-S2-U804 &  & $   -$ & $    ...$ & $    ...$ & $    ...$ & $                ...$ & $                ...$ & $                ...$ & $  16.69 \pm    0.05$ & $  15.83 \pm    0.04$ & $  14.92 \pm    0.09$ & $  13.37 \pm    0.07$ & $   9.14 \pm    0.11$ & II & $\surd$ \\
...-S2-U820 &  & $   -$ & $    ...$ & $    ...$ & $    ...$ & $  13.69 \pm    0.03$ & $  12.15 \pm    0.03$ & $  11.22 \pm    0.03$ & $   9.83 \pm    0.00$ & $   9.33 \pm    0.00$ & $   8.92 \pm    0.00$ & $   8.16 \pm    0.00$ & $   4.72 \pm    0.00$ & II & \\
...-S2-U1083 &  & $   -$ & $    ...$ & $    ...$ & $    ...$ & $                ...$ & $                ...$ & $                ...$ & $  14.67 \   0.01$ & $  14.24 \pm    0.02$ & $  13.99 \pm    0.08$ & $  13.38 \pm    0.08$ & $                ...$ & II & $\surd$ \\
...-S2-U1313 &  & $   -$ &  $    ...$ & $    ...$ & $    ...$ & $                ...$ & $                ...$ & $                ...$ & $  14.66 \   0.02$ & $  14.19 \pm    0.02$ & $  13.79 \pm    0.05$ & $  13.24 \pm    0.06$ & $                ...$ & II & \\
...-S2-U1640 &  & $   -$ &  $    ...$ & $    ...$ & $    ...$ & $  14.81 \pm    0.04$ & $  13.90 \pm    0.05$ & $  13.59 \pm    0.05$ & $  13.00 \pm    0.01$ & $  12.74 \pm    0.01$ & $  12.39 \pm    0.02$ & $  11.85 \pm    0.03$ & $   9.43 \pm    0.17$ & II & \\
...-S2-U1660 &  & $   -$ &  $    ...$ & $    ...$ & $    ...$ & $  15.40 \pm    0.06$ & $  14.70 \pm    0.09$ & $  14.19 \pm    0.09$ & $  14.05 \pm    0.01$ & $  14.03 \pm    0.02$ & $  13.72 \pm    0.09$ & $  13.00 \pm    0.17$ & $                ...$ & II & \\
...-S2-U1713 &  & $   -$ &  $    ...$ & $    ...$ & $    ...$ & $  15.86 \pm    0.10$ & $  15.07 \pm    0.11$ & $  14.86 \pm    0.15$ & $  14.53 \   0.01$ & $  14.47 \pm    0.02$ & $  14.29 \pm    0.13$ & $  13.36 \pm    0.22$ & $                ...$ & II & \\
...-S2-U2219 &  & $   -$ &  $    ...$ & $    ...$ & $    ...$ & $                ...$ & $  14.59 \pm    0.09$ & $                ...$ & $  13.58 \pm    0.01$ & $  13.34 \pm    0.01$ & $  12.80 \pm    0.04$ & $  11.99 \pm    0.05$ & $   9.47 \pm    0.14$ & II & \\
\hline \end{tabular}
\end{center}\end{sidewaystable*}

\begin{table}\tiny\begin{center}
\caption{Composition of the sub-samples described in Sect.~\ref{subsect:census_spatial}.}
\label{tab:samples_flag}
\begin{tabular}{lccc}\hline
NGC7129-...   &  $M>0.5\,M_\odot$ & $M>0.5\,M_\odot$     & $M>0.5\,M_\odot$     \\
       &                   & in core              & in core              \\
       &                   &                      & $A_{\rm V} < 5$\,mag \\ \hline
...-S3-X1 & $\surd$ &        &        \\
...-S3-X2 & $\surd$ &        &        \\
...-S3-X3 & $\surd$ &        &        \\
...-S3-X4 & $\surd$ &        &        \\
...-S3-X6 & $\surd$ & $\surd$ & $\surd$ \\
...-S3-X9 & $\surd$ &        &        \\
...-S3-X10 & $\surd$ & $\surd$ & $\surd$ \\
...-S3-X11 & $\surd$ & $\surd$ & $\surd$ \\
...-S3-X13 & $\surd$ & $\surd$ & $\surd$ \\
...-S3-X14 & $\surd$ &        &        \\
...-S3-X15 & $\surd$ & $\surd$ & $\surd$ \\
...-S3-X16 & $\surd$ & $\surd$ & $\surd$ \\
...-S3-X17 & $\surd$ & $\surd$ & $\surd$ \\
...-S3-X18 & $\surd$ & $\surd$ & $\surd$ \\
...-S3-X19 & $\surd$ & $\surd$ & $\surd$ \\
...-S3-X20 & $\surd$ & $\surd$ & $\surd$ \\
...-S3-X21 & $\surd$ &        &        \\
...-S3-X22 & $\surd$ & $\surd$ &        \\
...-S3-X23 & $\surd$ & $\surd$ & $\surd$ \\
...-S3-X25 & $\surd$ & $\surd$ &        \\
...-S3-X26 & $\surd$ & $\surd$ & $\surd$ \\
...-S3-X28 & $\surd$ &        &        \\
...-S3-X29 & $\surd$ & $\surd$ & $\surd$ \\
...-S3-X30 & $\surd$ &        &        \\
...-S3-X31 & $\surd$ & $\surd$ & $\surd$ \\
...-S3-X32 & $\surd$ & $\surd$ & $\surd$ \\
...-S3-X35 & $\surd$ & $\surd$ & $\surd$ \\
...-S3-X36 & $\surd$ & $\surd$ & $\surd$ \\
...-S3-X40 & $\surd$ & $\surd$ & $\surd$ \\
...-S3-X43 & $\surd$ & $\surd$ &        \\
...-S3-X45 & $\surd$ & $\surd$ &        \\
...-S3-X51 & $\surd$ & $\surd$ & $\surd$ \\
...-S3-X52 & $\surd$ &        &        \\
...-S3-X56 & $\surd$ &        &        \\
...-S2-X1 & $\surd$ &        &        \\
...-S2-X4 & $\surd$ &        &        \\
...-S2-X5 & $\surd$ &        &        \\
...-S2-X9 & $\surd$ &        &        \\
...-S3-U815 & $\surd$ & $\surd$ & $\surd$ \\
...-S3-U840 & $\surd$ &        &        \\
...-S3-U939 & $\surd$ & $\surd$ &        \\
...-S3-U1012 & $\surd$ & $\surd$ &        \\
...-S3-U1085 & $\surd$ & $\surd$ &        \\
...-S3-U1103 & $\surd$ & $\surd$ &        \\
...-S3-U1107 & $\surd$ & $\surd$ &        \\
...-S3-U1109 & $\surd$ & $\surd$ & $\surd$ \\
...-S3-U1194 & $\surd$ & $\surd$ &        \\
...-S3-U1246 & $\surd$ & $\surd$ &        \\
...-S2-U613 & $\surd$ &        &        \\
...-S2-U820 & $\surd$ &        &        \\
...-S2-U1640 & $\surd$ &        &        \\
\hline
 \end{tabular}\end{center}\end{table}

\begin{acknowledgements}
We would like to thank R.Gutermuth for having provided a table with the 
complete {\em Spitzer} photometry of NGC\,7129 and his YSO classification prior
to publication. We appreciate the constructive comments of the anonymous referee that have
improved the manuscript. 
The authors have received partial funding from the FP\,6 Marie Curie Research
Training Network {\em Constellation} under contract MRTN-CT-2006-035890. 
This paper is based on data obtained with {\em Chandra}, {\em Spitzer}, and 2\,MASS. 
BS wishes to thank E.Flaccomio, F.Albacete-Colombo and M.Caramazza for stimulating discussions.
\end{acknowledgements}

\bibliographystyle{aa} 
\bibliography{stelzer}

\end{document}